\documentclass[aps,prb,letterpaper,superscriptaddress,twocolumn,showpacs,floatfix,10pt]{revtex4-1}
\date{\today}
\usepackage{amsfonts}
\usepackage{amsmath}
\usepackage{amssymb}
\usepackage{graphicx}
\usepackage{epstopdf}
\usepackage{epsfig}
\usepackage{color}
\usepackage{mathtools}
\usepackage{hyperref}
\usepackage{tabularx}
\usepackage{subfigure}
\usepackage{anyfontsize}
\usepackage{placeins}
\usepackage{tikz}
\usepackage{bbm}
\usepackage{enumerate}
\usepackage[inline,final]{showlabels}
\usetikzlibrary{shapes,positioning,arrows,calc,intersections,through,patterns}
\showlabels{bibitem}

\begin{document}
\newcommand{\op}[1]{\operatorname{#1}}
\newcommand{\cchip}[2]{\chi\nobreak\!=\nobreak\!#1, \,\chi'\nobreak\!=\nobreak\!#2}
\newcommand{\aparam}[2]{a_1\nobreak\!=\nobreak\!#1, \,a_2\nobreak\!=\nobreak\!#2}
\newcommand{\up}{\uparrow}
\newcommand{\down}{\downarrow}
\newcommand{\rev}[1]{{#1}}
\def\bra#1{\langle#1\vert}
\def\ket#1{\vert#1\rangle}
\def\ketbra#1{\vert#1\rangle\langle#1\vert}

\newcommand \entwidth{4cm}
\newcommand \entheight{.8cm}
\newcommand \isovheight{1.0cm}
\newcommand \isovwidth{1.7cm}
\newcommand \isomheight{0.7cm}
\newcommand \isomwidth{2.3cm}
\newcommand \halfheight{.9cm}
\tikzstyle{site}=[circle,draw,fill=blue!20, inner sep=0pt,minimum size=10mm]
\tikzstyle{qop}=[circle,draw,fill=black!60!blue, inner sep=0pt,minimum size=3.5mm]
\tikzstyle{siteq}=[circle,draw,fill=blue!20, inner sep=0pt,minimum size=7mm]
\tikzstyle{sitehalf}=[circle,draw,fill=red!40, inner sep=0pt,minimum size=17mm]
\tikzstyle{sitenext}=[circle,draw,fill=blue!40, inner sep=0pt,minimum size=20mm]
\tikzstyle{sitec}=[circle,draw,fill=blue!40, inner sep=0pt,minimum size=10mm]
\tikzstyle{sitenimp}=[circle,preaction={fill=blue!40},draw, inner sep=0pt,minimum size=10mm,pattern=north west lines, pattern color=black!60!blue]
\tikzstyle{MPS}=[rectangle,draw,fill=violet, minimum size=15mm]
\tikzstyle{disentangler}=[trapezium,fill=cyan!40!blue!10,draw=cyan!75,trapezium angle=75,minimum height=\entheight,minimum width=\entwidth,trapezium stretches]
\tikzstyle{entangler}=[trapezium,fill=cyan!40!green!10,draw=cyan!75,trapezium angle=105,minimum height=\entheight,minimum width=\entwidth,trapezium stretches]
\tikzstyle{isovL}=[isosceles triangle,fill=red!40!magenta!20,draw=magenta!75,minimum height=\isovheight,minimum width=\isovwidth,isosceles triangle stretches, shape border uses incircle, shape border rotate=150]
\tikzstyle{isovR}=[isosceles triangle,fill=violet!40,draw=purple!75,minimum height=\isovheight,minimum width=\isovwidth,isosceles triangle stretches, shape border uses incircle, shape border rotate=30]
\tikzstyle{isovLc}=[isosceles triangle,fill=red!20!magenta!40,draw=magenta!75,minimum height=\isovheight,minimum width=\isovwidth,isosceles triangle stretches, shape border uses incircle, shape border rotate=-150]
\tikzstyle{isovRc}=[isosceles triangle,fill=violet!60,draw=purple!75,minimum height=\isovheight,minimum width=\isovwidth,isosceles triangle stretches, shape border uses incircle, shape border rotate=-30]
\tikzstyle{isovLf}=[isosceles triangle,fill=red!20!magenta!40,draw=magenta!75,minimum height=\isovheight,minimum width=\isovwidth,isosceles triangle stretches, shape border uses incircle, shape border rotate=-30]
\tikzstyle{isovRf}=[isosceles triangle,fill=violet!60,draw=purple!75,minimum height=\isovheight,minimum width=\isovwidth,isosceles triangle stretches, shape border uses incircle, shape border rotate=210]
\tikzstyle{isovLcf}=[isosceles triangle,fill=red!40!magenta!20,draw=magenta!75,minimum height=\isovheight,minimum width=\isovwidth,isosceles triangle stretches, shape border uses incircle, shape border rotate=30]
\tikzstyle{isovRcf}=[isosceles triangle,fill=violet!40,draw=purple!75,minimum height=\isovheight,minimum width=\isovwidth,isosceles triangle stretches, shape border uses incircle, shape border rotate=150]
\tikzstyle{sitehalfleft}=[semicircle,fill=green!20!blue!30,draw=blue!75,minimum size=.9cm, shape border rotate=90]
\tikzstyle{sitehalfright}=[semicircle,fill=blue!20!green!30,draw=blue!75,minimum size=.9cm, shape border rotate=270]
\tikzstyle{isometry}=[isosceles triangle,fill=blue!60!violet!20,draw=blue!75,minimum height=\isomheight,minimum width=\isomwidth,isosceles triangle stretches, shape border uses incircle, shape border rotate=90]
\tikzstyle{isometryconj}=[isosceles triangle,fill=blue!40!violet!40,draw=blue!75,minimum height=\isomheight,minimum width=\isomwidth,isosceles triangle stretches, shape border uses incircle, shape border rotate=270]
\tikzstyle{sitetopleft}=[semicircle,fill=red!20!orange!30,draw=blue!75,minimum size=\halfheight, shape border uses incircle, shape border rotate=45]
\tikzstyle{sitetopright}=[semicircle,fill=orange!20!red!30,draw=blue!75,minimum size=\halfheight, shape border uses incircle, shape border rotate=315]
\tikzstyle{sitebotleft}=[semicircle,fill=red!30!orange!20,draw=blue!75,minimum size=\halfheight, shape border uses incircle, shape border rotate=135]
\tikzstyle{sitebotright}=[semicircle,fill=orange!30!red!20,draw=blue!75,minimum size=\halfheight, shape border uses incircle, shape border rotate=225]
\tikzstyle{sitehalftop}=[semicircle,fill=green!30!blue!20,draw=blue!75,minimum size=\halfheight, shape border rotate=0]
\tikzstyle{sitehalfbot}=[semicircle,fill=blue!30!green!20,draw=blue!75,minimum size=\halfheight, shape border rotate=180]
\tikzstyle{leftisom}=[isosceles triangle,fill=blue!60!green!20,draw=blue!75,minimum height=\isomheight,minimum width=\isomwidth,isosceles triangle stretches, shape border uses incircle, shape border rotate=0]
\tikzstyle{rightisom}=[isosceles triangle,fill=blue!40!green!40,draw=blue!75,minimum height=\isomheight,minimum width=\isomwidth,isosceles triangle stretches, shape border uses incircle, shape border rotate=180]

\title{ Phase transitions of a 2D deformed-AKLT model  }

\author{Nicholas Pomata}
 \affiliation{C. N. Yang Institute for Theoretical Physics and Department of Physics and Astronomy, State University of New York at Stony Brook, NY 11794-3840, United States}

\author{Ching-Yu Huang}
 \affiliation{C. N. Yang Institute for Theoretical Physics and Department of Physics and Astronomy, State University of New York at Stony Brook, NY 11794-3840, United States}  
 \affiliation{Physics Division, National Center for Theoretical Science, Hsinchu 30013, Taiwan}

\author{Tzu-Chieh Wei}
 \affiliation{C. N. Yang Institute for Theoretical Physics and Department of Physics and Astronomy, State University of New York at Stony Brook, NY 11794-3840, United States}  

\vfill
\begin{abstract}
We study spin-2 deformed-AKLT models on the square lattice, 
specifically a two-parameter family
of $O(2)$-symmetric ground-state wavefunctions as defined by
Niggemann, Kl\"umper, and Zittartz,
who found previously that the phase diagram consists of a N\'eel-ordered phase
and a disordered phase which contains the AKLT point. Using
tensor-nework methods, we not only confirm the N\'eel phase but also find an
XY phase with quasi-long-range order and a region adjacent to it, within
the AKLT phase, with very large correlation length, and investigate the
consequences of a perfectly-factorizable point at the corner of that
phase.
\end{abstract}

\maketitle

%%%%%%%%%%%%%%%%%%%%%%%%%%
\section{ introduction}
\label{sec:intro}
Haldane's prediction concerning the finite spectral gap of the 1D
integer-spin antiferromagnetic Heisenberg chain and its featureless ground state
was quite unexpected\cite{Haldane1D}, as it seemed incompatible with the
thereom of Lieb, Schultz and Mattis \cite{LSM} on half-integer spin chains, with
the difference coming from the presence or absence of
a topological $\theta$ term. In order to understand the integer-spin case,
Affleck, Kennedy, Lieb and Tasaki (AKLT) constructed a state which admits no
local order parameter and which is the exact ground state of a
Hamiltonian whose finite gap can be proven rigorously.\cite{AKLT_PRL,AKLT_1988}
Their construction, which used valence bonds, was then generalized to two
dimensions, for example on the honeycomb and square lattices.
Recently, these two-dimensional valence-bond AKLT states and parent
Hamiltonians have also been recognized as examples of systems with
weak symmetry-protected topological order\cite{CZX,AKLTstrange,SREstrange},
and, somewhat unexpectedly, as a means to realize universal quantum
computation in a measurement-based approach.\cite{AKLT_QC_honeycomb,AKLT_QC_square}

Here we study a two-parameter family of wave functions on the square lattice
as constructed by Niggemann, Kl\"umper, and Zittartz (NKZ)
\cite{NKZspin2} and investigate the corresponding phase diagram. 
These wave functions, which contain the AKLT state as a special case,
are ground states of a class of two-site interacting frustration-free
spin-2
Hamiltonians on the square lattice, which have spin-flip and rotation symmetry
in the $z$ direction and are symmetric under lattice rotations, translations,
and reflections.\cite{NKZspin2}
We shall refer to these states as the ``deformed-AKLT'' family of states,
as they may be obtained by applying an on-site deformation to the AKLT state.

In their original work, through a combination of Monte-Carlo analysis and
approximation by an exactly-solvable classical model, Niggemann,
Kl\"umper, and Zittartz predicted that an
Ising-like transition divides
the two-parameter phase diagram into a N\'eel-ordered phase and a disordered
phase. This matches their result for the spin-$\frac{3}{2}$ model,
which Hieida et al.\cite{Hieida} further confirmed by applying a progenitor
of the CTMRG approach that we describe in Appendix~\ref{app:CTMRG}.
As in the preceding work by Huang, Wagner,
and Wei\cite{AKLTspin32},
we apply tensor-network analyses to this system in order to better 
understand and characterize these phases. In addition,
following the evidence for an 
XY-like phase in that work, we seek to determine whether or not the disordered
``phase'' further divides into multiple phases,
which we strongly expect to find since the
disordered region of the phase diagram contains both the AKLT point,
which possesses symmetry-protected topological (SPT) order,
in its interior, and a product state at its boundary.

Among the phases, the ordered phase can be easily characterized by spontaneous
symmetry breaking using a staggered $S_z$ as a local order parameter; as this
order parameter can be directly calculated by tensor-network methods,
we can accurately locate the boundary between this phase and the
AKLT phase. For the featureless valence-bond AKLT phase,
we can use simulated modular $S$ and $T$ matrices to distinguish its SPT order
from other phases.\cite{tnST,LevinGuSPT,HungWenSPT}
We also isolate, in a region surrounding the product-state
point at the origin of the parameter space, a critical phase with distinctive
properties that we can examine in terms of the conformal field theory of the
classical XY model.  By doing so we reveal
robust evidence for the existence of such a phase and for a Kosterlitz-Thouless
transition between it and the SPT-ordered AKLT phase. This is in contrast to
the spin-3/2 case\cite{AKLTspin32}, which new evidence
 presented in Sec.~\ref{sec:honeycomb}
suggests does not contain a truly critical, or quasi-long-range ordered,
XY phase, but instead only has a region of
very long correlation length. Such pseudo-quasi-long-range order also exists on
the square lattice, in a region of the AKLT phase adjacent to the
true XY phase.  Its existence is related to the suppression of $S_z=\pm 2$
components in this region, resulting in approximate spin-1
behavior for which the Berry
phase from the topological $\theta$ term almost suppresses isolated
tunneling processes.\cite{Haldane2D}

\begin{figure}[t]
\includegraphics[width=0.5\textwidth]{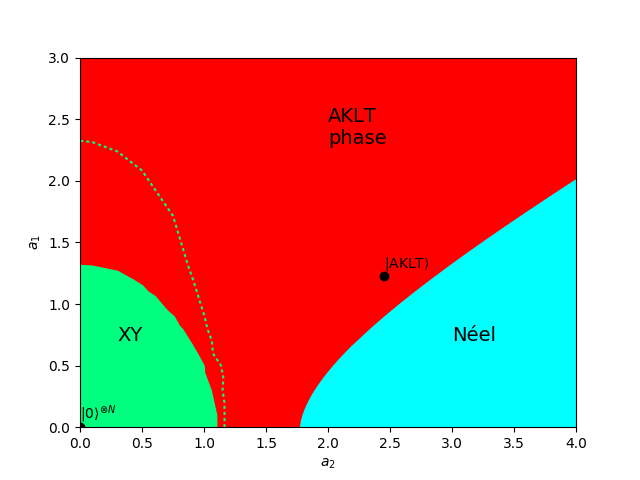}
\caption{The phase diagram of the square-lattice deformed-AKLT model with
deformation parameterized by $a_2$ and $a_1$ as given in
\eqref{eqn:deformation-definition}.
N\'eel indicates the N\'eel-ordered phase, with boundary determined as in
Fig.~\ref{fig:curveVBSNeel}; XY indicates the XY-like phase with
quasi-long-range order, with boundary estimated by interpolating from the
data in Fig.~\ref{fig:XYmaps20}b; and AKLT indicates the AKLT phase, with
the isotropic AKLT point indicated as $|\text{AKLT}\rangle$. Likewise the
product state at the origin of parameter space is noted as
$|0^{\otimes N}\rangle$.
The green dotted line demarks the pseudo-quasi-long-range-ordered region;
points on this line have correlation length $\xi \sim 10^3$
estimated from TNR data by interpolating the parameter where the classical
central charge takes the value $c \simeq 0.35$ after 10 RG steps, as indicated
by Fig.~\ref{fig:VBScorr}.
}
\label{fig:squarephase}
\end{figure}

We  also examine the possibility of a third disordered phase, a trivial phase
adiabatically connected to the product state at the origin
$a_1=a_2=0$ of the two-parameter
space (as shown in  Fig.~\ref{fig:squarephase}). With any single fixed bond-dimension sweep of the phase diagram,
we find that the trivial phase occupies only a very small region near the
origin; as we increase the bond dimension of the tensor-network
algorithm being used, we find that that region shrinks,
suggesting that this ``phase'' might not be anything more than an
isolated point in the phase diagram.

In Sec.~\ref{sec:state}, we begin by describing the family of states we will
be working with and their inherent properties, in addition to how
tensor-network algorithms can apply to them.
Then in Sec.~\ref{sec:phases} we will describe the phases that
we expect to find in the phase diagram of the system on the square
lattice, and detail our results, as obtained
using the tensor-network renormalization (TNR) and higher-order tensor
renormalization group (HOTRG) methods and summarized in the phase
diagram in Fig.~\ref{fig:squarephase}.
Finally, in Sec.~\ref{sec:honeycomb}, we return to the honeycomb lattice
to re-evaluate the evidence for the XY phase there.

\section{ The valence-bond state  }
\label{sec:state}

To define the deformed-AKLT state, we write 
a general AKLT state, which will be a tensor-network state 
with bond dimension $\chi=2$ on an arbitrary lattice and introduce a
continuously-parameterized deformation.

We start with some lattice with coordination number $q$. 
On each link we place a state of two spin-$\frac{1}{2}$ virtual spins
such that each vertex has $q$ such spins. We then produce the
physical degree of freedom by applying a projector $\mathbb{P}_q$ from the $q$
spins $|\eta_i \rangle$  onto the spin-$q/2$ subspace:
\begin{align}
\mathbb{P}_q = \sum_{\eta_1, \eta_2,...,\eta_q} c_s|s \rangle \langle \eta_1, \eta_2,...,\eta_q|,
\end{align} 
where $s=\sum_i\eta_i$ is the physical index, $\eta_i=\pm\frac{1}{2}$ represent
the virtual spins in their $S_z$ basis, and $c_s$ are Clebsch-Gordan
coefficients.  This yields the AKLT state
\begin{align}
 |\psi_\text{AKLT} \rangle= \bigotimes_{v\in V}(\mathbb{P}_q)_v  \bigotimes_{l\in L} | \psi^- \rangle_{l}, 
\end{align}
where the singlet states
$| \psi^- \rangle = |\! \up  \down  \rangle - | \! \down  \up  \rangle $
are placed on every link $l$ of the lattice.

We then apply a diagonal, spin-flip-invariant deformation
\begin{equation}
D(\vec{a})=\sum_{s=-q/2} ^{q/2} \frac{a_{|s|}}{c_s} |s \rangle \langle s|
\label{eqn:deformation-definition}
\end{equation}
in the $S^z$ basis to the physical indices. Then we arrive at a family
of deformed-AKLT states,
\begin{align}
|\Psi(\vec{a})_{\rm deformed} \rangle \propto D(\vec{a})^{ \otimes N}  |\psi_\text{AKLT} \rangle. 
\end{align} 
For the remainder of this work, we will fix $a_0 = 1$ (or $a_{\frac{1}{2}}=1$
for half-integer-spin cases).
We thus, for example, end up with two independent parameters in
the spin-2 case and only one independent parameter in the spin-3/2 case.

In short, the deformed-AKLT family of wave functions can be written as 
\begin{align}
\label{eqn:deformed}
|\Psi(\vec{a})_{\rm deformed} \rangle = \bigotimes_{v\in V} \left( D(\vec{a})\mathbb{P}_q  \right)_{v} \bigotimes_{l\in L} | \psi^- \rangle_{l}, 
\end{align} 
where the operator
$D(\vec{a}) \mathbb{P}_q$ maps the virtual spaces (which represent the
entanglement between the virtual spins) at each vertex $v$ to the
physical space.

We can modify the original two-site AKLT Hamiltonian\cite{AKLTgap} to obtain a
parent Hamiltonian which locally annihilates this state:
\begin{align}
\label{eqn:Ha}
H(\vec{a})\equiv &\sum_{\langle i,j\rangle} D(\vec{a})^{-1}_i \!\otimes\! D(\vec{a})^{-1}_j \, h_{ij}^{(\rm AKLT)} \, D(\vec{a})^{-1}_i \!\otimes\! D(\vec{a})^{-1}_j,\\
h_{ij}^{(\rm AKLT)} &\equiv \frac{1}{28}\left(S_{ij} + \frac{7}{10}S_{ij}^2+\frac{7}{45}S_{ij}^3+\frac{1}{90}S_{ij}^4\right)\notag\\
S_{ij} &\equiv \vec{S}_i\cdot\vec{S}_j\notag
%\\\rev H(\vec{a})|\Psi(\vec{a})\rangle &\color{red}= \left(\bigotimes_{k\notin\{i,j\}}D(\vec{a})_k\right)D(\vec{a})^{-1}_iD(\vec{a})^{-1}_jh_{ij}^{(\rm AKLT)}|\psi_\text{AKLT}\rangle = 0
\end{align}
As $h_{ij}^{(\rm AKLT)}$ annihilates the AKLT state, it follows
that $H(\vec{a})$ annihilates the deformed AKLT state.

Additionally, Niggeman, Kl\"umper, and Zittartz constructed
a more general, five-parameter family of two-site, frustration-free
Hamiltonians, invariant under lattice symmetries as well as on-site spin-flip
and $S_z$ invariance.
We note however that the above Hamiltonian is not well-defined when any
component $a_i$ of $\vec{a}$ is zero (under which circumstance we
would need to increase the rank of the two-site Hamiltonian - impossible with a
continuous deformation).
We shall also consider below AKLT-like states constructed using
maximally-entangled two-qubit states other than $|\psi^-\rangle$  as the valence bonds.

\subsection{Tensor network representation, bond states, and symmetry}
\label{sec:symmetry}
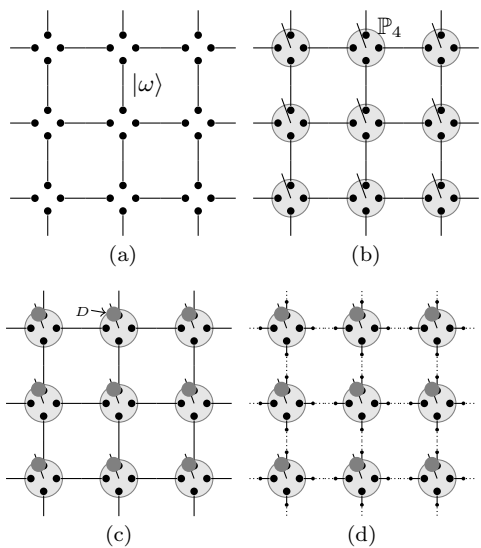
\begin{figure}[h]
  \centering
  \begin{subfigure}
    \centering
    \begin{tikzpicture}
      \foreach \i in {0,...,2}
      \foreach \j in {0,...,2} {
        \node[coordinate] (\i\j) at (\i,\j) {};
        \foreach \a in {0,90,180,270} {
          \node[circle,fill=black,inner sep=0pt,minimum size=1mm] (\i\j\a) at ($(\i\j)+(\a:.17)$) {};
          \draw (\i\j\a) -- +(\a:.33);
        }
      }
      \node[right] (label) at ($.5*(11)+.5*(12)$) {$|\omega\rangle$};
    \node[below] at (current bounding box.south) {\fontsize{8}{8}\selectfont (a)};
    \end{tikzpicture}
  \end{subfigure}
  \begin{subfigure}
    \centering
    \begin{tikzpicture}
      \foreach \i in {0,...,2}
      \foreach \j in {0,...,2} {
        \node[circle,fill=black!10,draw=gray,inner sep=3pt,minimum size=5mm] (\i\j) at (\i,\j) {};
        \draw (\i\j.center) -- +(110:.35);
        \foreach \a in {0,90,180,270} {
          \node[circle,fill=black,inner sep=0pt,minimum size=1mm] (\i\j\a) at ($(\i\j)+(\a:.17)$) {};
          \draw (\i\j\a) -- +(\a:.33);
        }
      }
      \node[circle,fill=none,draw=none,inner sep=0pt,label={[label distance=.26mm]45:$\mathbb{P}_4$}] at (12) {};
    \node[below] at (current bounding box.south) {\fontsize{8}{8}\selectfont (b)};
    \end{tikzpicture}
  \end{subfigure}
  \begin{subfigure}
    \centering
    \begin{tikzpicture}
      \foreach \i in {0,...,2}
      \foreach \j in {0,...,2} {
        \node[circle,fill=black!10,draw=gray,inner sep=3pt,minimum size=5mm] (\i\j) at (\i,\j) {};
        \draw (\i\j.center) -- +(110:.35);
        \foreach \a in {0,90,180,270} {
          \node[circle,fill=black,inner sep=0pt,minimum size=1mm] (\i\j\a) at ($(\i\j)+(\a:.17)$) {};
          \draw (\i\j\a) -- +(\a:.33);
        }
        \node[circle,fill=black!50,inner sep=0pt,minimum size=2mm] (\i\j q) at ($(\i\j)+(110:.2)$) {};
      }
      \draw[<-] (12q) -- node[left] {\tiny$D$} +(-.3,.05) ;
    \node[below] at (current bounding box.south) {\fontsize{8}{8}\selectfont (c)};
    \end{tikzpicture}
  \end{subfigure}
  \begin{subfigure}
    \centering
    \begin{tikzpicture}
      \foreach \i in {0,...,2}
      \foreach \j in {0,...,2} {
        \node[circle,fill=black!10,draw=gray,inner sep=3pt,minimum size=5mm] (\i\j) at (\i,\j) {};
        \draw (\i\j.center) -- +(110:.35);
        \foreach \a in {0,90,180,270} {
          \node[circle,fill=black,inner sep=0pt,minimum size=1mm] (\i\j\a) at ($(\i\j)+(\a:.17)$) {};
          \draw (\i\j\a) -- ++(\a:.18) node[circle,fill=black,inner sep=0pt,minimum size=.5mm] (\i\j\a b){};
          \draw[densely dotted] (\i\j\a b) -- +(\a:.15);
        }
        \node[circle,fill=black!50,inner sep=0pt,minimum size=2mm] (\i\j q) at ($(\i\j)+(110:.2)$) {};
      }
    \node[below] at (current bounding box.south) {\fontsize{8}{8}\selectfont (d)};
    \end{tikzpicture}
  \end{subfigure}
  \caption{The valence-bond state and tensor-network state pictures of the
  spin-2 deformed-AKLT state on the square lattice.
  (a) A singlet state (or, more generally, a bond state $|\omega\rangle$
  composed from two virtual spin-$\frac{1}{2}$ degrees of freedom) is placed
  on each edge of the lattice.
  (b) The AKLT state is formed by placing a spin-2 projector $\mathbb{P}_4$ on
  each site; the spin-2 indices of the projectors indicate the physical
  degrees of freedom.
  (c) From there the deformation $D(\vec{a})$ is applied to each site, resulting
  in the deformed-AKLT state.
  (d) In order to represent this in the typical manner of a PEPS, we may for
  example perform a Schmidt decomposition on the bond states; the Schmidt
  indices then becomes the ``bond'' indices of the tensor network, and the
  contraction of the resulting objects and the deformation matrix with the
  five indices of the projection matrix yields the on-site tensor.}
  \label{fig:aklt_structure_square}
\end{figure}
Given a state in the valence-bond picture we have just presented, it is natural
to represent it as a tensor network state (TNS), namely a projected entangled
pair state (PEPS). For those who are not familiar with tensor
network states, we recommend several of the cited review and pedagogical
papers\cite{Vidal_iPEPS_2009,PerezGarciaMPS,PerezGarciaPEPS,BridgemanChubb,
MPS_PEPS_phases}.
In this representation we place on each lattice site
a rank-$(q+1)$ tensor with one physical index (the spin on the site) and
$q$ virtual indices (corresponding to the $q$ virtual spins at the site,
or more precisely their Schmidt index);
pairs of virtual indices of adjacent sites are contracted over in a tensor trace
(tTr) to yield the physical state. In an AKLT system we begin with a
rank-$(q+1)$
tensor on each site (the projector $\mathbb{P}_q$) as well as a rank-2 tensor
on each link (the virtual singlets). To get a PEPS description we may 
assign each singlet to a neighboring site and contract it with the
corresponding index of that site's projector, although the way in
which the bonds are defined is essentially a gauge choice and
thus can be easily varied. From this we obtain a PEPS description of a general
deformed-AKLT state by contracting the AKLT physical indices with the
deformation matrix $D(\vec{a})$.

We may also alter the state by replacing the singlet state $|\psi^-\rangle$
in the above description with a more general bond state $|\omega\rangle$.
In particular we may use the Bell states
\begin{align}
& | \phi^+ \rangle= |\uparrow \uparrow \rangle+| \downarrow \downarrow \rangle \notag \\
& | \phi^- \rangle = | \uparrow \uparrow \rangle-|\downarrow \downarrow \rangle =  I \otimes \sigma^z | \phi^+ \rangle  \notag \\
& | \psi^+ \rangle = | \uparrow \downarrow \rangle+| \downarrow \uparrow \rangle =  I \otimes \sigma^x | \phi^+ \rangle  \notag \\
& | \psi^- \rangle = | \uparrow \downarrow \rangle-|\downarrow \uparrow\rangle =  I \otimes i \sigma^y | \phi^+ \rangle,
\end{align} 
where $ \sigma^k,   k\in \{0,x,y,z\}$ are Pauli matrices and $ \sigma^0=  \mathbb{I}$; we may refer to the states $|\phi^\pm\rangle$ as ``ferromagnetic''
bond states and the states $|\psi^\pm\rangle$ as ``antiferromagnetic'' bond
states, due to the behavior of the respective systems in the ordered regime
as discussed below.
This construction is shown graphically in Fig.~\ref{fig:aklt_structure_square}.

When working on a bipartite lattice, we may change from one such
bond state to another
by applying $SU(2)$ transformations $U_A$ and $U_B$ which commute with the
deformation $D(\vec{a})$ to all of the sites of
the sublattices $A$ and $B$, respectively. Due to the $SU(2)$-invariance of
the projector $\mathbb{P}_q$, this is equivalent to performing the
transformation $|\omega\rangle \mapsto
U_A^{(1/2)}\otimes U_B^{(1/2)}|\omega\rangle$ to every bond state.
Therefore, if we start with the singlet $|\psi^-\rangle$ as our bond state,
we may then convert it to
\begin{enumerate}[i]
\item $|\phi^+\rangle$ by applying $U_A = R_y^{-\frac{\pi}{2}}$
and $U_B = R_y^\frac{\pi}{2}$ to the $A$ and $B$ sublattices, respectively;
\item $|\phi^-\rangle$ by applying $U_A = R_x^{-\frac{\pi}{2}}$ and
$U_B = R_x^{\frac{\pi}{2}}$; and
\item $|\psi^+\rangle$ by applying $U_A = R_z^{-\frac{\pi}{2}}$ and
$U_B = R_z^{\frac{\pi}{2}}$,
\end{enumerate}
where the $SU(2)$ rotation $R_j^\phi \equiv e^{-i\phi S_j}$.
Thus, given physical data from any of these four systems, we may easily
produce the corresponding
information about any of the other. If for example
 we find it simpler to manipulate the tensors we use in the case
$\ket\omega = \ket{\phi^+}$, we can apply any conclusions
we draw about that case,
such as boundaries of phase diagrams, to the more standard case of
$\ket\omega=\ket{\psi^-}$.

However, on a lattice which is not bipartite, this mapping is not generally
possible; in fact only the $\phi^+$ and $\phi^-$ bond states can be identified
with each other (by applying $R_z^\frac{\pi}{2}$ to every site), and we
expect in general to get three distinct phase diagrams from these four
bond states.

As both the tensors used to build the AKLT state, that is
the projector $\mathbb{P}_q$ and the
singlet state $\psi^-$, are invariant under $SU(2)$ transformations,
the state itself maintains a global $SU(2)$ invariance. However, the deformation
$D(\vec{a})$ 
breaks this symmetry down to a subgroup isomorphic to $O(2)$ which
can be characterized by its action on the $xy$ plane, on which rotations are
generated by $S_z$ and reflections are produced by spin-flips such
as $R^\pi_x = e^{\pi i S_x}$ and $R^\pi_y = e^{\pi i S_y}$.

Now consider states $\ket{\Psi_\omega}$ with different bond states,
obtained by applying $U_A$ and $U_B$ to the
state $\ket{\Psi_{\psi^-}}$. If $g \in O(2)$ preserves $\ket{\Psi_{\psi^-}}$,
then $\ket{\Psi_\omega}$ will be preserved by applying
$U_AgU_A^\dagger$ to sublattice A and $U_BgU_B^\dagger$ to sublattice
B.\footnote{This cannot be done when the lattice is not bipartite;
and in fact the choice of bond state in combination with the deformation
breaks the $SU(2)$ symmetry of the AKLT state down to $\mathbb{Z}_2\times
\mathbb{Z}_2$ when the bond state is ferromagnetic.}
For the antiferromagnetic bond state $\psi^+$, $U_A$ and $U_B$ commute with
$U(1)$ rotations and are exchanged by $O(2)$ reflections, so that the symmetry
applied in that case will still preserve the state (assuming there are an even
number of sites).

When performing numerical analysis, it may be useful, for data collection
and/or for numerical stability, to explicitly preserve the global on-site
symmetry\cite{Singh1,HeST} by ensuring that the tensors produced in each
step of the renormalization procedures remain invariant. (However, due to
limitations in our code as of when these data were collected, we have not
preserved $O(2)$ itself but rather some adequately large finite subgroup
thereof, either $\mathbb{Z}_2\times \mathbb{Z}_2$, $D_{40}$, or $D_{80}$.)

\subsection{Representation as a (pseudo)classical model}
\label{sec:pseudoclassical}

\begin{figure}[h]
  \centering
  \begin{subfigure}
    \centering
    \begin{tikzpicture}[scale=0.7, every node/.style={transform shape}]
      \foreach \i in {0,...,2} 
      \foreach \j in {0,...,4} {
      \node[siteq] (\i\j b) at (1.5*\i,1.5*\j) {};
      \foreach \a in {0,90} \draw (\i\j b) -- +(\a:.75);
      \foreach \a in {0,90} \draw (\i\j b) -- +(\a:-.75);
      \node[siteq] (\i\j k) at ($(\i\j b)+(.2,-.6)$) {};
      \draw (\i\j k) -- (\i\j b.center);
      \foreach \a in {0,90} \draw (\i\j k) -- +(\a:.75);
      \foreach \a in {0,90} \draw (\i\j k) -- +(\a:-.75);
      %\foreach \a in {180,270} \draw (\i+1,\j+1) -- +(\a:1);
      }
      \node[qop] (Aop) at ($.6*(00b)+.4*(00k)$) {};
      \draw [-,line width=.4mm] (Aop) to (00b.center);
      \node[siteq] (00k2) at (00k) {};
      \draw [-,line width=.4mm] (00k2) to (Aop.center);
      \node[qop] (Bop) at ($.6*(24b)+.4*(24k)$) {};
      \draw [-,line width=.4mm] (Bop) to (24b.center);
      \node[siteq] (24k2) at (24k) {};
      \draw [-,line width=.4mm] (24k2) to (Bop.center);
      \foreach \i in {0,...,2} {
      \draw[dotted,thick] ($.5*(\i 0b)+.5*(\i 0k)+(0,-1)$) -- +(0,-.25);
      \draw[dotted,thick] ($.5*(\i 4b)+.5*(\i 4k)+(0,1)$) -- +(0,.25);
      }
      \foreach \j in {0,...,4} {
      \draw[dotted,thick] ($.5*(0\j b)+.5*(0\j k)+(-1,0)$) -- +(-.25,0);
      \draw[dotted,thick] ($.5*(2\j b)+.5*(2\j k)+(1,0)$) -- +(.25,0);
      }
    \node[below] at (current bounding box.south) {\fontsize{11.43}{11.43}\selectfont (a)};
    \end{tikzpicture}
  \end{subfigure}
  \begin{subfigure}
    \centering
    \begin{tikzpicture}[scale=0.7, every node/.style={transform shape}]
      \foreach \i in {0,...,2} 
      \foreach \j in {0,...,4} {
      \node[sitec] (\i\j) at (1.5*\i,1.5*\j) {};
      \foreach \a in {0,90} \draw[double] (\i\j) -- +(\a:.75);
      \foreach \a in {0,90} \draw[double] (\i\j) -- +(\a:-.75);
      %\foreach \a in {180,270} \draw (\i+1,\j+1) -- +(\a:1);
      }
      \node[sitenimp] (Aop) at (00) {};
      \node[sitenimp] (Bop) at (24) {};
      \foreach \i in {0,...,2} {
      \draw[dotted,thick] ($(\i 0)+(0,-1)$) -- +(0,-.25);
      \draw[dotted,thick] ($(\i 4)+(0,1)$) -- +(0,.25);
      }
      \foreach \j in {0,...,4} {
      \draw[dotted,thick] ($(0\j)+(-1,0)$) -- +(-.25,0);
      \draw[dotted,thick] ($(2\j)+(1,0)$) -- +(.25,0);
      }
    \node[below] at (current bounding box.south) {\fontsize{11.43}{11.43}\selectfont (b)};
    \end{tikzpicture}
  \end{subfigure}
  \caption{Correlations in a PEPS and the corresponding representation in a
  classical model. (a) The  correlation function $\langle\Psi|AB|\Psi\rangle$
  of two one-site operators in a quantum state $\Psi$ represented by a PEPS.
  (b) The correlation function $\langle\mathcal{O}_A\mathcal{O}_B\rangle$
  of two classical ``operators,'' shaded, which replace the weight matrix at a
  site
  with a different tensor. In this case the classical model is the ``doubled
  vertex model'' and the operators in it are determined by contracting the
  quantum operator with bra and ket tensors.}
\label{fig:quantumclassical}
\end{figure}
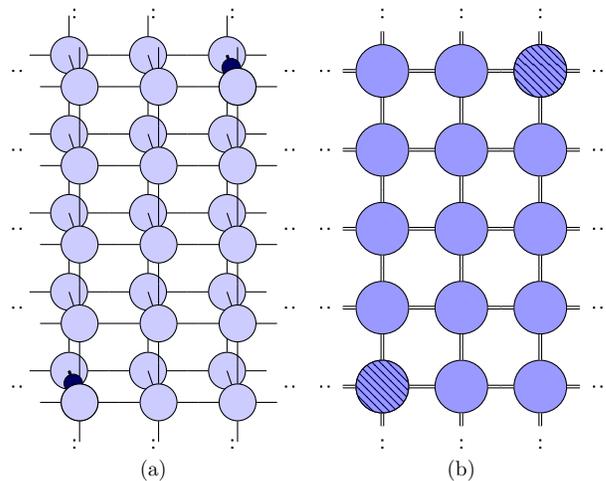
Much as we have shown, a two-dimensional quantum state can often be represented
as a
tensor network state (TNS),  whose coefficients in a fixed basis are expressed
as a contraction of a tensor network, that is, a tensor trace:
\begin{align}
|\psi\rangle=\sum_{s_1,s_2,\cdots s_m\cdots}\text{tTr}(A^{s_1}A^{s_2}\cdots A^{s_m}\cdots)|s_1 s_2\cdots s_m\cdots\rangle, 
\label{TNS}
\end{align} 
where $A^s_{\alpha,\beta,\gamma,\ldots}$ is a local tensor with a physical index
$s$ and internal or bond indices $\alpha,\beta,\gamma,\ldots$, and $\text{tTr}$
denotes tensor contraction of all the connected inner indices according to the
underlying lattice structure. TNS defined on two- or higher-dimensional lattices
are often referred to as PEPS.

The norm squared of a TNS is given by 
\begin{align}
\langle \psi |  \psi \rangle =\text{tTr} (\mathbb{T}^1  \mathbb{T}^2  \mathbb{T}^3\cdots \mathbb{T}^m\cdots   ), 
\label {ST_TNS}
\end{align} 
where we form the local {\it doubled tensor\/} $\mathbb{T}^i$ by
merging a bra layer and a ket layer, contracting the physical indices of
corresponding pairs of tensors $A$ and $A^*$:
\begin{align}
\mathbb{T}\equiv \sum _s  (A^s_{\alpha,\beta, \gamma,\delta,\ldots })  \times  (A^s_{\alpha',\beta', \gamma',\delta',\ldots }) ^*.
\label{eqn:doubletensor}
\end{align}

In this way, a quantum model maps into something resembling a classical
vertex model on the same lattice, in which the doubled tensor plays the role
of the weight matrix in the corresponding classical model\footnote{However,
it is not guaranteed that the tensor will correspond to a true classical model
as it may not contain strictly real, nonnegative entries.}. As in
Fig.~\ref{fig:quantumclassical}, we can often translate observable quantities
describing the
quantum state into observable quantities describing the classical
model, which helps us get information about the former from the latter.
We will refer to this as the ``doubled vertex'' model.

However, in two and higher dimensions it is in general computationally
intractible to exactly calculate the tensor trace, that is, to contract the
whole tensor network, for reasonably large system sizes.
Several approximation schemes have been proposed as solutions in this context,
such as the iPEPS algorithm\cite{Vidal_iPEPS_2009}, the corner transfer matrix
method (CTMRG)\cite{Nishino_CTMRG_1997,OrusCTM}, and coarse-graining
approaches\cite{Levin_TRG_2007,Xiang_TRG_2008,Wen_Gu_Levin_TRG_2008,TNR}, all
of which tackle the contraction problem essentially by truncating information
and thus scaling down the computational complexity to the polynomial level.

In Appendix~\ref{app:methods},
we will discuss those methods we have used, namely the corner transfer matrix,
quantum-state renormalization group, higher-order tensor renormalization group, 
tensor network renormalization, and loop-TNR methods.

\FloatBarrier
\section{Results}
\label{sec:phases}

Here we describe how we characterize the distinct phases that
appear in this two-parameter family of states, as shown in
Fig.~\ref{fig:squarephase}, and then
present the numerical results arising
from this analysis.
\subsection{The N\'eel-ordered phase}
\label{sec:ordered}

\begin{figure}[h!]
\includegraphics[width=.5\textwidth]{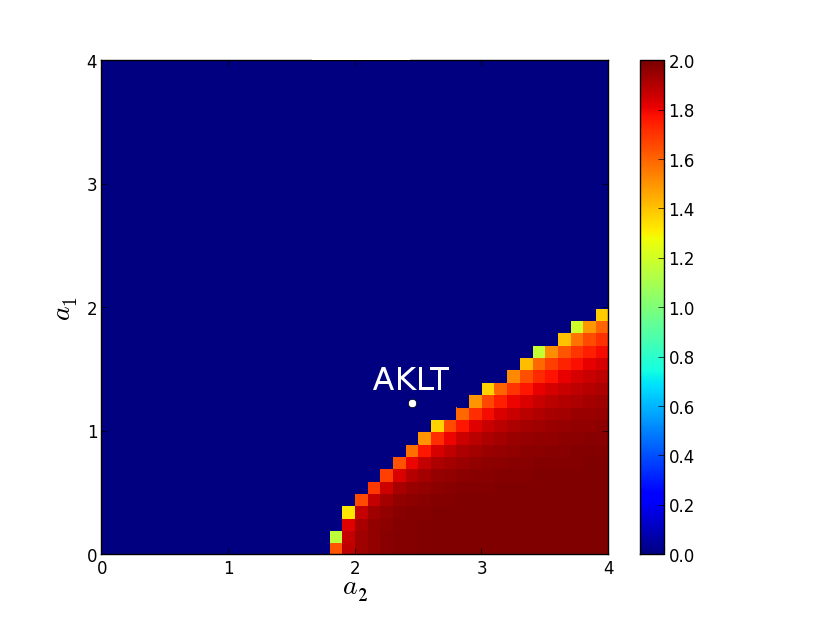}
\caption{Applying HOTRG with $\chi=40$, and extracting the magnetization 
$\langle S_z\rangle$, we find a sharp phase transition from a disordered
region (containing the AKLT, XY, and product-state phases) into an ordered
region, with the magnetization rapidly increasing to 2.}
\label{fig:square_Mz}
\end{figure}
In the limit $a_2\to \infty$ (equivalently, $a_0,a_1 \to 0$, where
the deformation becomes a projection onto $S_z=\pm 2$) the tensors
$T = Q(\vec{a})\mathbb{P}_4$ effectively become the mapping
$|2\rangle\langle\uparrow\uparrow\uparrow\uparrow\!\!| + 
|\!-2\rangle\langle\downarrow\downarrow\downarrow\downarrow\!\!|$.
Assuming the standard bond state $\psi^-$, the deformed-AKLT state
will then be a cat
state with two dominant configurations $|\!+\!2,-2,\ldots,+2,-2\rangle$ and
$|\!-\!2,+2,\ldots,-2,+2\rangle$. Thus, as we approach this limit  we expect a
phase where these states
will, in the thermodynamic limit, exhibit spontaneous symmetry breaking
to N\'eel-ordered states.

We can detect this order using the staggered ordered parameter $(-1)^{n+m}S_z$.
In Fig.~\ref{fig:square_Mz}, we see that this order parameter obtains an
expectation value within a well-defined region surrounding the $a_2\to \infty$
limit (and nowhere else).

\subsection{The AKLT phase}
\label{sec:AKLT}

\begin{figure}[h!]
\includegraphics[width=0.5\textwidth]{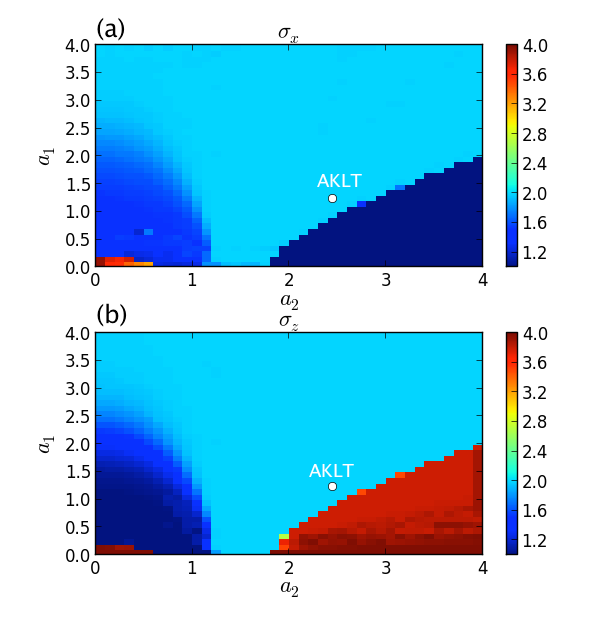}
\caption{ The trace of the simulated modular matrix $T$ of symmetry twists (a)
$\sigma_x$ and (b) $\sigma_z$, calculated using HOTRG with bond dimension
$\chi=30$ after 10 RG steps,
plotted over all the regions of the phase diagram we have studied.
These quantities sharply define the AKLT-N\'eel phase transition,
and approximately define the KT transition.}
  \label{fig:tnST_square}
\end{figure}
The isotropic AKLT state is known to have nontrivial but weak symmetry-protected
topological order, preserved under translations combined with on-site
$SU(2)$ transformations\cite{AKLTstrange,SREstrange,CZX} or a suitable subgroup
thereof - for our purposes,
rotations and reflections in the $xy$ plane combining to form $O(2)$.
As discussed in Sec.~\ref{sec:symmetry}, the deformations we are
considering commute with these symmetries, so we expect the AKLT point
$(a_2,a_1,a_0)=(\sqrt{6}, \sqrt{\frac{3}{2}},1 )$ to be contained within a
larger disordered-antiferromagnet phase behaving as a nontrivial weak SPT phase
under these symmetries.

In order to detect this phase we use simulated modular matrices of
Huang and Wei\cite{tnST,HungWenSPT}, which originated from the idea that
gauging SPT order yields intrinsic topological order\cite{LevinGuSPT}.
Applying $R^\pi_i$ to the physical index of a site tensor
is equivalent to applying the Pauli matrix $\sigma^i$ to the virtual indices.
Because of this we can extract simulated $S$ and $T$ matrices by
representing symmetry twists in the Hamiltonian with strings of $\sigma^i$
operators applied to virtual bonds. At the AKLT point, the modular matrices
arising from symmetry twists $\sigma^x$, $\sigma^y$, or $\sigma^z$ should be
\begin{align}
  \label{nontri_oZ2}
  S =  \begin{pmatrix}
    1 & 0&0 &0 \\
    0& 0&1 &0 \\
     0 & 1&0 &0\\
     0 & 0&0 &1
 \end{pmatrix},  \ \
 T =  \begin{pmatrix}
  1 & 0&0 &0 \\
  0& 1&0 &0 \\
   0 & 0&0 &1\\
   0 & 0&1 &0
 \end{pmatrix}, 
\end{align}
so that $\op{tr}(S) = \op{tr}(T) = 2$. As we expect this to be a constant
of an SPT phase, we may follow these quantities and use
$\op{tr}(T) = 2$ as an indicator of nontrivial SPT order; when $\op{tr}(T)$
is 1 or 4, meanwhile, we identify a trivial or symmetry-breaking phase.

In Fig.~\ref{fig:tnST_square}, we see that, in precisely the ordered region
indicated by Fig.~\ref{fig:square_Mz}, the traces of the
modular $T$ matrices corresponding to each of the symmetry twists
$\sigma_x$ and $\sigma_z$ take the trivial values of 1 and 4,
respectively. In most of the disordered region, meanwhile, both of these
traces equal 2, with a very sharp transition between these two regimes. Where
we find $\op{tr}(T_x) = \op{tr}(T_y) = 2$, we are within the SPT-ordered
AKLT phase. We will discuss the region in which these traces appear to
continuously vary shortly.
\subsection{Characteristics of the transition between N\'eel-ordered and AKLT phases}

\begin{figure}[h!]
\includegraphics[width=0.5\textwidth]{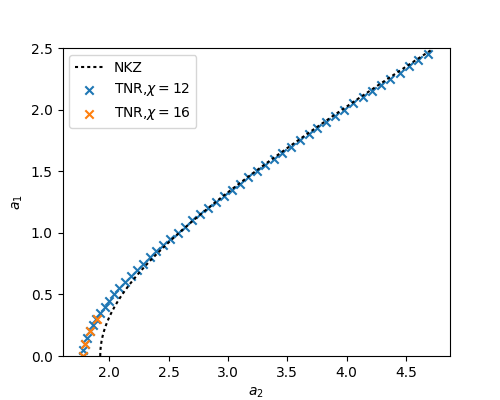}
\caption{The line of transition between the N\'eel and AKLT phases, as
determined by sweeping $a_2$ given fixed $a_1$ with TNR for nontrivial
central charge at scales of up to 12 coarse-graining steps, primarily using
bond dimension $\cchip{12}{10}$ but with $\cchip{16}{12}$ for confirmation.
We find reasonable agreement with the findings of NKZ in the asymptotic limit
$a_1,a_2\to \infty$, but for $a_1\to 0$ we find disagreement, confirmed by
increasing bond dimension as in Table~\ref{tab:VBSNeel}, that exceeds their
estimates of error.}
\label{fig:curveVBSNeel}
\end{figure}

\begin{figure}[h!]
\includegraphics[width=0.5\textwidth]{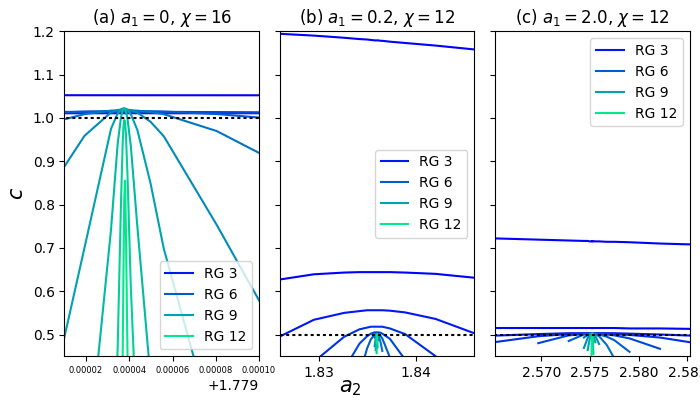}
\caption{Scanning in the neighborhood of the AKLT-N\'eel critical line,
we find as we coarse-grain that the curve of estimated $c$ versus $a_2$
forms an increasingly narrow
peak, with height $c=1$ at $a_1=0$ and $c=1/2$ elsewhere.}
\label{fig:cVBSNeel}
\end{figure}

\begin{figure}[h!]
\includegraphics[width=0.5\textwidth]{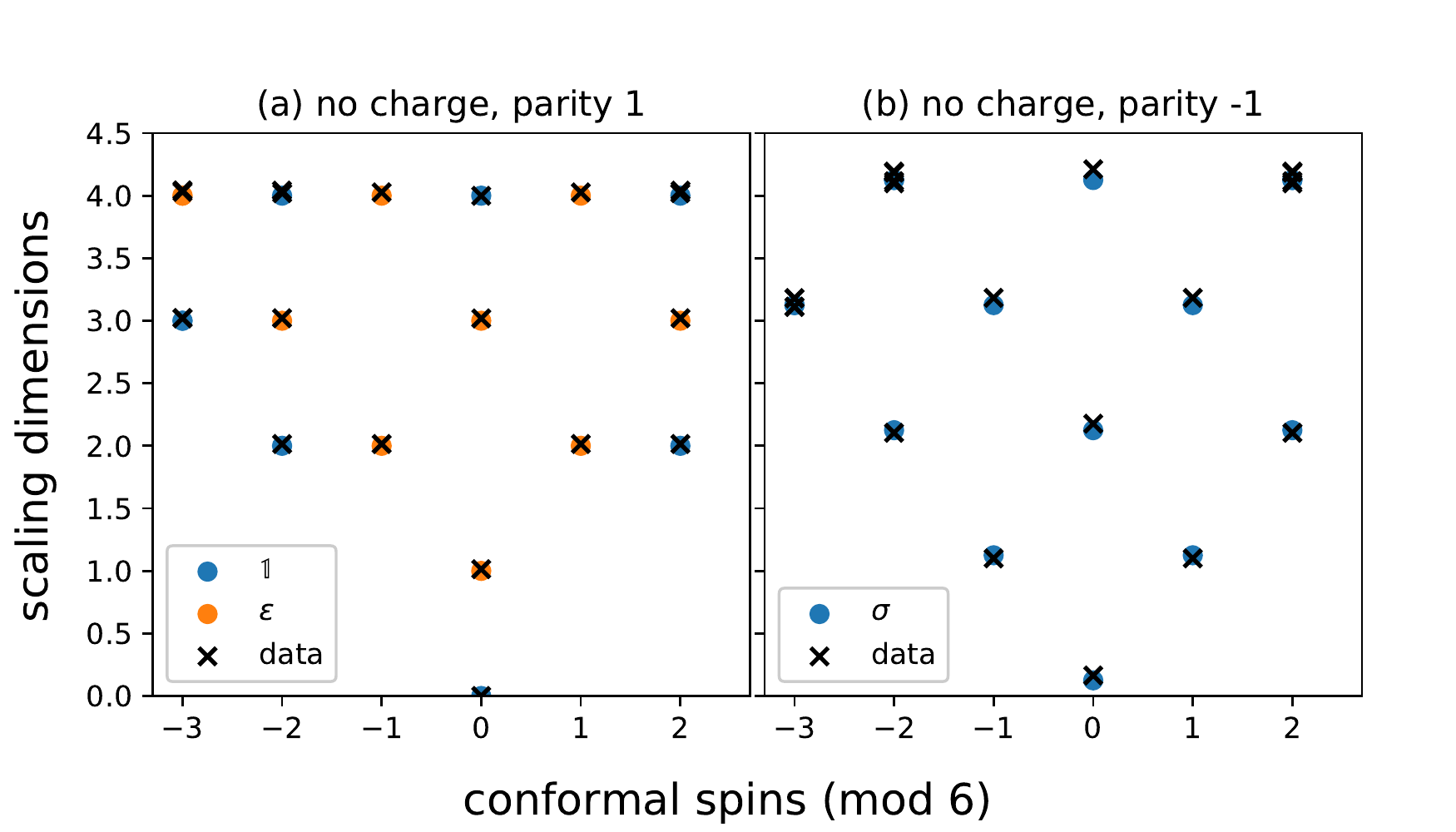}
\caption{The conformal tower at the critical line between the AKLT and
N\'eel phases, specifically $a_1 = 1.0$, $a_2 = 2.575228$, with
$\cchip{12}{10}$, after 7 RG steps.
We find excellent replication of the conformal
tower of the Ising CFT. Here we have marked theoretical values for the
conformal towers of primary operators
$\mathbbm{1}$, $\epsilon$, and $\sigma$. We find no scaling
operators with nontrivial $U(1)$ charge; what we find instead, with the
spin operator $\sigma$ in the parity -1 sector, is that (arbitrary) $O(2)$
reflections play the role of spin-flips in the Ising model.}
\label{fig:CFTVBSNeelIsing}
\end{figure}

The disordered and N\'eel-ordered phases described above were previously
identified by Niggeman, Kl\"umper, and Zittartz in their original
work, using Monte-Carlo methods. They additionally claim that the critical line
separating these two phases has Ising-like critical exponents and is located at
$a_2^2 = (3.0\pm0.1)a_1^2 + (3.7\pm0.3)$. We evaluate this claim using TNR:
given a value of $a_1$, we analyze several candidate values of $a_2$,
coarse-graining until the estimated value of $c$ passes below a threshold,
and then take a refined selection of $a_2$s around the value which
had the greatest $c$. By using this method to seek a point at which $c$
maintains an asymptotic value up to 12 coarse-graining steps,
we can resolve $a_2$ to several parts per hundred thousand,
as Fig.~\ref{fig:cVBSNeel} demonstrates. Fig.~\ref{fig:curveVBSNeel}
compares results from two values of the bond dimension and from the
estimates in the original work. Moreover, this analysis neatly
confirms the ``Ising-like'' nature of the transition; in 
Fig.~\ref{fig:CFTVBSNeelIsing}, we see with TNR that the IR limit of the
doubled vertex model along
the transition exactly matches the Ising CFT with $c=\frac{1}{2}$, with
spin-flips $R^\pi_\phi$ playing the role of spin-flips in the Ising model.
However in the $a_1\to 0$ limit we find $c$ becomes 1, also demonstrated in
Fig.~\ref{fig:cVBSNeel}.

\begin{table}
\[
\footnotesize
\begin{array}{r ||  c c c c | c }
  a_1 & \chi=12, & \chi=16, & \chi=20, & \chi=24, & \text{NKZ}\\
  & \chi'=10 & \chi'=12 & \chi'=14 & \chi'=16\\
  \hline
  0.0 & 1.774789(6) & 1.779038(1) & 1.779243(1) & 1.779348(1) & 1.92(8)\\
  0.1 & 1.795103(6) & 1.798905(3) & - & - & 1.93(8)\\
  0.2 & 1.835964(6) & 1.839119(6) & - & - & 1.95(8)\\
  0.3 & 1.892065(6) & 1.894952(6) & - & - & 1.99(8)
\end{array}
\]
\caption{We use TNR to determine the critical line between the AKLT and
N\'eel-ordered phases, increasing bond dimension from $\chi=12$ to $\chi=16$
at four points and then to $\chi=20$ and $\chi=24$ at one point
to determine the accuracy
of our estimates. Although we determine that the bias is much greater than
the uncertainty of these estimates, we find that it appears nonetheless to
be within $\Delta a_2 < 0.01$ for our least accurate, $\chi=12$ estimates,
and within $\Delta a_2 < 0.001$ for our $\chi=16$ estimates. We also find that
the error appears to decrease for increasing $a_1$, which also reduces
the difference between the original Niggemann, Kl\"umper, and Zittartz (NKZ)
estimates and ours until they are
within appropriate error of each other.}
\label{tab:VBSNeel}
\end{table}

\subsection{The XY-like phase}
\label{sec:XY-properties}
In a region near the origin $a_1=a_2=0$ of the phase diagram,
we will find that the state has infinite correlation length (or equivalently,
quasi-long-range order), as was reported for the analogous model on the
honeycomb lattice.\cite{AKLTspin32} This quasi-long-range-ordered region
will explain much of the anomalous behavior observed in
Fig.~\ref{fig:tnST_square}. In this phase, the doubled vertex model of
\eqref{eqn:doubletensor} is described in the infrared limit by the
continuously-parametrized field theory of the compactified free boson, much like
the XY model in its low-temperature phase. We will begin by describing this
 conformal field theory (CFT), following Fendley \cite{Fendley}
and Di Francesco, Mathieu, and S\'en\'echal\cite{DiFrancesco}.

The compactified-free-boson CFT has central charge 1 and
is characterized by a bosonic field $\phi$ whose values are angles and which
has some coupling constant $g$.\footnote{In the XY model, under the original
approximation scheme of Kosterlitz and Thouless\cite{Kosterlitz},
$g = \frac{k_BT}{4\pi J}$, based on the critical exponents they derive.}
The field $\phi$ itself is not a valid operator on the CFT due to its
logarithmic
divergences; however the theory admits derivative and vertex operators,
represented in terms of the holomorphic and antiholomorphic components
$\phi~=~\varphi~+~\bar\varphi$ as
\begin{equation}
\begin{array}{c|c|c}
\text{Field}&\Delta&s\\\hline
\partial\varphi&1&1\\
\bar\partial\bar\varphi&1&-1\\
V_{e,m}&\frac{e^2}{2g} + \frac{m^2g}{2}&em
\end{array}
\end{equation}
where $\Delta = h+\bar{h}$ is the scaling dimension and $s=h-\bar{h}$ is the 
conformal spin, $h$ and $\bar{h}$ being the holomorphic and antiholomorphic
conformal dimensions. The vertex operators $V_{e,m} \propto \  :\mathrel{e^{i(e+gm)\varphi}e^{i(e-gm)\bar\varphi}}:$ are indexed by an
``electric charge'' $e \in \mathbb{Z}$ and ``magnetic charge'' $m\in\mathbb{Z}$.
Here $e$ is a $U(1)$ charge, which is to say that a global rotation
$\phi \mapsto \phi+\phi_0$
will send $V_{e,m}\mapsto e^{ie\phi_0}V_{e,m}$; magnetic charge indicates
vortex winding number: in a configuration produced by inserting
$V_{e,m}(z)$, there is a branch cut from $z$ to infinity (or to another vortex)
around which $\phi$ picks up $2\pi m$.

In the full global on-site $O(2)$ symmetry of the compactified free boson or of
the XY model, the charge-$\pm k$ representations
$v \mapsto e^{\pm i k\phi}v$
of the subgroup $U(1)~\subset~O(2)$, of rotations
$|s\rangle \mapsto e^{-i \phi S_z}|s\rangle$, pair to form
doublets; an additional pair of 1D representations have no $U(1)$ charge but
are either even or odd under reflections. Under this symmetry, the derivative
operators belong to the rotation-invariant reflection-odd representation;
the two electric operators $V_{\pm e,0}$ for $e \neq 0$ form a doublet; the four
electromagnetic operators $V_{\pm e, \pm m}$ for $e,m \neq 0$ form a pair of
doublets; and the magnetic operators $V_{0,\pm m}$ for $m > 0$ are exchanged
under reflections, so that $V_{0,m}+V_{0,-m}$ belongs to the
trivial representation and $V_{0,m}-V_{0,-m}$ belongs to the
reflection-odd representation. In particular, $V_{0,1}+V_{0,-1}$
has the smallest scaling dimension of any
$O(2)$-invariant primary operator other than the identity, and will therefore
induce a phase transition when it becomes relevant. As
$\Delta_{0,\pm 1}=\frac{g}{2}$, and a scaling operator is relevant
when $\Delta < d$, this occurs at coupling $g=2d=4$.

\begin{figure} [h!]
\includegraphics[width=0.5\textwidth]{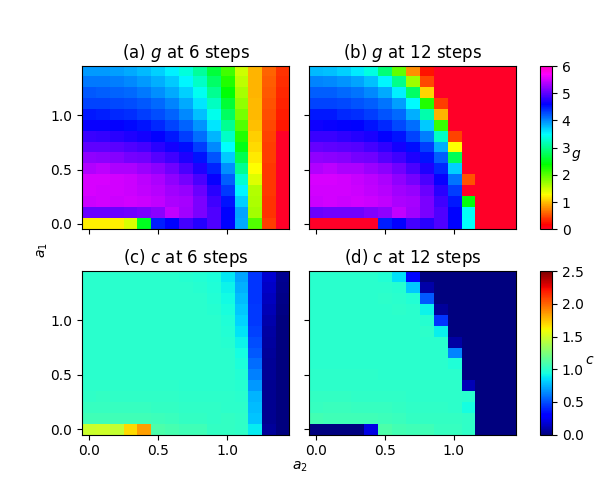}
\caption{Basic estimates of $g$ and $c$ from TNR, $\cchip{20}{14}$.
(a) After six coarse-graining steps we can see that the region $g \geq 4$ is
bounded by a curve which intersects the axes at roughly $a_1 = 1.2$ and
$a_2=1.0$.  As $a_2$ is held constant and $a_1$ decreases, $g$ increases to as
much as 5.5 before falling off towards the $a_2$ axis.
(b) After 12 coarse-graining steps, $g$ has remained approximately constant
within the region where $g \geq 4$, but it
has decreased, sometimes substantially,
where $g < 4$. (c) After six RG steps $c$ maintains a value very close to 1 in
a region roughly corresponding to $g > 2.5$. (d) This is also true after 12
RG steps, but that region has shrunk due to changes in the estimated value of
$g$. Between estimates for $g$ and for $c$, we end up with an XY-like phase
that's quite well-defined away from the pseudo-quasi-long-range ordered region.}
\label{fig:XYmaps20}
\end{figure}

\begin{figure}[h!]
\includegraphics[width=0.5\textwidth]{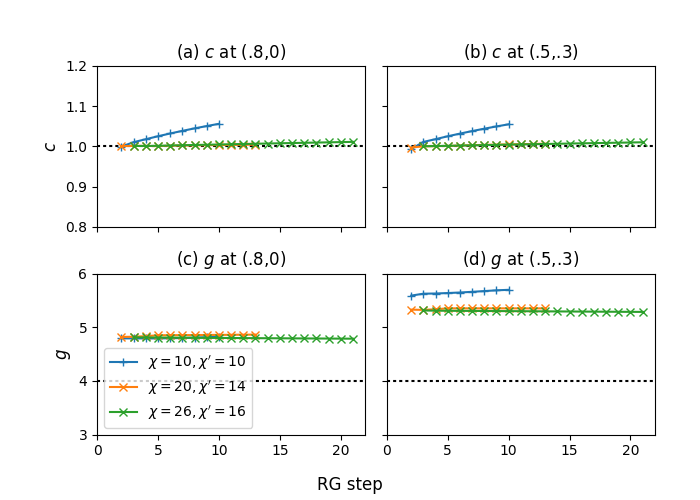}
\caption{Following the estimated quantities $g$ and $c$ with increasing
system size - measured in number of coarse-graining steps - at
different bond dimensions. (a),(b) For both $a_1\nobreak\!=\nobreak\!.8, \,a_2\nobreak\!=\nobreak\!0$ and $a_1\!=\!.5,\,a_2\!=.\!3$,
$c$ converges as the bond dimension increases, and by $\chi\!=\!26$ maintains
a value of $1.005\pm.005$ as coarse-graining increases the length scale.
(c) At $a_1\!=\!.8,\,a_2\!=\!0$,
$g$ converges with increasing bond dimension, with a value of $4.80\pm .01$
at $\chi\!=\!26$
which is steady under coarse-graining. (d) At $a_1\!=\!.5,\,a_2\!=\!.3$, $g$ converges
with increasing bond dimension, with a value of $5.30\pm .01 $ at $\chi\!=\!26$
which is steady under coarse-graining.
(Here, and in any other such presentations of data from TNR, we use
$+$ to mark data obtained while preserving $\mathbb{Z}_2\times\mathbb{Z}_2$
and $\times$ to mark data obtained while preserving
$D_{2N}$, typically $D_{80}$; the latter is usually more consistent and
demonstrates more stability.)}
\label{fig:XYcompare-interior}
\end{figure}

We will conclude that this CFT describes the doubled vertex model, including a
transition at $g=4$. By performing TNR (see Appendix~\ref{sec:TNR}), we can
approximate the value of the classical central charge $c$ and the coupling $g$
in much of the XY phase and estimate
the contours of that phase. We find a region in which the
estimated value of $c$ converges to approximately 1 and the estimated value of
$g$ converges to varying values: $g \simeq 4$ on the boundary of this region
and increases to a value of about 5.5 going inward towards the origin.
In Fig.~\ref{fig:XYmaps20} we use TNR with bond dimensions 
$\cchip{20}{14}$ (which regulate
the size of renormalized degrees of freedom in each half-step and intermediate
step, respectively) to define this region:
Its outer boundary intersects the $a_1$ axis at approximately 1.2 and the
$a_2$ axis at approximately 1.1; its inner boundary is unclear. As
will be discussed later, the results of this scan are not conclusive in the
inner region, which requires analysis with higher bond dimensions.

Up to a certain point, however, the values from this analysis prove robust when
we increase bond dimension, as shown for two values from the interior of the XY
region in Fig.~\ref{fig:XYcompare-interior}. Furthermore, in
Appendix~\ref{app:tower}, we analyze the conformal tower at these points
in order to get a convincing confirmation that the doubled vertex model
has as its infrared limit a compactified-free-boson CFT.

\begin{figure}[h!]
\includegraphics[width=.5\textwidth]{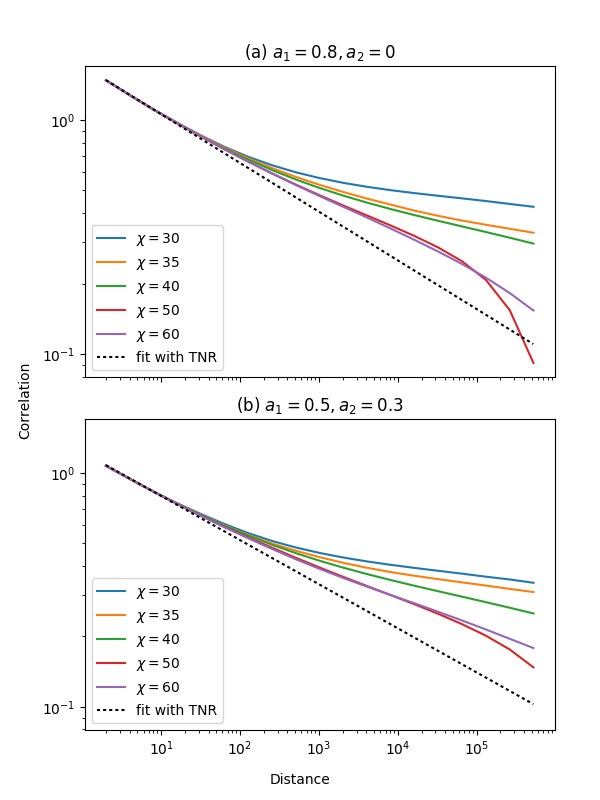}
\caption{Correlation functions of $S_x$ from HOTRG, on a $2^{21}\times 2^{21}$
torus, compared with power-law
estimates of the form $C(L) = C_0L^{-\eta}$, where $\eta=2\Delta=1/g$ is
determined by the TNR estimates in Fig.~\ref{fig:XYcompare-interior} and
$C_0$ is estimated to provide the best tangent line. In both cases
we see the measured curve approaching the power-law estimate with
increasing bond dimension. (a) At $a_1=.8,a_2=0$, the TNR estimate
gives $\eta=0.1887$, and we estimate $C_0=1.711$. (b) At $a_1=.5,a_2=.3$, the
TNR estimate gives $\eta=0.2083$, and we estimate $C_0=1.2309$.}
\label{fig:XYcorr}
\end{figure}

In Appendix~\ref{app:XY-theory}, we attempt
to explain this using the spin-coherent-state
picture as presented by Haldane\cite{Haldane2D} (see also
Auerbach\cite{Auerbach}). Within that framework,
$S_x$ and $S_y$, as classical observables in the sense of
Fig.~\ref{fig:quantumclassical}, should be proportional to the
primary operators
$V_{1,0}\pm V_{-1,0}$, which have scaling dimension $\frac{1}{2g}$ and conformal
spin 0 and which transform into each other under the $k=1$ irrep of the
on-site $O(2)$ symmetry. In particular, we expect the correlation-function
behavior
\begin{equation}
\langle S_\phi(\vec{r})S_\phi(\vec{r}')\rangle \sim |\vec{r}-\vec{r}'|^{2-d-\eta} = |\vec{r}-\vec{r}'|^{-\frac{1}{g}},
\end{equation}
where $S_\phi$ is the combination $S_x\cos\phi + S_y\sin\phi$. As 
$S_\phi$ is a quantum observable as well, if this power-law relation holds
it implies quasi-long-range ordered behavior of the quantum model.

To this end we have attempted to extract
the correlations of $S_x$ at these points. Our current TNR methods
cannot efficiently calculate correlation functions at the
bond dimensions of our probe, so we have instead used HOTRG; but this method
cannot replicate long-range behavior of critical systems with finite bond
dimension, so we instead try to replicate the $\chi\to\infty$ limit by
increasing the bond dimension.  In Fig.~\ref{fig:XYcorr} we demonstrate
that the correlation functions thus obtained approach 
a curve $C_0r^{-\eta}$, where $\eta$ is the critical exponent expected from the
previously-discussed TNR studies and the coefficient $C_0$ is chosen to fit
the HOTRG data. This demonstrates that the quasi-long-range
order of the doubled vertex model does in fact reflect quasi-long-range order
of the quantum state. In Appendix~\ref{app:delta}, we additionally study
the critical exponent $\delta$.

% In Appendix~\ref{app:spin1}, we
%attempt to further apply this analysis by explaining the
%pseudo-quasi-long-ranged behavior in terms of suppression of tunneling
%effects.

\subsection{The Kosterlitz-Thouless transition between the XY and AKLT phases}
\label{sec:KT}

\begin{figure}[h!]
\includegraphics[width=.5\textwidth]{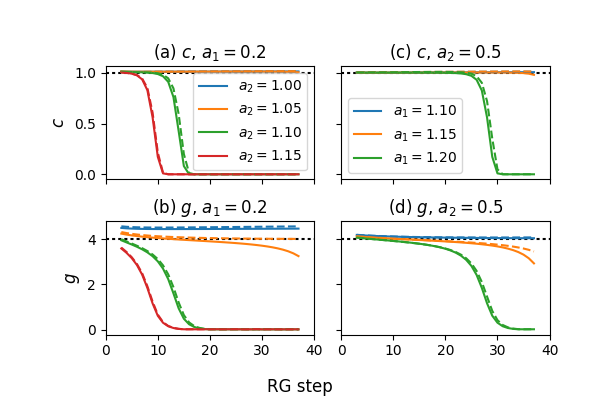}
\caption{We use TNR at $\cchip{20}{14}$ (dashed lines) and $\cchip{30}{20}$
(solid lines) to analyze the long-range behavior of the system in the
immediate vicinity of the KT-like transition at $a_1=0.2$ (a),(b) and
$a_2=0.5$ (c),(d). We examine values of our estimates for the classical central
charge $c$ (a),(c) and the coupling $g$ (b),(d) after successive coarse-graining
steps, and conclude:
  (i) While corrections to behavior from increasing bond dimension are
  present, they are unlikely to influence estimates of the location of the
  transition by more than about $|\Delta\vec{a}| \simeq 0.05$, which indicates
  that the $\chi=20$ behavior of Fig.~\ref{fig:XYmaps20} is largely accurate.
  (ii) The idea that a coupling $g<4$ induces the appearance of a length
  scale is readily confirmed; in fact in (b) we see that, at $a_1=0.2,a_2=1.05$,
  the system appears to flow to a non-trivial fixed point when $g$ barely fails
  to cross below 4 at $\chi=20$, but at $\chi=30$ a small correction to 
  shorter-ranged values of $g$ induces this crossing and causes a violation
  of scale invariance. Moreover, $c$ only deviates noticeably from 1 at
  length scales where $g$ is rapidly flowing towards 0, but invariably does
  so under those conditions.
  (iii) In (b) we see much more substantial corrections to $g$ at shorter
  length scales than in (d), likely related to the much stronger persistence
  of pseudo-quasi-long-range behavior near the points analyzed in (d) than
  (b) - specifically, the same perturbations which we expect to induce the
  transition for $g<4$ may also lead to corrections to the coupling for $g>4$,
  such that the suppression of these perturbations that increases the
  persistence of pseudo-quasi-long-range order near the $a_1$ axis may
  also reduce corrections to $g$ near the transition in that region.
}
\label{fig:KTcomp}
\end{figure}

\begin{figure}[h!]
\includegraphics[width=.5\textwidth]{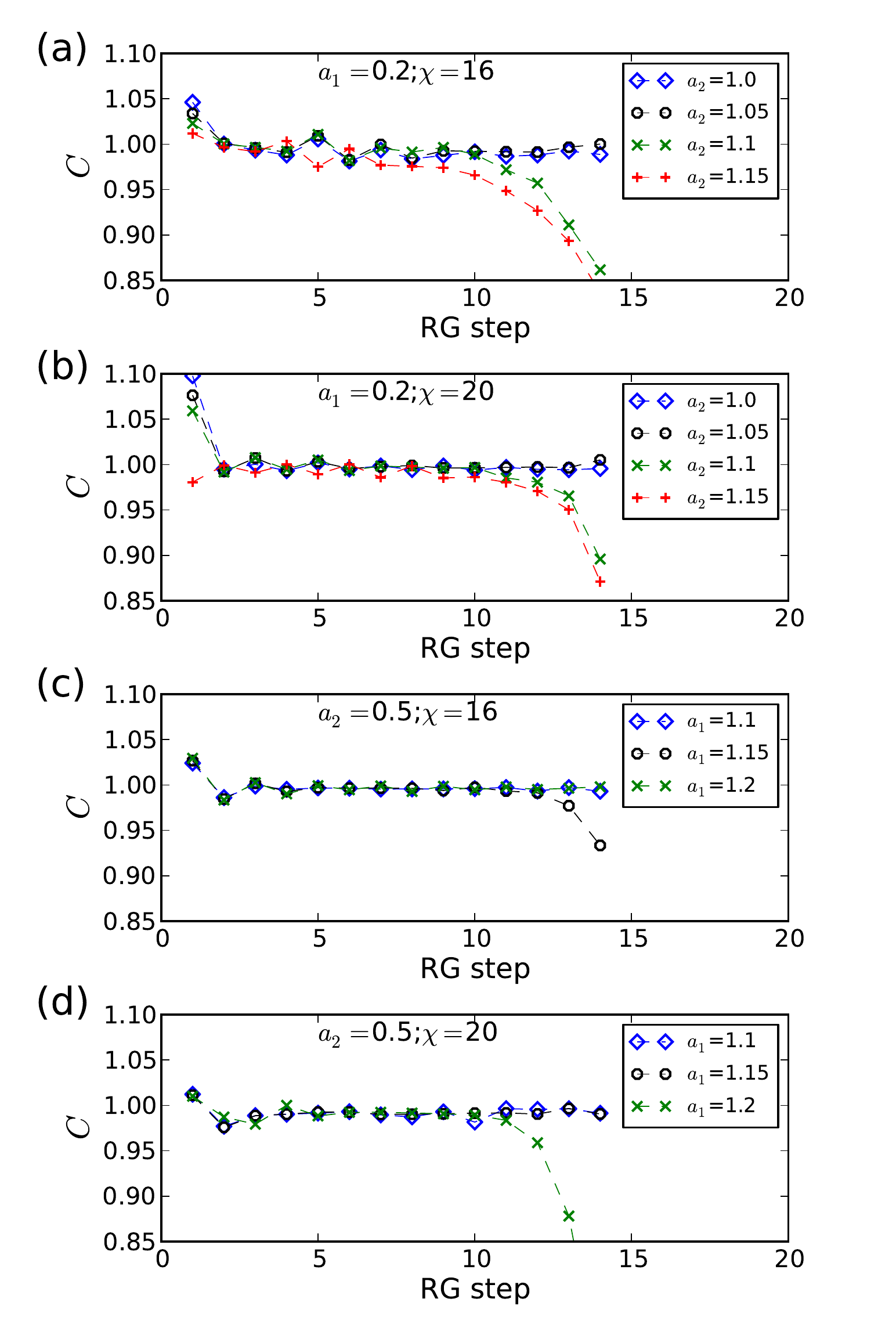}
\caption{We use loop-TNR to analyze the same points in the phase diagram
as in Fig.~\ref{fig:KTcomp}, estimating $c$ for $\chi=16$ and $\chi=20$.
Although the results are noisier than those obtained with TNR (possibly partly
since we did not preserve symmetries when using loop-TNR).
By $\chi=20$ the results are largely similar to the TNR results: when we
consider the KT transition at $a_1=0.2$, we observe that $c$ appears stable
for $a_2=1$ and $a_2=1.05$ but not $a_2=1.1$ or $a_2=1.15$; meanwhile, when
we consider the KT transition at $a_2=0.5$, we observe that $c$ appears stable
for $a_1=1.1$ and $a_1=1.15$ but not for $a_1=1.2$.}
\label{fig:loopKT}
\end{figure}

We have hypothesized that the XY phase is stable only when the long-range
value of the coupling $g$ is at least 4. Fig.~\ref{fig:XYmaps20} gives some
evidence for this at selected length scales and fixed bond dimension. In
Fig.~\ref{fig:KTcomp}, we take some points that we can predict to be near
the phase transition and observe how estimates for $c$ and $g$ behave at very
large length scales. The result confirms the expected behavior of $c$ and $g$ as
asymptotic values. Moreover, a substantial increase in bond dimension from that
used for Fig.~\ref{fig:XYmaps20} does not substantially alter the result,
lending confidence to our conclusion. In Fig.~\ref{fig:loopKT} we apply
loop-TNR and find results compatible with those obtained from
Evenbly and Vidal's TNR algorithm.
In Appendix~\ref{app:KT} we examine many points around the
transition to try to see a more thorough picture.

\begin{figure}[h!]
\includegraphics[width=.5\textwidth]{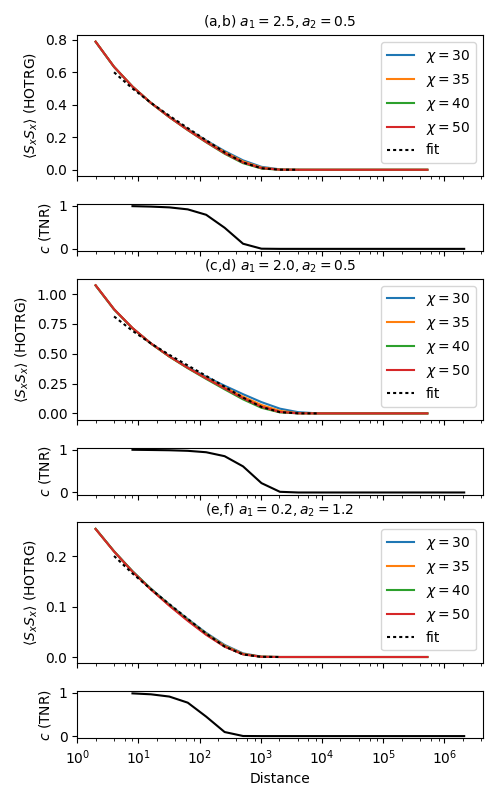}
\caption{Using HOTRG, we examine the correlation function
$\langle S_xS_x\rangle$ at several points inside the AKLT phase, varying
bond dimension to confirm convergence and fitting the highest-bond-dimension
curves to the exponential-decay form $C(r) = C_0 r^{-\eta}e^{-r/\xi}$.
Additionally, by interpolating from TNR data with $\cchip{20}{14}$, we
estimate the classical central charge at the length scale of the
correlation length $\xi$ determined by fitting and find
that the sharp falloff in estimated $c$ roughly predicts the correlation
length, with $c(\xi) \sim 0.35$. (Note that, as we obtain $c$ by comparing
transfer matrices $M$ of different length, our estimates of $c$ do not
correspond to a precise length; here we use $c(L)$ to refer to the $c$
obtained by comparing $M(3L)$ and $M(2L)$.)
(a),(b) At $a_1=2.5,a_2=0.5$, toward the outside of the pseudo-quasi-long-range
ordered region, we obtain $C_0 \simeq 0.853$, $\eta \simeq 0.247$,
and $\xi \simeq 365$, with $c(\xi) \simeq 0.30$.
(c),(d) At $a_1=2.0,a_2=0.5$, deeper within the pseudo-quasi-long-range ordered
region, we obtain $C_0 \simeq 1.113$, $\eta \simeq 0.223$,
and $\xi \simeq 693$, with $c(\xi) \simeq 0.44$.
(e),(f) At $a_1=1.2,a_2=0.2$, near the $a_2$ axis and thus further from
pseudo-quasi-long-range ordered effects,
we obtain $C_0 \simeq 0.287$, $\eta \simeq 0.245$,
and $\xi \simeq 203$, with $c(\xi) \simeq 0.21$.
}
\label{fig:VBScorr}
\end{figure}

We can also use correlation functions to confirm that points on either side of
the supposed transition lie in, respectively, critical and non-critical phases.
Contrasting Fig.~\ref{fig:XYcorr} with
corresponding data from outside of the transition as in Fig.~\ref{fig:VBScorr},
we find that in the latter case the correlation function converges to a form
that exponentially decays to machine epsilon.

We see that, within what we label the XY phase, our estimates of $g$
converge with successive coarse-graining operations, with small negative
corrections that grow as we approach the transition,
at which $g$ converges to a value
of approximately 4. Beyond the transition, the estimated $g$ falls to zero,
either starting below 4 and quickly dropping, or starting above 4, slowly
declining until it passes below 4, and then quickly dropping,
in both cases indicating that scale-invariance fails and so the doubled
vertex model departs from the conformal invariance of the XY
phase\footnote{Recall that
our estimates of $g$ are inversely proportional to our estimates of the first
scaling dimension, determined from the logarithm of the ratio of
the first two  eigenvalues of the transfer matrix
(or the first two with $U(1)$ charges 0 and 1, respectively).
Thus, when we observe the estimated value of $g$ falling to 0 under
coarse-graining, we can typically assume that this implies the opening of a gap.
These cases should not be confused
with actual free-boson CFTs with $g < 1$, which are equivalent to
$g > 1$ theories under T-duality, which exchanges ``magnetic'' vertex operators
with ``electric'' operators and takes $g \leftrightarrow 1/g$ -
in which case the vertex operators with the least scaling dimension are now
``magnetic'' rather than ``electric''.}. Either way, the estimate
for $c$ plateaus at approximately 1 and diverges noticeably from that value
only when $g$ is significantly less than 4 at the length scale in question.
In fact, in Fig.~\ref{fig:VBScorr} we will find that the length scale at which
$c$ measurably decays approximately predicts the correlation length.

\subsection{The pseudo-quasi-long-range ordered region}
\label{sec:pqlro}

Informed by the observation that the decay of $c$ roughly predicts
correlation length, we note that we can generally estimate $c$ to be
about 1 much further into the AKLT phase for small $a_2$
than it does for small $a_1$; in particular this behavior is most evident in
Figs.~\ref{fig:XYmaps20} and \ref{fig:KTlines}.
This suggests an extended region of
``pseudo-quasi-long-range order'' (behavior which imitates that of the
quasi-long-range ordered XY phase up to length scales large enough that they
may be experimentally impractical)
near the $a_1$ axis. In fact in Fig.~\ref{fig:squarephase}, we delineate
a region ``above'' the XY phase that is nearly as large, which we believe to
have correlation length of about 1000 times the lattice spacing or more.
(Such behavior has, notably, interfered with attempts to delineate the XY phase
using methods other than TNR, even at relatively high bond dimensions.)
Based on the
analysis of Haldane\cite{Haldane2D}, we believe this is because the system in
this region resembles a spin-1 antiferromagnet well enough that tunneling
processes of odd
winding number, including those that induce the KT transition, are almost
suppressed and can therefore only weakly break scale invariance. We
present this argument in somewhat more detail in Appendix~\ref{app:spin1}.

We may also confirm the KT transition using the physical critical
exponent $\delta$, as in Appendix~\ref{app:delta}, or 
use the corner entropy to approximate the boundary of the XY
phase, as in Appendix~\ref{app:corner}.  The results are consistent with those
from TNR, although not as precise.

\subsection{The elusive product-state phase}
\label{sec:prodstate}
\begin{figure}[h!]
\includegraphics[width=.5\textwidth]{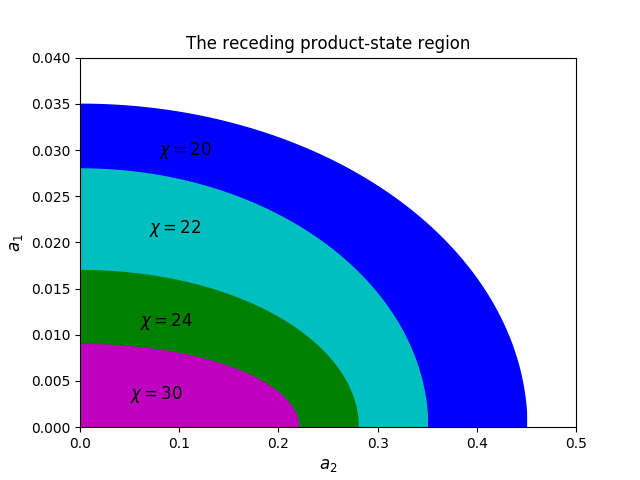}
\caption{A very rough estimate of the bounds of the product-state region
obtained using TNR at various bond dimensions. For a given bond dimension
$\chi$, points within the regions labeled with $\chi$ or $\chi_2>\chi$
are expected to correspond to parameters for which TNR will report trivial
behavior at that bond dimension; outside of that region XY-like behavior
is expected at that bond dimension. Most of the data used to
establish these boundaries can be found in Figs. \ref{fig:prodpts} and
\ref{fig:prodKT}.}
\label{fig:prodchi}
\end{figure}

\begin{figure}[h!]
\centering
\includegraphics[width=.5\textwidth]{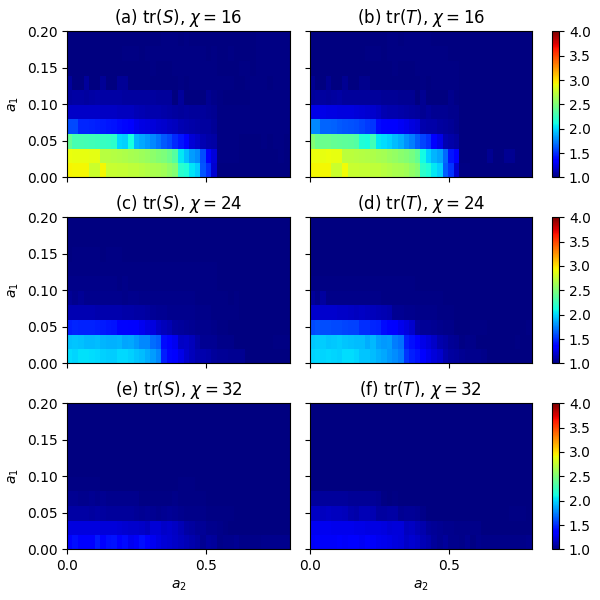}
\caption{Traces of the simulated modular $S$ and $T$ matrices obtained by
inserting virtual symmetry matrices
$\sigma_z$, using HOTRG with bond dimensions
(a),(b) $\chi\nobreak\!=\nobreak\!16$ (c),(d) $\chi\nobreak\!=\nobreak\!24$ (e),(f) $\chi\nobreak\!=\nobreak\!32$. We see that both traces
appear to be
close to 1 in much of the XY phase, but rise toward the trivial value of
4 in a small region within $a_1\!<\!.04, \,a_2\!<\!.4$, similar to
the estimates in Fig.~\ref{fig:prodKT}. However, as the bond dimension grows,
the extent of this region appears to shrink, much as in Fig.~\ref{fig:prodchi}.}
\label{fig:prodST}
\end{figure}

\begin{figure}[h!]
\includegraphics[width=.5\textwidth]{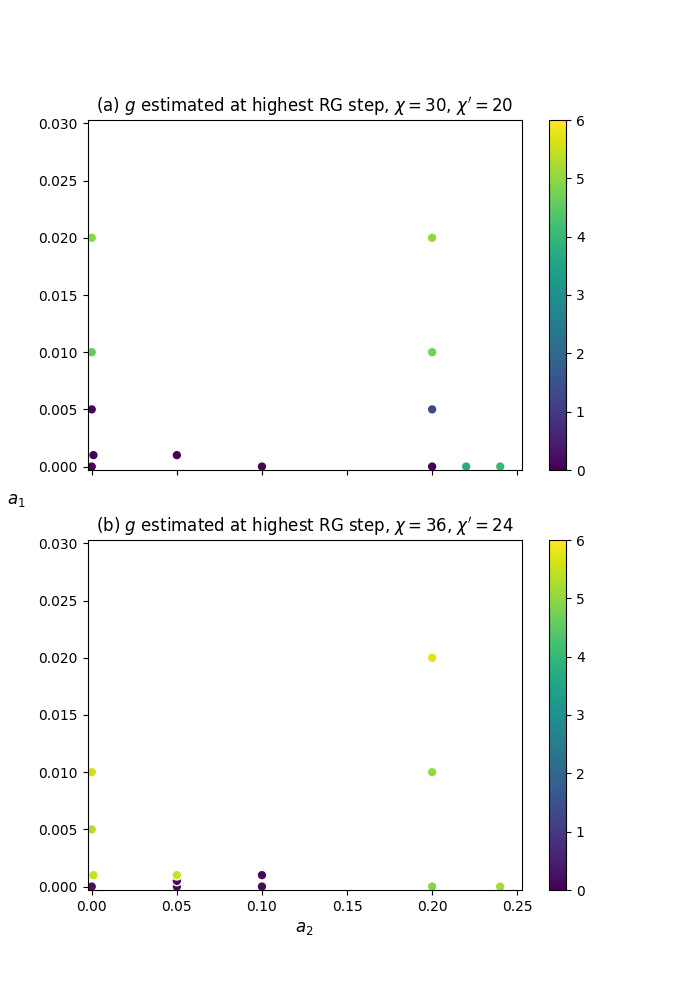}
\caption{We perform 12 coarse-graining steps of TNR at several points near
the origin, estimating $g$ after the final coarse-graining step and using
that to determine the behavior in this region. (a) For $\cchip{30}{20}$,
$g$ consistently falls as we approach the origin within the area shown,
inducing a KT-like transition into an apparent trivial phase in a region
which intersects the $a_1$ axis at $a_1 \simeq 0.01$ and the $a_2$ axis at
$a_2 \simeq 0.2$. (b) For $\cchip{36}{24}$, however, there does not
appear to be a well-defined product-state phase, at least for $a_1\geq 0.001$.
Rather, we observe fluctuations in the estimated value of $g$ with no
discernable pattern and with non-trivial limiting values of $g$ found
as close to the origin as $(a_1,a_2) = (0.001,0.001)$.}
\label{fig:prodpts}
\end{figure}

When $a_1 = a_2 = 0$, however, the deformation projects the state onto the
product $|0,0,\ldots,0,0\rangle$. Na\"ively, we expect to find a gapped,
entirely trivial phase in a region around this.  But the amplitude of
$|0,0,\ldots,0,0\rangle$ in the AKLT state is precisely
equal to the partition function
of a six-vertex model (specifically at the ``square-ice'' point
$a=b=c=1$), whose degrees of freedom are reflected in the
virtual, or entanglement, degrees of freedom of the PEPS
representation of this state.  On the square
lattice this model has been well-studied and is known to be critical,
behaving in the infrared as the compactified free boson CFT with coupling
$g=\frac{1}{3}$.  At this point,
therefore, the doubled vertex model is critical with $c=2$ despite the
quantum state being trivial in every way.

In TNR studies of the region of the phase diagram surrounding this
point, when we approach the origin 
\textit{with fixed bond dimension}
we find that the system appears to encounter a second
KT-like transition, visible in Fig.~\ref{fig:XYmaps20}: $g$ appears to fall
from a maximum value, ultimately reaching a value of 4 after which the behavior
ceases to be critical, with $c$ ultimately falling to 0 rather than remaining
stable. This behavior is shown in more detail in Appendix~\ref{app:KT}.
We may also observe such behavior by analyzing simulated $S$ and $T$ traces
as in Fig.~\ref{fig:prodST}, which appear to reach a trivial value of 4 in
this region.

However, when we increase the bond dimension, we find that the boundary of
this transition recedes towards the origin as in Fig.~\ref{fig:prodchi},
and points which appear to have
finite correlation length at lower bond dimension tend to obtain a central
charge of 1 at higher bond dimensions.
In Appendix~\ref{app:linecomp} we analyze individual points
near the origin of the phase diagram and find that results are highly sensitive
to bond dimension. In fact, when we
increase the bond dimension from $\cchip{30}{20}$ to $\cchip{36}{24}$,
rather than straightforward behavior in a well-defined region we find noisy
fluctuations in $g$ (some of which do pass below 4), as in
Fig.~\ref{fig:prodpts}. Some results from $\cchip{42}{28}$ are also
presented in Appendix~\ref{app:linecomp}; they do little to clarify this
picture.

The principal alternatives we should consider are that:
\begin{enumerate}
  \item The product-state phase has finite, but small, extent and is defined
  by a KT-like transition much as the data in Fig.~\ref{fig:XYmaps20} suggests;
  \item The product-state phase does not exist; rather, the coupling $g$
  \begin{enumerate}
    \item peaks along some curve, before falling to a limiting value,
    likely 4, approaching the origin;
    \item keeps rising to a limiting value of 6 or greater approaching the
    origin; or
    \item keeps rising to $\infty$, much as at the ferromagnetic Heisenberg
    point of the XXZ model.\cite{Fendley}
  \end{enumerate}
\end{enumerate}
It is currently unclear which of these is most likely, but evidence does suggest
the absence of a separate gapped and completely trivial phase.

To this lack of evidence we add that the arguments we have used to
justify our expectation of a trivial phase are inherently flawed.
The parent Hamiltonian\cite{NKZspin2} derived by Niggemann, Kl\"umper,
and Zittartz does not
extend to the limiting case $a_1=a_2=0$, nor does the formulation in
\eqref{eqn:Ha}: in order to project out spin values, a parent Hamiltonian
will generally have to increase its rank, which is impossible to do via
continuous deformation. For example, at all points in the interior of the
phase diagram, the two-site parent Hamiltonian will annihilate
$\ket{12}-\ket{21}$ (see Eq.~11 of Niggemann, Kl\"umper, and
Zittartz\cite{NKZspin2}); therefore, so must the limit of any sequence of
such Hamiltonians. For $a_1\to 0$ or $a_2\to 0$, this introduces the
possibility that
some sites may have (respectively) $S_z=1$ or $S_z=2$ with nonzero probability
in the ground space of the limiting Hamiltonian, even though they have been
projected out of the deformed
state.

We also note that, if we were to include the product state at $a_1=a_2=0$ in the
phase diagram, we would be implicitly suggesting that nearby states in
the phase diagram could be obtained by perturbing this product state, which
should in turn imply that it is enclosed in the phase diagram by a phase
of finite correlation length. However, we suggest that such ``perturbations''
may instead have quasi-long-range correlations and therefore should, in the
thermodynamic limit, dominate the original (product) state at any magnitude.
We conclude, therefore, that if we wish to connect the point $a_1=a_2=0$ with
the surrounding points of the phase diagram, we should do so with extreme
caution.

\section{Re-examining the XY phase on the honeycomb lattice}
\label{sec:honeycomb}

\begin{figure}[h!]
\includegraphics[width=0.5\textwidth]{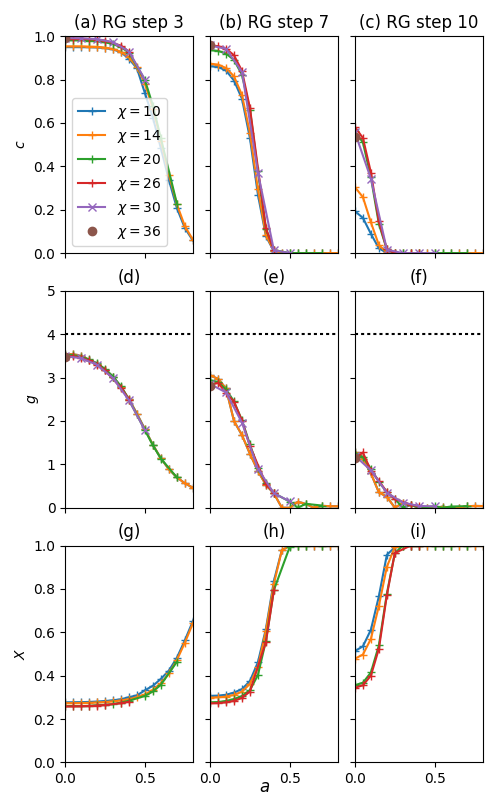}
\caption{We use TNR at various bond dimensions to analyze the honeycomb-lattice
system.
(a) At linear system size $O(10)$, a region extending to $a \simeq 0.4$ appears
to have long-range behavior with $c=1$.
(b) For system length $O(100)$, this has shrunk to
within $a \simeq 0.1$. (c) By system length $O(1000)$, all points clearly do
not exhibit quasi-long-range ordered behavior.
(d) The value of $g$ estimated at small system sizes is close to 3.5 at
$a=0$, and smoothly falls off leaving the ``pseudo-quasi-long-range ordered''
region. (e),(f) At all points the estimate of $g$ gradually falls off to 0.
(g),(h),(i) TNR estimates of the Chen-Gu-Wen $X$-ratio \cite{ChenGuWen} rise towards
the AKLT-phase value of 1. Within the pseudo-quasi-long-range ordered region,
however, they may appear to take a different, nontrivial intermediate value 
up to fairly large length scales.}
\label{fig:hexregion}
\end{figure}

\begin{figure}[h!]
\includegraphics[width=0.5\textwidth]{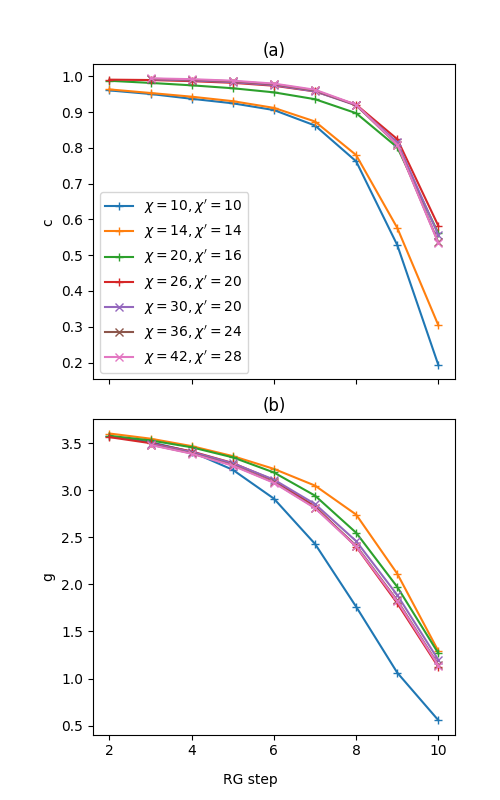}
\caption{We use TNR to analyze the honeycomb system at $a=0$, the most likely
candidate for critical behavior.
(a) Although increasing bond dimension tends to raise the estimated
value of the classical central charge $c$ towards 1, as a function of
the length scale $c$ appears to converge to
a decaying form at the highest bond dimensions tested.
(b) When we estimate the coupling $g$, we find that its initial value always
appears to be close to $3.6 < 4.0$, and that it never
appears to be stable under coarse-graining.
We additionally note that several successive increases in bond dimension, from
$\cchip{26}{20}$ to ultimately $\cchip{42}{28}$,
do not substantially affect the estimates for either $c$ or $g$.
}
\label{fig:hexpoint}
\end{figure}
In Huang, Wagner and Wei\cite{AKLTspin32}, the analogous model for the
honeycomb lattice was
examined using tensor-network methods, and it was concluded that a
quasi-long-range ordered phase 
exists close to the value $a=0$ of the perturbation parameter.
Having refined our analysis to more sensitively
judge the properties and boundaries of such a phase, as in
Sec.~\ref{sec:XY-properties}, we return to
this model. The numerical methods are the same as for the
square-lattice model (save for a
correction in order to account for the anisotropy induced by blocking
honeycomb-lattice sites into square-lattice sites).

In Fig.~\ref{fig:hexregion}, we observe that successive applications
of TNR coarse-graining, with increasing bond dimension, fail to maintain
evidence of a ``transition'' to a quasi-long-range ordered region; rather,
we observe that what appears to be an XY phase shrinks in size as
RG proceeds. We nowhere and at no scale estimate $g\geq 4$, and what
was previously believed to be a phase with quasi-long-range order appears to
give way to a region with pseudo-quasi-long-range order similar to the
analogous region of the AKLT phase on the square lattice. Much as we claim
that, when $a_2\sim 0$, the square-lattice state approximates a spin-1
antiferromagnet in which isolated vortices of winding number 1 (modulo 2) are
suppressed by square-lattice symmetries, we claim that, when $a\sim 0$,
the honeycomb-lattice state approximates a spin-$\frac{1}{2}$ antiferromagnet
in which isolated vortices of winding number 1 and 2 (modulo 3) are
suppressed by honeycomb-lattice symmetries. If this is true, we expect
that XY couplings as low as $g=\frac{4}{9}$ should become ``approximately
stable,'' reproducing pseudo-quasi-long-range ordered behavior. 
In Fig.~\ref{fig:hexpoint}, we see that even at the point $a=0$, as we increase
the bond dimension we consistently observe behavior
consistent with finite, but large, correlation length.

\section{Discussion} \label{sec:Conclusion} 
We have used tensor-network methods to explore the two-parameter phase diagram
of the deformed-AKLT model on the square lattice. In addition to confirming
our expectations about the AKLT and N\'eel phases, we find a well-defined
quasi-long-range ordered phase with properties resembling those of the classical
XY model, including a Kosterlitz-Thouless (KT)-like transition.
Evenbly and Vidal's
TNR algorithm\cite{TNR} gives us a way to effectively and accurately extract a
substantial amount of information about this behavior; prior to its
development, we may not have even been able to conclusively demonstrate the
phase's existence, as was the case when tensor-network methods were previously
applied to a similar question.\cite{AKLTspin32}
Although we have \textit{not} been
able to efficiently use TNR to directly compute correlation functions, it has
yielded predictions about critical exponents $\eta$ and $\delta$ that we
have been able to roughly confirm with HOTRG. We also find from our
analysis that a ``pseudo-quasi-long-range ordered'' region of persistently large
correlation length extends from part of that transition. We explain this
by arguing that isolated tunneling processes are approximately suppressed,
a claim which could benefit from more rigorous analysis.

We have also re-examined the honeycomb case. Using the analysis that we have
applied to the XY phase of the square-lattice model, we have
found that the region previously identified
as an XY phase is instead a pseudo-quasi-long-range ordered region of the
AKLT phase; that model has no true XY phase.

We also find some peculiar behavior when the parameter $a_1$ is very small.
Aside from an apparent crossover in the AKLT/N\'eel transition in this limit,
we find that there is a region close to the origin ($a_2<0.3$, $a_1\ll 0.1$)
where the system's behavior is no longer evident. Although we have largely
exhausted our resources in attempting to determine the exact behavior in that
region using current methods, we may be able to extract more information
either by refining our techniques, for example by taking further advantage of
the symmetry\cite{Singh1}, or by analyzing the $a_1=0$ line specifically
with approaches that may only apply there. In doing so we would wish to
determine whether or not this region contains a distinct
phase with no long-range order, if so, what the nature of the transition is,
and if not, what the system's behavior is as $a_1$ and $a_2$ are both reduced
to 0.

Future work may examine the mechanism of the Kosterlitz-Thouless transition,
including the origin of the coupling which we have labeled $g$ and the role of
tunneling processes. Extensions to this system, such as deformations of 
spin-$2m$ AKLT states, a spin-1 AKLT-like state, or the kagome-lattice AKLT
state, may also give us more information about the underlying physics.
The state's behavior along the $a_1=0$ axis, and how it relates to the
behavior in the interior of the phase diagram as $a_1\to 0$,
may have much to tell us about the behavior of the XY phase near or in the
``product-state'' region.

\nocite{Kosterlitz}

\acknowledgments
The authors would like to acknowledge useful discussions with
Alexander Abanov, Ian Affleck, and
especially Cenke Xu, who suggested the physical picture for the
pseudo-quasi-long-range ordered region.
This work was partially supported by
the National Science Foundation under Grant No. PHY 1620252 and Grant
No. PHY 1314748.

\appendix
\FloatBarrier
\section {Review of  numerical methods }
\label{app:methods}

In this work, we employ several numerical tensor-network methods; here,
we briefly discuss each of them. We use different methods for their varying
strengths: that is, which quantities we can use them to (efficiently)
calculate, and which regimes they are accurate in.

\subsection{The corner transfer matrix method}
\label{app:CTMRG}

\begin{figure}
\includegraphics[width=0.45\textwidth]{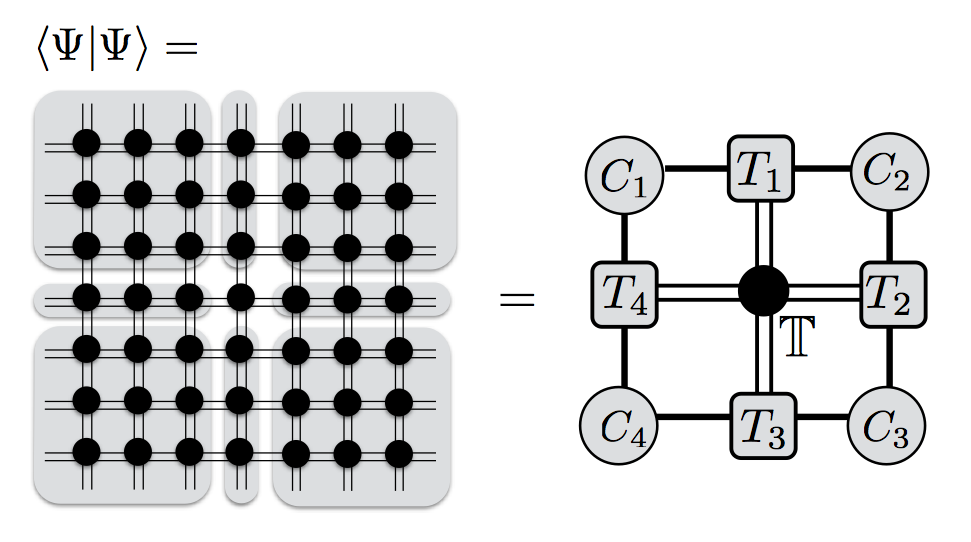}
\caption{In CTMRG, we represent each quadrant of the plane, as well as
rays extending from a given site, as individual tensors.
Typically we represent the
entire norm squared through a contraction of the four quadrant tensors,
four ray tensors, and one double tensor representing a single site. Additional
rows and columns can be inserted as needed.}
\label{fig:normal_CTM}
\end{figure}
We first describe the corner transfer matrix renormalization group (CTMRG),
which was first introduced by Baxter.\cite{Baxter}
It was further developed
and applied as a numerical method for analyzing classical statistical systems
by Nishino and Okunishi.\cite{Nishino_CTMRG_1997} More recently, it was
adapted to the contraction of tensor networks by Orus.\cite{OrusCTM}.

CTMRG represents a classical partition function, or equivalently the norm of
a quantum state as in \eqref{eqn:doubletensor}, using a tensor environment
$\{ C_1,C_2,C_3,C_4,T_1,T_2,T_3,T_4 \}$, where the $C_i$ represent quadrants
of the plane and the $T_i$ represent rays extending from the ``central'' site
whose environment we are considering, as in Fig.~\ref{fig:normal_CTM}.
To obtain these environment tensors, we initialize them randomly, and then
iteratively add rows and columns of sites, which we add to the environment
tensors, truncating pairs of bonds to prevent exponential growth of the bond
dimension.
Once the environment tensors have been determined, we can evaluate the partition
function and the
expectation values of local observables of a classical statistical system,
or the norm and the expectation values of a
quantum state represented by a tensor network state, as the entire network
eventually reduces to only a few sites, on which we can directly calculate
the necessary tensor trace.

In addition to computing expectation values, we can use CTMRG to
directly calculate correlation length, as follows. We may insert columns of
sites (or, more precisely, contracted $\{T_1,\mathbb{T},T_3\}$ triples) to
``lengthen'' the system and obtain correlation functions of large (horizontal)
distances; when we do the contracted network begins to resemble the partition
function of a one-dimensional statistical system, the columns 
$M = \op{tTr}(T_1\mathbb{T}T_3)$ being transfer matrices.  We note that
convergence of the environment tensors suggests that the columns at either end
approximate the dominant left and right eigenvectors of these transfer matrices.
If $\lambda_i$ are the eigenvalues ordered by magnitude, this means that
for some operator $\mathcal{O}$ with
$\op{tTr}(T_1A\mathcal{O}A^* T_3)=M_\mathcal{O}$, similar to the correspondence
in Fig.~\ref{fig:quantumclassical},
\begin{align}
\langle \mathcal{O}(0)\mathcal{O}(L)\rangle
&= \frac{v_LM_\mathcal{O}M^{L-1}M_\mathcal{O}v_R}{v_LM^{L+1}v_R} \notag\\
&= C_0 + C_1\left(\frac{\lambda_1}{\lambda_0}\right)^L + O\left((\lambda_2/\lambda_0)^L\right),
\end{align}
where constants $C_0$ and $C_1$ can be easily determined from $M_\mathcal{O}$
and the eigenvalues and eigenvectors of $M$. Thus the correlation function
decays exponentially with decay constant $\op{log}(\lambda_0/\lambda_1)$.

We may also use CTMRG to extract a quantity which we refer to as the ``corner
entropy''. We may imagine a classical partition function as representing a
Euclidean path integral of a 1D quantum system\cite{TNRMERA}, with the upper and
lower half-planes representing the ket and bra vectors, respectively. In this
case, once we have completed the CTMRG procedure and represented the partition
function as the contraction $\op{tTr}(C_1C_2C_3C_4)$, we should expect that
the bond between $C_1$ and $C_4$ should approximate the ``physical'' state
of the entire left half of the line
and the bond between $C_2$ and $C_3$ should likewise approximate the
``physical'' state of the right half of the line. Then we expect
the tensor trace of these four tensors with the bond between $C_1$ and $C_4$
left open to approximate the (non-normalized) reduced density matrix of the
left half of the system - that is, its spectrum should resemble a truncated
version of the actual reduced density matrix.
Then the entropy of the spectrum of the open
tensor trace $\rho_\text{corner} = \op{tr}(C_1C_2C_3C_4)$
should approximate the entanglement entropy between the left and
right rays. We thus call the quantity
$S_\text{corner}=-\!\op{tr}(\rho_\text{corner}\log \rho_\text{corner})$
the ``corner entropy.''

In a technique closely related to CTMRG, which we refer to as the quantum-state
corner transfer matrix method or quantum-state renormalization group
(QSRG), we may use the tensor environment
to represent the quantum state rather than the norm-squared.\cite{QSRG}
Each environment
tensor has a ``physical'' index which represents the physical content of the
state in the respective region of the lattice and which is renormalized
along with the virtual bonds as rows and columns of sites are added. This
method can be useful for approximating entanglement spectra.

\subsection{Higher-order TRG}
\label{sec:HOTRG}

\begin{figure}
\centering
\newcommand \isomdist{1.6cm}
\newcommand \sitedist{.5cm}
\newcommand \figwidth{4.3cm}
\newcommand \sitespace{\figwidth+\sitedist+.3cm}
\begin{tikzpicture}[scale=0.85, every node/.style={transform shape}]
  \clip (-\figwidth,-\figwidth) rectangle (\figwidth,\figwidth);
  \node[coordinate] (11) at (0,0) {};
  \node[coordinate, left = \sitespace of 11] (01) {};
  \node[coordinate, right = \sitespace of 11] (21) {};
  \node[coordinate, above = \sitespace of 11] (10) {};
  \node[coordinate, left = \sitespace of 10] (00) {};
  \node[coordinate, right = \sitespace of 10] (20) {};
  \node[coordinate, below = \sitespace of 11] (12) {};
  \node[coordinate, left = \sitespace of 12] (02) {};
  \node[coordinate, right = \sitespace of 12] (22) {};
  \node[site, above left = \sitedist of 11] (A11tl) {};
  \node[site, above right = \sitedist of 11] (A11tr) {};
  \node[site, below left = \sitedist of 11] (A11bl) {};
  \node[site, below right = \sitedist of 11] (A11br) {};
  \node[site, below left = \sitedist of 10] (A10bl) {};
  \node[site, below right = \sitedist of 10] (A10br) {};
  \node[site, above left = \sitedist of 12] (A12tl) {};
  \node[site, above right = \sitedist of 12] (A12tr) {};
  \node[site, above right = \sitedist of 01] (A01tr) {};
  \node[site, below right = \sitedist of 01] (A01br) {};
  \node[site, above left = \sitedist of 21] (A21tl) {};
  \node[site, below left = \sitedist of 21] (A21bl) {};
  \node[site, above left = \sitedist of 22] (A22tl) {};
  \node[site, above right = \sitedist of 02] (A02tr) {};
  \node[site, below left = \sitedist of 20] (A20bl) {};
  \node[site, below right = \sitedist of 00] (A00br) {};
  \node[isometry, above = \isomdist of 11] (u11t) {};
  \node[isometryconj, below = \isomdist of 11] (u11b) {};
  \node[leftisom, right = \isomdist of 11] (u11r) {};
  \node[rightisom, left = \isomdist of 11] (u11l) {};
  \node[isometry, above = \isomdist of 12] (u12t) {};
  \node[isometryconj, below = \isomdist of 10] (u10b) {};
  \node[leftisom, right = \isomdist of 01] (u01r) {};
  \node[rightisom, left = \isomdist of 21] (u21l) {};
  \draw (A11tr) -- (u11t.south-|A11tr)
        (A11tl) -- (u11t.south-|A11tl)
        (A11tl) -- (A11tl-|u11l.east)
        (A11bl) -- (A11bl-|u11l.east)
        (A11bl) -- (u11b.north-|A11bl)
        (A11br) -- (u11b.north-|A11br)
        (A11br) -- (A11br-|u11r.west)
        (A11tr) -- (A11tr-|u11r.west)
        (A12tr) -- (u12t.south-|A12tr)
        (A12tl) -- (u12t.south-|A12tl)
        (A21tl) -- (A21tl-|u21l.east)
        (A21bl) -- (A21bl-|u21l.east)
        (A10bl) -- (u10b.north-|A10bl)
        (A10br) -- (u10b.north-|A10br)
        (A01br) -- (A01br-|u01r.west)
        (A01tr) -- (A01tr-|u01r.west);
  \draw (u11t.apex) to (u10b.apex);
  \draw (u12t.apex) to (u11b.apex);
  \draw (u21l.apex) to (u11r.apex);
  \draw (u11l.apex) to (u01r.apex);
  \draw (A11tr) to (A11tl);
  \draw (A11tr) to (A11br);
  \draw (A11tl) to (A11bl);
  \draw (A11br) to (A11bl);
\end{tikzpicture}
\caption{Spin-blocking with HOTRG, wherein projective truncations are used to
reduce pairs of vertical bonds, and then pairs of horizontal bonds, into single
bonds of comparable bond dimension. Four sites are then turned into a single
site, with the interior bond of the projective truncation used as the new bond
of the coarse-grained ``site''.}
\label{fig:HOTRG}
\end{figure}
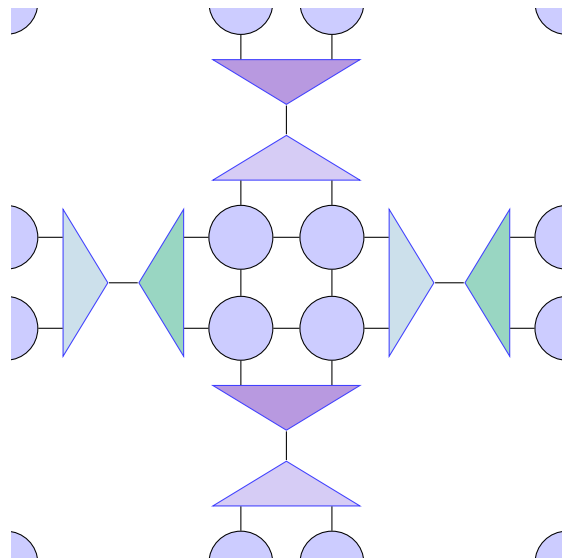
The remaining methods we will describe are coarse-graining methods,
implementations of real-space renormalization inspired by Kadanoff's
spin-blocking method. In these methods, we combine blocks of site tensors
into a single unit and apply some method to discard (hopefully irrelevant)
information so that the bond space of the blocked tensor remains manageable.
The resulting tensor represents a site of a coarse-grained lattice.
In this way, we ``renormalize'' an $N\times N$ partition function
into an $N/2\times N/2$ partition function and repeat until the
partition function can be directly evaluated. We typically think of this as
an implementation of the renormalization group in which the couplings
manifest as tensor elements.

The simplest of the methods that we will use is
Xie et al.'s higher-order tensor renormalization group 
\cite{Xiang_HOTRG_2012} (HOTRG). The fundamental
idea is projective truncation: We take a pair of bonds, each contracted between
two different tensors, and insert an orthogonal projector
\begin{equation}
{\mathbb{P}^{ij}}_{i'j'} = {W^{ij}}_k{{W^\dagger}_{i'j'}}^k,
\end{equation}
where the tensor $W$ is an isometry between the product of the vector spaces
denoted by indices $i$ and $j$, and the vector space denoted by index $k$ which
will represent renormalized degrees of freedom. The
isometries are typically determined iteratively, using the principle that
given an environment matrix $E$, when seeking a unitary matrix $W$, the
contraction $\op{Tr}(EW)$ is minimized by $W = U^\dagger V$, where 
$E = UDV^\dagger$ is the singular value decomposition of $E$.
Once the unitaries have been
determined, they are typically then contracted with other tensors in such a
way that the new virtual index $k$ becomes a bond index.

In HOTRG, we use projective truncation to coarse-grain bonds by pairing
them together - first vertically, then horizontally. This means that an
$N\times M$-site lattice becomes an $N\times M/2$-site lattice, and then
a $N/2\times M/2$-site lattice. The way that bonds are collected to accomplish
this is demonstrated in Fig.~\ref{fig:HOTRG}.

HOTRG provides a straightforward method for consolidating the information that
is ``located'' at adjacent sites or edges, and as such allows for a relatively
controlled computational cost in calculating expectation values and
correlation functions.  This also provides a way to renormalize string defects
expressed as matrix product operators,
such as those used to calculate modular $S$ and $T$ matrices in topological
models \cite{HeST} or simulated modular $S$ and $T$ matrices in
SPT models \cite{tnST}, by collecting defect tensors that lie along adjacent
bonds and coarse-graining them along with the bonds.

\subsection{Tensor Network Renormalization}
\label{sec:TNR}

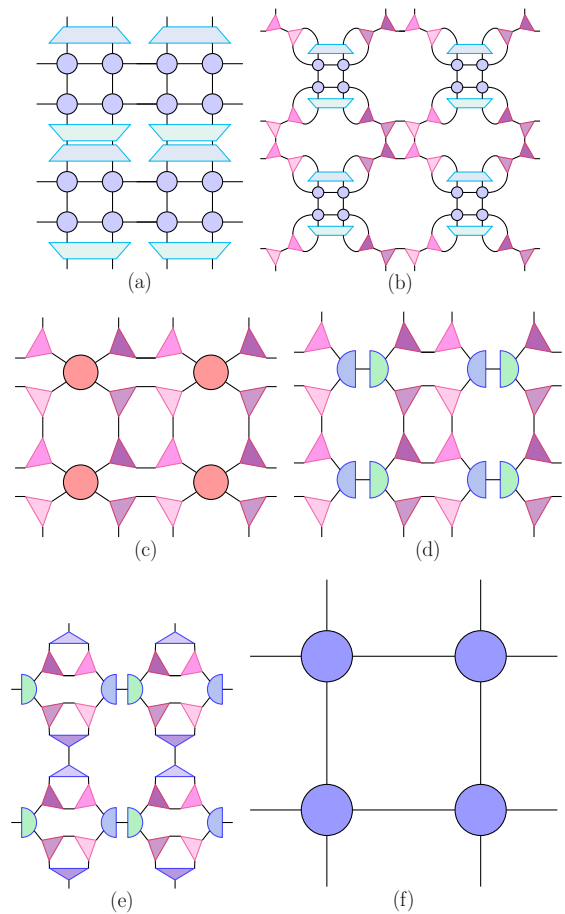
\begin{figure}
\centering
% Effective font sizes will be 8 - use 8/(scale)
\begin{subfigure} % Step 1: Disentangle
  \centering
    \begin{tikzpicture}[scale=0.27, every node/.style={transform shape}]
      \node[disentangler] (u1) at (0,0) {};
      \node[entangler,below = 4cm of u1] (uc1) {};
      \node[site,below = .5cm of u1.200] (Atl1) {};
      \node[site,above = .5cm of uc1.160] (Abl1) {};
      \node[site,below = .5cm of u1.340] (Atr1) {};
      \node[site,above = .5cm of uc1.20] (Abr1) {};
      \node[disentangler,below=5cm of u1] (u2) {};
      \node[entangler,below = 4cm of u2] (uc2) {};
      \node[site,below = .5cm of u2.200] (Atl2) {};
      \node[site,above = .5cm of uc2.160] (Abl2) {};
      \node[site,below = .5cm of u2.340] (Atr2) {};
      \node[site,above = .5cm of uc2.20] (Abr2) {};
      \node[disentangler,right=1.6cm of u1] (u3) {};
      \node[entangler,below = 4cm of u3] (uc3) {};
      \node[site,below = .5cm of u3.200] (Atl3) {};
      \node[site,above = .5cm of uc3.160] (Abl3) {};
      \node[site,below = .5cm of u3.340] (Atr3) {};
      \node[site,above = .5cm of uc3.20] (Abr3) {};
      \node[disentangler,below=5cm of u3] (u4) {};
      \node[entangler,below = 4cm of u4] (uc4) {};
      \node[site,below = .5cm of u4.200] (Atl4) {};
      \node[site,above = .5cm of uc4.160] (Abl4) {};
      \node[site,below = .5cm of u4.340] (Atr4) {};
      \node[site,above = .5cm of uc4.20] (Abr4) {};
      \draw (uc1.160) to (Abl1) ;
      \draw (uc1.20) to (Abr1) ;
      \draw (uc1.200) to (u2.160);
      \draw (uc1.340) to (u2.20);
      \draw (u1.200) to (Atl1) ;
      \draw (u1.340) to (Atr1) ;
      \draw (u1.160) to +(0,+.5);
      \draw (u1.20) to +(0,+.5);
      \draw (Abl1) to (Abr1);
      \draw (Atl1) to (Atr1);
      \draw (Abl1) to (Atl1);
      \draw (Atr1) to (Abr1);
      \draw (Abl1) to +(-1.5,0);
      \draw (Abr1) to (Abl3);
      \draw (Atl1) to +(-1.5,0);
      \draw (Atr1) to (Atl3);
      \draw (uc2.160) to (Abl2) ;
      \draw (uc2.20) to (Abr2) ;
      \draw (uc2.200) to +(0,-.5);
      \draw (uc2.340) to +(0,-.5);
      \draw (u2.200) to (Atl2) ;
      \draw (u2.340) to (Atr2) ;
      \draw (Abl2) to (Abr2);
      \draw (Atl2) to (Atr2);
      \draw (Abl2) to (Atl2);
      \draw (Atr2) to (Abr2);
      \draw (Abl2) to +(-1.5,0);
      \draw (Abr2) to (Abl4);
      \draw (Atl2) to +(-1.5,0);
      \draw (Atr2) to (Atl4);
      \draw (uc3.160) to (Abl3) ;
      \draw (uc3.20) to (Abr3) ;
      \draw (uc3.200) to (u4.160);
      \draw (uc3.340) to (u4.20);
      \draw (u3.200) to (Atl3) ;
      \draw (u3.340) to (Atr3) ;
      \draw (u3.160) to +(0,+.5);
      \draw (u3.20) to +(0,+.5);
      \draw (Abl3) to (Abr3);
      \draw (Atl3) to (Atr3);
      \draw (Abl3) to (Atl3);
      \draw (Atr3) to (Abr3);
      \draw (Abl3) to +(-1.5,0);
      \draw (Abr3) to +(1.5,0);
      \draw (Atl3) to +(-1.5,0);
      \draw (Atr3) to +(1.5,0);
      \draw (uc4.160) to (Abl4) ;
      \draw (uc4.20) to (Abr4) ;
      \draw (uc4.200) to +(0,-.5);
      \draw (uc4.340) to +(0,-.5);
      \draw (u4.200) to (Atl4) ;
      \draw (u4.340) to (Atr4) ;
      \draw (Abl4) to (Abr4);
      \draw (Atl4) to (Atr4);
      \draw (Abl4) to (Atl4);
      \draw (Atr4) to (Abr4);
      \draw (Abl4) to +(-1.5,0);
      \draw (Abr4) to +(1.5,0);
      \draw (Atl4) to +(-1.5,0);
      \draw (Atr4) to +(1.5,0);
    \node[below] at (current bounding box.south) {\fontsize{29.6}{29.6}\selectfont (a)};
    \end{tikzpicture}
\end{subfigure}
\begin{subfigure} % Step 2: Merge corner degrees of freedom
  \centering
    \begin{tikzpicture}[scale=0.15, every node/.style={transform shape}]
      \node[disentangler] (u1) at (0,0) {};
      \node[disentangler,below = 10.5cm of u1] (u2) {};
      \node[disentangler,right = 9cm of u1] (u3) {};
      \node[disentangler,below = 10.5cm of u3] (u4) {};
      \node[entangler,below = 4cm of u1] (uc1) {};
      \node[site,below = .5cm of u1.200] (Atl1) {};
      \node[site,above = .5cm of uc1.160] (Abl1) {};
      \node[site,below = .5cm of u1.340] (Atr1) {};
      \node[site,above = .5cm of uc1.20] (Abr1) {};
      \node[isovL,above left = 2.0cm and 1.5cm of Atl1] (vL1) {};
      \node[isovLc,below left = 2.0cm and 1.5cm of Abl1] (vLc1) {};
      \node[isovR,above right = 2.0cm and 1.5cm of Atr1] (vR1) {};
      \node[isovRc,below right = 2.0cm and 1.5cm of Abr1] (vRc1) {};
      \node[isovLf,above left = .2cm and 1cm of vL1] (vLf1) {};
      \node[isovLcf,below left = .2cm and 1cm of vLc1] (vLcf1) {};
      \node[isovRf,above right = .2cm and 1cm of vR1] (vRf1) {};
      \node[isovRcf,below right = .2cm and 1cm of vRc1] (vRcf1) {};
      \node[entangler,below = 4cm of u2] (uc2) {};
      \node[site,below = .5cm of u2.200] (Atl2) {};
      \node[site,above = .5cm of uc2.160] (Abl2) {};
      \node[site,below = .5cm of u2.340] (Atr2) {};
      \node[site,above = .5cm of uc2.20] (Abr2) {};
      \node[isovL,above left = 2.0cm and 1.5cm of Atl2] (vL2) {};
      \node[isovLc,below left = 2.0cm and 1.5cm of Abl2] (vLc2) {};
      \node[isovR,above right = 2.0cm and 1.5cm of Atr2] (vR2) {};
      \node[isovRc,below right = 2.0cm and 1.5cm of Abr2] (vRc2) {};
      \node[isovLf,above left = .2cm and 1cm of vL2] (vLf2) {};
      \node[isovLcf,below left = .2cm and 1cm of vLc2] (vLcf2) {};
      \node[isovRf,above right = .2cm and 1cm of vR2] (vRf2) {};
      \node[isovRcf,below right = .2cm and 1cm of vRc2] (vRcf2) {};
      \node[entangler,below = 4cm of u3] (uc3) {};
      \node[site,below = .5cm of u3.200] (Atl3) {};
      \node[site,above = .5cm of uc3.160] (Abl3) {};
      \node[site,below = .5cm of u3.340] (Atr3) {};
      \node[site,above = .5cm of uc3.20] (Abr3) {};
      \node[isovL,above left = 2.0cm and 1.5cm of Atl3] (vL3) {};
      \node[isovLc,below left = 2.0cm and 1.5cm of Abl3] (vLc3) {};
      \node[isovR,above right = 2.0cm and 1.5cm of Atr3] (vR3) {};
      \node[isovRc,below right = 2.0cm and 1.5cm of Abr3] (vRc3) {};
      \node[isovLf,above left = .2cm and 1cm of vL3] (vLf3) {};
      \node[isovLcf,below left = .2cm and 1cm of vLc3] (vLcf3) {};
      \node[isovRf,above right = .2cm and 1cm of vR3] (vRf3) {};
      \node[isovRcf,below right = .2cm and 1cm of vRc3] (vRcf3) {};
      \node[entangler,below = 4cm of u4] (uc4) {};
      \node[site,below = .5cm of u4.200] (Atl4) {};
      \node[site,above = .5cm of uc4.160] (Abl4) {};
      \node[site,below = .5cm of u4.340] (Atr4) {};
      \node[site,above = .5cm of uc4.20] (Abr4) {};
      \node[isovL,above left = 2.0cm and 1.5cm of Atl4] (vL4) {};
      \node[isovLc,below left = 2.0cm and 1.5cm of Abl4] (vLc4) {};
      \node[isovR,above right = 2.0cm and 1.5cm of Atr4] (vR4) {};
      \node[isovRc,below right = 2.0cm and 1.5cm of Abr4] (vRc4) {};
      \node[isovLf,above left = .2cm and 1cm of vL4] (vLf4) {};
      \node[isovLcf,below left = .2cm and 1cm of vLc4] (vLcf4) {};
      \node[isovRf,above right = .2cm and 1cm of vR4] (vRf4) {};
      \node[isovRcf,below right = .2cm and 1cm of vRc4] (vRcf4) {};
      \draw (uc1.160) to (Abl1) ;
      \draw (uc1.20) to (Abr1) ;
      \draw (u1.160)  .. controls ++(0,.8) and ++(.5,.3).. (vL1.right corner);
      \draw (u1.20)  .. controls ++(0,.8) and ++(-.5,.3).. (vR1.left corner);
      \draw (u1.200) to (Atl1) ;
      \draw (u1.340) to (Atr1) ;
      \draw (uc1.200)  .. controls ++(0,-.8) and ++(.5,-.3).. (vLc1.left corner);
      \draw (uc1.340)  .. controls ++(0,-.8) and ++(-.5,-.3).. (vRc1.right corner);
      \draw (Abl1) to (Abr1);
      \draw (Atl1) to (Atr1);
      \draw (Abl1) to (Atl1);
      \draw (Atr1) to (Abr1);
      \draw (Atl1.west) .. controls  ++(-1,0) and ++(0.0,-1).. (vL1.left corner);
      \draw (Atr1.east) .. controls  ++(1,0) and ++(0.0,-1).. (vR1.right corner);
      \draw (Abl1.west) .. controls  ++(-1,0) and ++(0.0,1).. (vLc1.right corner);
      \draw (Abr1.east) .. controls  ++(1,0) and ++(0.0,1).. (vRc1.left corner);
      \draw (vL1.apex) to (vLf1.apex);
      \draw (vR1.apex) to (vRf1.apex);
      \draw (vLc1.apex) to (vLcf1.apex);
      \draw (vRc1.apex) to (vRcf1.apex);
      \draw (uc2.160) to (Abl2) ;
      \draw (uc2.20) to (Abr2) ;
      \draw (u2.160)  .. controls ++(0,.8) and ++(.5,.3).. (vL2.right corner);
      \draw (u2.20)  .. controls ++(0,.8) and ++(-.5,.3).. (vR2.left corner);
      \draw (u2.200) to (Atl2) ;
      \draw (u2.340) to (Atr2) ;
      \draw (uc2.200)  .. controls ++(0,-.8) and ++(.5,-.3).. (vLc2.left corner);
      \draw (uc2.340)  .. controls ++(0,-.8) and ++(-.5,-.3).. (vRc2.right corner);
      \draw (Abl2) to (Abr2);
      \draw (Atl2) to (Atr2);
      \draw (Abl2) to (Atl2);
      \draw (Atr2) to (Abr2);
      \draw (Atl2.west) .. controls  ++(-1,0) and ++(0.0,-1).. (vL2.left corner);
      \draw (Atr2.east) .. controls  ++(1,0) and ++(0.0,-1).. (vR2.right corner);
      \draw (Abl2.west) .. controls  ++(-1,0) and ++(0.0,1).. (vLc2.right corner);
      \draw (Abr2.east) .. controls  ++(1,0) and ++(0.0,1).. (vRc2.left corner);
      \draw (vL2.apex) to (vLf2.apex);
      \draw (vR2.apex) to (vRf2.apex);
      \draw (vLc2.apex) to (vLcf2.apex);
      \draw (vRc2.apex) to (vRcf2.apex);
      \draw (uc3.160) to (Abl3) ;
      \draw (uc3.20) to (Abr3) ;
      \draw (u3.160)  .. controls ++(0,.8) and ++(.5,.3).. (vL3.right corner);
      \draw (u3.20)  .. controls ++(0,.8) and ++(-.5,.3).. (vR3.left corner);
      \draw (u3.200) to (Atl3) ;
      \draw (u3.340) to (Atr3) ;
      \draw (uc3.200)  .. controls ++(0,-.8) and ++(.5,-.3).. (vLc3.left corner);
      \draw (uc3.340)  .. controls ++(0,-.8) and ++(-.5,-.3).. (vRc3.right corner);
      \draw (Abl3) to (Abr3);
      \draw (Atl3) to (Atr3);
      \draw (Abl3) to (Atl3);
      \draw (Atr3) to (Abr3);
      \draw (Atl3.west) .. controls  ++(-1,0) and ++(0.0,-1).. (vL3.left corner);
      \draw (Atr3.east) .. controls  ++(1,0) and ++(0.0,-1).. (vR3.right corner);
      \draw (Abl3.west) .. controls  ++(-1,0) and ++(0.0,1).. (vLc3.right corner);
      \draw (Abr3.east) .. controls  ++(1,0) and ++(0.0,1).. (vRc3.left corner);
      \draw (vL3.apex) to (vLf3.apex);
      \draw (vR3.apex) to (vRf3.apex);
      \draw (vLc3.apex) to (vLcf3.apex);
      \draw (vRc3.apex) to (vRcf3.apex);
      \draw (uc4.160) to (Abl4) ;
      \draw (uc4.20) to (Abr4) ;
      \draw (u4.160)  .. controls ++(0,.8) and ++(.5,.3).. (vL4.right corner);
      \draw (u4.20)  .. controls ++(0,.8) and ++(-.5,.3).. (vR4.left corner);
      \draw (u4.200) to (Atl4) ;
      \draw (u4.340) to (Atr4) ;
      \draw (uc4.200)  .. controls ++(0,-.8) and ++(.5,-.3).. (vLc4.left corner);
      \draw (uc4.340)  .. controls ++(0,-.8) and ++(-.5,-.3).. (vRc4.right corner);
      \draw (Abl4) to (Abr4);
      \draw (Atl4) to (Atr4);
      \draw (Abl4) to (Atl4);
      \draw (Atr4) to (Abr4);
      \draw (Atl4.west) .. controls  ++(-1,0) and ++(0.0,-1).. (vL4.left corner);
      \draw (Atr4.east) .. controls  ++(1,0) and ++(0.0,-1).. (vR4.right corner);
      \draw (Abl4.west) .. controls  ++(-1,0) and ++(0.0,1).. (vLc4.right corner);
      \draw (Abr4.east) .. controls  ++(1,0) and ++(0.0,1).. (vRc4.left corner);
      \draw (vL4.apex) to (vLf4.apex);
      \draw (vR4.apex) to (vRf4.apex);
      \draw (vLc4.apex) to (vLcf4.apex);
      \draw (vRc4.apex) to (vRcf4.apex);
      \draw (vLf1.right corner) to ++(-.5,0);
      \draw (vLf1.left corner) to ++(0,.5);
      \draw (vRf1.left corner) to (vLf3.right corner);
      \draw (vRf1.right corner) to ++(0,.5);
      \draw (vLcf1.left corner) to ++(-.5,0);
      \draw (vLcf1.right corner) to (vLf2.left corner);
      \draw (vRcf1.left corner) to (vRf2.right corner);
      \draw (vRcf1.right corner) to (vLcf3.left corner);
      \draw (vLf2.right corner) to ++(-.5,0);
      \draw (vRf2.left corner) to (vLf4.right corner);
      \draw (vLcf2.left corner) to ++(-.5,0);
      \draw (vLcf2.right corner) to ++(0,-.5);
      \draw (vRcf2.left corner) to ++(0,-.5);
      \draw (vRcf2.right corner) to (vLcf4.left corner);
      \draw (vLf3.right corner) to ++(-.5,0);
      \draw (vLf3.left corner) to ++(0,.5);
      \draw (vRf3.left corner) to ++(.5,0);
      \draw (vRf3.right corner) to ++(0,.5);
      \draw (vLcf3.right corner) to (vLf4.left corner);
      \draw (vRcf3.left corner) to (vRf4.right corner);
      \draw (vRcf3.right corner) to ++(.5,0);
      \draw (vRf4.left corner) to ++(.5,0);
      \draw (vLcf4.right corner) to ++(0,-.5);
      \draw (vRcf4.left corner) to ++(0,-.5);
      \draw (vRcf4.right corner) to ++(.5,0);
    \node[below] at (current bounding box.south) {\fontsize{53.3}{53.3}\selectfont (b)};
    \end{tikzpicture}
\end{subfigure}
\begin{subfigure} % Step 3: Condense intermediate tensor
  \centering
    \begin{tikzpicture}[scale=0.27, every node/.style={transform shape}]
      \node[sitehalf] (B1) at (0,0) {};
      \node[sitehalf,below = 3.7cm of B1] (B2) {};
      \node[sitehalf,right = 4.7cm of B1] (B3) {};
      \node[sitehalf,below = 3.7cm of B3] (B4) {};
      \node[isovLf,above left = .3cm and 1.0cm of B1] (vL1) {};
      \node[isovRf,above right = .3cm and 1.0cm of B1] (vR1) {};
      \node[isovLcf,below left = .3cm and 1.0cm of B1] (vLc1) {};
      \node[isovRcf,below right = .3cm and 1.0cm of B1] (vRc1) {};
      \node[isovLf,above left = .3cm and 1.0cm of B2] (vL2) {};
      \node[isovRf,above right = .3cm and 1.0cm of B2] (vR2) {};
      \node[isovLcf,below left = .3cm and 1.0cm of B2] (vLc2) {};
      \node[isovRcf,below right = .3cm and 1.0cm of B2] (vRc2) {};
      \node[isovLf,above left = .3cm and 1.0cm of B3] (vL3) {};
      \node[isovRf,above right = .3cm and 1.0cm of B3] (vR3) {};
      \node[isovLcf,below left = .3cm and 1.0cm of B3] (vLc3) {};
      \node[isovRcf,below right = .3cm and 1.0cm of B3] (vRc3) {};
      \node[isovLf,above left = .3cm and 1.0cm of B4] (vL4) {};
      \node[isovRf,above right = .3cm and 1.0cm of B4] (vR4) {};
      \node[isovLcf,below left = .3cm and 1.0cm of B4] (vLc4) {};
      \node[isovRcf,below right = .3cm and 1.0cm of B4] (vRc4) {};
      \draw (B1) to (vL1.apex);
      \draw (B1) to (vLc1.apex);
      \draw (B1) to (vR1.apex);
      \draw (B1) to (vRc1.apex);
      \draw (B2) to (vL2.apex);
      \draw (B2) to (vLc2.apex);
      \draw (B2) to (vR2.apex);
      \draw (B2) to (vRc2.apex);
      \draw (B3) to (vL3.apex);
      \draw (B3) to (vLc3.apex);
      \draw (B3) to (vR3.apex);
      \draw (B3) to (vRc3.apex);
      \draw (B4) to (vL4.apex);
      \draw (B4) to (vLc4.apex);
      \draw (B4) to (vR4.apex);
      \draw (B4) to (vRc4.apex);
      \draw (vL1.right corner) to ++(-.5,0);
      \draw (vL1.left corner) to ++(0,.5);
      \draw (vR1.left corner) to (vL3.right corner);
      \draw (vR1.right corner) to ++(0,.5);
      \draw (vLc1.left corner) to ++(-.5,0);
      \draw (vLc1.right corner) to (vL2.left corner);
      \draw (vRc1.left corner) to (vR2.right corner);
      \draw (vRc1.right corner) to (vLc3.left corner);
      \draw (vL2.right corner) to ++(-.5,0);
      \draw (vR2.left corner) to (vL4.right corner);
      \draw (vLc2.left corner) to ++(-.5,0);
      \draw (vLc2.right corner) to ++(0,-.5);
      \draw (vRc2.left corner) to ++(0,-.5);
      \draw (vRc2.right corner) to (vLc4.left corner);
      \draw (vL3.right corner) to (vR1.left corner);
      \draw (vL3.left corner) to ++(0,.5);
      \draw (vR3.left corner) to ++(.5,0);
      \draw (vR3.right corner) to ++(0,.5);
      \draw (vLc3.right corner) to (vL4.left corner);
      \draw (vRc3.left corner) to (vR4.right corner);
      \draw (vRc3.right corner) to ++(.5,0);
      \draw (vR4.left corner) to ++(.5,0);
      \draw (vLc4.right corner) to ++(0,-.5);
      \draw (vRc4.left corner) to ++(0,-.5);
      \draw (vRc4.right corner) to ++(.5,0);
    \node[below] at (current bounding box.south) {\fontsize{29.6}{29.6}\selectfont (c)};
    \end{tikzpicture}
\end{subfigure}
\begin{subfigure} % Step 4: Split intermediate tensor
  \centering
    \begin{tikzpicture}[scale=0.27, every node/.style={transform shape}]
      \node[sitehalfleft] (CL1) at (0,0) {};
      \node[sitehalfright, right = .7cm of CL1] (CR1) {};
      \node[sitehalfleft, below = 3.8cm of CL1] (CL2) {};
      \node[sitehalfright, right = .7cm of CL2] (CR2) {};
      \node[sitehalfleft, right = 5.5cm of CL1] (CL3) {};
      \node[sitehalfright, right = .7cm of CL3] (CR3) {};
      \node[sitehalfleft, below = 3.8cm of CL3] (CL4) {};
      \node[sitehalfright, right = .7cm of CL4] (CR4) {};
      \node[isovLf,above left = .6cm and .6cm of CL1] (vL1) {};
      \node[isovRf,above right = .6cm and .6cm of CR1] (vR1) {};
      \node[isovLcf,below left = .6cm and .6cm of CL1] (vLc1) {};
      \node[isovRcf,below right = .6cm and .6cm of CR1] (vRc1) {};
      \node[isovLf,above left = .6cm and .6cm of CL2] (vL2) {};
      \node[isovRf,above right = .6cm and .6cm of CR2] (vR2) {};
      \node[isovLcf,below left = .6cm and .6cm of CL2] (vLc2) {};
      \node[isovRcf,below right = .6cm and .6cm of CR2] (vRc2) {};
      \node[isovLf,above left = .6cm and .6cm of CL3] (vL3) {};
      \node[isovRf,above right = .6cm and .6cm of CR3] (vR3) {};
      \node[isovLcf,below left = .6cm and .6cm of CL3] (vLc3) {};
      \node[isovRcf,below right = .6cm and .6cm of CR3] (vRc3) {};
      \node[isovLf,above left = .6cm and .6cm of CL4] (vL4) {};
      \node[isovRf,above right = .6cm and .6cm of CR4] (vR4) {};
      \node[isovLcf,below left = .6cm and .6cm of CL4] (vLc4) {};
      \node[isovRcf,below right = .6cm and .6cm of CR4] (vRc4) {};
      \draw (CL1) to (CR1);
      \draw (CL2) to (CR2);
      \draw (CL3) to (CR3);
      \draw (CL4) to (CR4);
      \draw (CL1) to (vL1.apex);
      \draw (CL1) to (vLc1.apex);
      \draw (CR1) to (vR1.apex);
      \draw (CR1) to (vRc1.apex);
      \draw (CL2) to (vL2.apex);
      \draw (CL2) to (vLc2.apex);
      \draw (CR2) to (vR2.apex);
      \draw (CR2) to (vRc2.apex);
      \draw (CL3) to (vL3.apex);
      \draw (CL3) to (vLc3.apex);
      \draw (CR3) to (vR3.apex);
      \draw (CR3) to (vRc3.apex);
      \draw (CL4) to (vL4.apex);
      \draw (CL4) to (vLc4.apex);
      \draw (CR4) to (vR4.apex);
      \draw (CR4) to (vRc4.apex);
      \draw (vL1.right corner) to ++(-.5,0);
      \draw (vL1.left corner) to ++(0,.5);
      \draw (vR1.left corner) to (vL3.right corner);
      \draw (vR1.right corner) to ++(0,.5);
      \draw (vLc1.left corner) to ++(-.5,0);
      \draw (vLc1.right corner) to (vL2.left corner);
      \draw (vRc1.left corner) to (vR2.right corner);
      \draw (vRc1.right corner) to (vLc3.left corner);
      \draw (vL2.right corner) to ++(-.5,0);
      \draw (vR2.left corner) to (vL4.right corner);
      \draw (vLc2.left corner) to ++(-.5,0);
      \draw (vLc2.right corner) to ++(0,-.5);
      \draw (vRc2.left corner) to ++(0,-.5);
      \draw (vRc2.right corner) to (vLc4.left corner);
      \draw (vL3.right corner) to (vR1.left corner);
      \draw (vL3.left corner) to ++(0,.5);
      \draw (vR3.left corner) to ++(.5,0);
      \draw (vR3.right corner) to ++(0,.5);
      \draw (vLc3.right corner) to (vL4.left corner);
      \draw (vRc3.left corner) to (vR4.right corner);
      \draw (vRc3.right corner) to ++(.5,0);
      \draw (vR4.left corner) to ++(.5,0);
      \draw (vLc4.right corner) to ++(0,-.5);
      \draw (vRc4.left corner) to ++(0,-.5);
      \draw (vRc4.right corner) to ++(.5,0);
    \node[below] at (current bounding box.south) {\fontsize{29.6}{29.6}\selectfont (d)};
    \end{tikzpicture}
\end{subfigure}
\begin{subfigure} % Step 5: Merge vertical bonds
  \centering
    \begin{tikzpicture}[scale=0.22, every node/.style={transform shape}]
      \node[sitehalfright] (CR1) at (0,0) {};
      \node[sitehalfleft, right = 3.9cm of CR1] (CL1) {};
      \node[sitehalfright, below = 6.4cm of CR1] (CR2) {};
      \node[sitehalfleft, right = 3.9cm of CR2] (CL2) {};
      \node[sitehalfright, right = 5.5cm of CR1] (CR3) {};
      \node[sitehalfleft, right = 3.9cm of CR3] (CL3) {};
      \node[sitehalfright, below = 6.4cm of CR3] (CR4) {};
      \node[sitehalfleft, right = 3.9cm of CR4] (CL4) {};
      \node[isometryconj] (wc1) at ($ (CL1)!.5!(CR1) + (0,-3.1cm) $) {};
      \node[isometry] (w1) at ($ (CL1)!.5!(CR1) + (0,3.1cm) $) {};
      \node[isometryconj] (wc2) at ($ (CL2)!.5!(CR2) + (0,-3.1cm) $) {};
      \node[isometry] (w2) at ($ (CL2)!.5!(CR2) + (0,3.1cm) $) {};
      \node[isometryconj] (wc3) at ($ (CL3)!.5!(CR3) + (0,-3.1cm) $) {};
      \node[isometry] (w3) at ($ (CL3)!.5!(CR3) + (0,3.1cm) $) {};
      \node[isometryconj] (wc4) at ($ (CL4)!.5!(CR4) + (0,-3.1cm) $) {};
      \node[isometry] (w4) at ($ (CL4)!.5!(CR4) + (0,3.1cm) $) {};
      \node[isovLf,above left = .6cm and .6cm of CL1] (vL1) {};
      \node[isovRf,above right = .6cm and .6cm of CR1] (vR1) {};
      \node[isovLcf,below left = .6cm and .6cm of CL1] (vLc1) {};
      \node[isovRcf,below right = .6cm and .6cm of CR1] (vRc1) {};
      \node[isovLf,above left = .6cm and .6cm of CL2] (vL2) {};
      \node[isovRf,above right = .6cm and .6cm of CR2] (vR2) {};
      \node[isovLcf,below left = .6cm and .6cm of CL2] (vLc2) {};
      \node[isovRcf,below right = .6cm and .6cm of CR2] (vRc2) {};
      \node[isovLf,above left = .6cm and .6cm of CL3] (vL3) {};
      \node[isovRf,above right = .6cm and .6cm of CR3] (vR3) {};
      \node[isovLcf,below left = .6cm and .6cm of CL3] (vLc3) {};
      \node[isovRcf,below right = .6cm and .6cm of CR3] (vRc3) {};
      \node[isovLf,above left = .6cm and .6cm of CL4] (vL4) {};
      \node[isovRf,above right = .6cm and .6cm of CR4] (vR4) {};
      \node[isovLcf,below left = .6cm and .6cm of CL4] (vLc4) {};
      \node[isovRcf,below right = .6cm and .6cm of CR4] (vRc4) {};
      \draw (CR1) to ++(-1cm,0);
      \draw (CR2) to ++(-1cm,0);
      \draw (CL1) to (CR3);
      \draw (CL2) to (CR4);
      \draw (CL3) to ++(1cm,0);
      \draw (CL4) to ++(1cm,0);
      \draw (CL1) to (vL1.apex);
      \draw (CL1) to (vLc1.apex);
      \draw (CR1) to (vR1.apex);
      \draw (CR1) to (vRc1.apex);
      \draw (CL2) to (vL2.apex);
      \draw (CL2) to (vLc2.apex);
      \draw (CR2) to (vR2.apex);
      \draw (CR2) to (vRc2.apex);
      \draw (CL3) to (vL3.apex);
      \draw (CL3) to (vLc3.apex);
      \draw (CR3) to (vR3.apex);
      \draw (CR3) to (vRc3.apex);
      \draw (CL4) to (vL4.apex);
      \draw (CL4) to (vLc4.apex);
      \draw (CR4) to (vR4.apex);
      \draw (CR4) to (vRc4.apex);
      \draw (vR1.left corner) to (vL1.right corner);
      \draw (vR1.right corner) to (w1.left corner);
      \draw (vRc1.left corner) to (wc1.right corner);
      \draw (w2.left corner) to (vR2.right corner);
      \draw (vRc1.right corner) to (vLc1.left corner);
      \draw (vR2.left corner) to (vL2.right corner);
      \draw (vRc2.left corner) to (wc2.right corner);
      \draw (vRc2.right corner) to (vLc2.left corner);
      \draw (vL1.right corner) to (vR1.left corner);
      \draw (vL1.left corner) to (w1.right corner);
      \draw (vLc1.right corner) to (wc1.left corner);
      \draw (w2.right corner) to (vL2.left corner);
      \draw (vLc2.right corner) to (wc2.left corner);
      \draw (wc1.apex) to (w2.apex);
      \draw (w1.apex) to ++(0,.5);
      \draw (wc2.apex) to ++(0,-.5);
      \draw (vR3.left corner) to (vL3.right corner);
      \draw (vR3.right corner) to (w3.left corner);
      \draw (vRc3.left corner) to (wc3.right corner);
      \draw (w4.left corner) to (vR4.right corner);
      \draw (vRc3.right corner) to (vLc3.left corner);
      \draw (vR4.left corner) to (vL4.right corner);
      \draw (vRc4.left corner) to (wc4.right corner);
      \draw (vRc4.right corner) to (vLc4.left corner);
      \draw (vL3.right corner) to (vR3.left corner);
      \draw (vL3.left corner) to (w3.right corner);
      \draw (vLc3.right corner) to (wc3.left corner);
      \draw (w4.right corner) to (vL4.left corner);
      \draw (vLc4.right corner) to (wc4.left corner);
      \draw (wc3.apex) to (w4.apex);
      \draw (w3.apex) to ++(0,.5);
      \draw (wc4.apex) to ++(0,-.5);
    \node[below] at (current bounding box.south) {\fontsize{36.4}{36.4}\selectfont (e)};
    \end{tikzpicture}
\end{subfigure}
\begin{subfigure} % Step 6: Contract final tensor
  \centering
    \begin{tikzpicture}[scale=0.34, every node/.style={transform shape}]
      \node[sitenext] (A1) at (0,0) {};
      \node[sitenext,below = 4cm of A1] (A2) {};
      \node[sitenext,right = 4cm of A1] (A3) {};
      \node[sitenext,below = 4cm of A3] (A4) {};
      \draw (A1) to (A2);
      \draw (A1) to (A3);
      \draw (A2) to (A4);
      \draw (A3) to (A4);
      \draw (A1) to ++(0,3);
      \draw (A1) to ++(-3,0);
      \draw (A2) to ++(0,-3);
      \draw (A2) to ++(-3,0);
      \draw (A3) to ++(0,3);
      \draw (A3) to ++(3,0);
      \draw (A4) to ++(0,-3);
      \draw (A4) to ++(3,0);
    \node[below] at (current bounding box.south) {\fontsize{23.5}{23.5}\selectfont (f)};
    \end{tikzpicture}
\end{subfigure}
\caption{The TNR procedure:\cite{TNR3}
(a) Disentangler pairs $u^\dagger u$ are added across pairs of
vertical bonds. As the disentanglers are unitary, this is a resolution of the
identity.
(b) Projective truncations $v_L^\dagger v_L$ and $v_R^\dagger v_R$,
which have intermediate bond dimension $\chi'$, approximate the identity on
nearby pairs of vertical and horizontal bonds at the
corners of blocks of sites. $v_L$, $v_R$, and $u$ are iteratively optimized
to maximize how well we approximate $2\times 2$ site blocks with the
corresponding objects.
(c) The sub-networks consisting of $2\times 2$ blocks of site tensors and
the disentanglers and isometries that share bonds are contracted to yield
intermediate tensors.
(d) These intermediate tensors are split using a truncated
singular value decomposition, wherein the truncation signifies eliminating
all but $\chi$ of the $\chi^{\prime 2}$ singular values so that the dimension
of the new bond is $\chi$.
(e) A projective truncation $w^\dagger w$ with intermediate bond dimension
$\chi$ is placed across pairs of vertical bonds.
(f) The collection of eight tensors whose boundary is the virtual bonds of
dimension $\chi$ formed in the last two steps is contracted to yield the site
tensor at the next level of renormalization.}
\label{fig:EVTNR}
\end{figure}
CTMRG and coarse-graining methods like HOTRG, while highly effective in many
situations, typically cannot effectively analyze systems with high correlation
length. In order to extract information about systems that do have high
correlation length, which is to say primarily those that exhibit or approximate
critical behavior, we must use more complex techniques, primarily the method
of tensor network renormalization (TNR) introduced by Evenbly and Vidal.
\cite{TNR,TNRMERA,TNR3,TNRHauru,TNRlocal} The algorithm is
based on Vidal's concept of entanglement renormalization,
and we will only be able to represent it in the heuristic terms presented in
Fig.~\ref{fig:EVTNR}.

While it is possible to use TNR to extract physical data such as expectation
values, the methods to do so are typically quite computationally expensive,
prohibitively so at the bond dimensions we have been primarily using.
Instead we focus on the method \cite{TNRHauru} of using Cardy's
formula for the
partition function on a torus in terms of conformal data\cite{Cardyoperator},
in which we diagonalize an $n$-site transfer matrix with a one-site rotational
twist and find that its eigenvalues are
\begin{equation}
\lambda_\alpha \simeq e^{-\frac{2\pi}{n}(\Delta_\alpha - \frac{c}{12})+nf+\frac{2\pi i}{n}s_\alpha},
\end{equation}
where $\alpha$ indexes scaling operators with scaling dimension $\Delta_\alpha$
and conformal spin $s_\alpha$, $c$ is the central charge,
$f$ is a nonuniversal free-energy constant, and equality only holds as
irrelevant perturbations vanish: that is, as coarse-graining takes us
towards the thermodynamic limit. Note that we must analyze the transfer matrix
at multiple sizes to isolate $c$ and $f$; that adding sites to the transfer
matrix increases both the number of eigenvalues for which we may expect this
approximation to be valid and the range of conformal spins which it
distinguishes; and that when we apply this analysis to a 2D quantum system,
the universal data $c$, $\Delta_\alpha$, and $s_\alpha$ that we extract
describe the infrared limit of the doubled vertex model and not
the quantum model. However, with caution we can draw limited
conclusions from it, as follows. Correlators of local
operators in the quantum state can as in Fig.~\ref{fig:quantumclassical}
be expressed as correlators of local operators in the doubled vertex model,
which means at least that quasi-long-range order in
the quantum model implies quasi-long-range order in
the doubled vertex model. Thus when we find that the doubled vertex model is
\textit{not} critical, we determine that the quantum model has finite
correlation length and is likely gapped. Conversely, when we find that the
doubled vertex model \textit{is} critical, we seek operators on the quantum
model whose correlations
correspond to those predicted by the classical scaling dimensions; when we
find them, we conclude that the quantum model has infinite correlation length
and thus is gapless.\cite{Hastings1,Hastings2}
\subsection{Loop optimization for tensor network renormalization}

\begin{figure}
\includegraphics[width=0.45\textwidth]{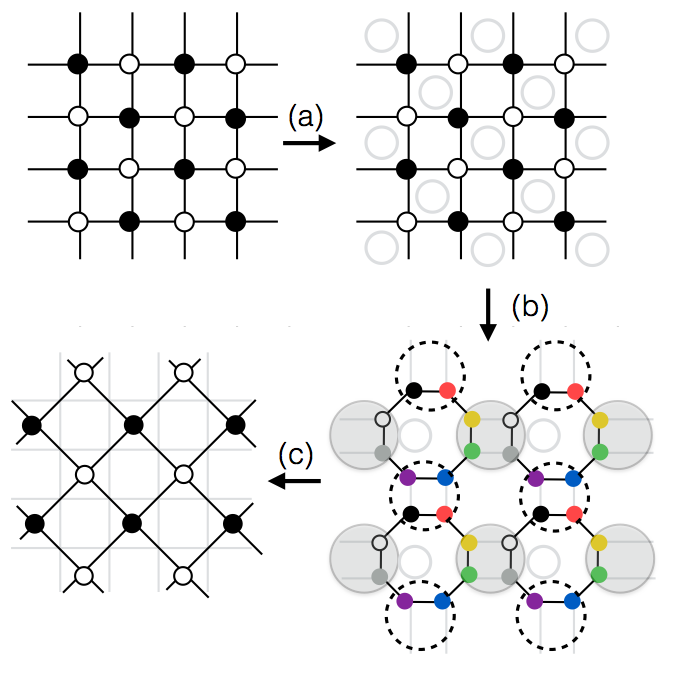}
\caption{ Schematic procedure of the Loop-TNR procedure in 2d. 
(a) We remove local entanglement on alternating plaquettes, labeled with grey
circles, by inserting projectors. 
(b) We convert these square plaquettes into octagon plaquettes made up of eight
rank-3 tensors and optimize them.
(c) We form new tensors by contracting the tensors in the grey and dotted
circles, as in the standard tensor renormalization group (TRG)
procedure proposed by Levin and Nave. }
  \label{fig:loopTNR}
\end{figure}

As with the above TNR procedure, loop optimization for tensor network
renormalization (loop-TNR), proposed by 
Yang, Gu, and Wen\cite{loopTNR}, has the goal of removing local entanglement in
order to obtain a renormalization-group fixed point in tensor form.
This approach is a real-space renormalization procedure based on the existing
TRG procedure \cite{Levin_TRG_2007} of Levin and Nave, modified by
removing short-range entanglement at the start of the procedure and
optimizing intermediate tensors in loops.
We may apply it to classical and quantum systems in much the same circumstances
as Evenbly and Vidal's TNR. 
The details of the algorithm, as sketched in Fig.~\ref{fig:loopTNR}, are
as follows:

\begin{enumerate}
\item Starting with a tensor contraction on a square lattice as in
Fig.~\ref{fig:loopTNR}(a), we put each
tensor into a canonical form by inserting ``projectors'' which filter out local
entanglement onto the bonds around plaquettes. These projectors are
constructed iteratively using QR decomposition.

\begin{figure}
  \includegraphics[width=0.3\textwidth]{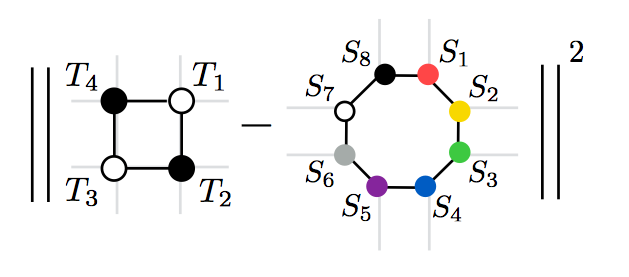}
  \caption{The cost function for optimizing loop-TNR tensors}
  \label{fig:costFunction}
\end{figure}

\item We then deform the tensor network from a square lattice to a
square-octagon lattice by performing a truncated singular-value decomposition,
as in Levin and Nave's TRG, to approximate each rank-4 $T_i$ tensor as the
contraction of a pair of rank-3 tensors $S_i$. This is shown in
Fig.~\ref{fig:loopTNR}(b).
\item We then optimize the $S_i$ in an octogonal loop by minimizing the cost
function in Fig.~\ref{fig:costFunction}.

\item Finally, as in standard TRG, we contract the four tensors around each
square plaquette of the square-octagon lattice [Fig.~\ref{fig:loopTNR}(c)].
These plaquettes become the
sites of a new, coarse-grained square lattice, with half as many sites as the
original lattice.
\end{enumerate}

\subsection{Explicit form of the PEPS tensors} 
For completeness we state the exact form of the tensors which comprise the
state primarily considered in this work, i.e., the spin-2 state, situated on
a lattice with coordination number $q=4$.
If the state has deformation $a_1,a_2$ and bond state $|\psi^-\rangle$,
the tensors have nonzero elements
\begin{align}
\label{square_tps}
&A^{2  }_{\up\up\up\up} = a_{2},  \;   A^{1 }_{\up\up\up\down}= A^{1  }_{\up\up\down\up} =A^{1  }_{\up\down\up\up} =A^{1  }_{\down\up\up\up} =a_{1} \notag \\
&  A^{-2}_{\down\down\down\down} = a_{2},  \;   A^{-1}_{\down\down\down\up}= A^{-1}_{\down\down\up\down} =A^{-1}_{\down\up\down\down} =A^{-1}_{\up\down\down\down} =a_{1}  \notag \\
&A^{0  }_{\up\up\down\down} =A^{0}_{\up\down\down\up}= A^{0 }_{\down\down\up\up} =A^{0}_{\down\up\up\down}=A^{0 }_{\up\down\up\down}=A^{0 }_{\down\up\down\up} =a_{0} \\
&  B^{2}_{\down\down\down\down} = a_{2},  \;   B^{1}_{\down\down\down\up}= B^{1}_{\down\down\up\down} =B^{1}_{\down\up\down\down} =B^{1}_{\up\down\down\down} =-a_{1}  \notag \\
&B^{-2  }_{\up\up\up\up} = a_{2},  \;   B^{-1 }_{\up\up\up\down}= B^{-1  }_{\up\up\down\up} =B^{-1  }_{\up\down\up\up} =B^{-1  }_{\down\up\up\up} =-a_{1} \notag \\
&B^{0 }_{\up\up\down\down} =B^{0}_{\up\down\down\up}= B^{0 }_{\down\down\up\up} =B^{0}_{\down\up\up\down}=B^{0 }_{\up\down\up\down}=B^{0 }_{\down\up\down\up} = a_{0} ,
\end{align} 

where tensors $A$ and $B$ are placed on alternating sublattices.

\section{The general Hamiltonian and the spin-1 analogy}
\label{app:ham}
\subsection{Niggeman, Kl\"umper, and Zittartz's Hamiltonian}
\label{app:NKZ}
In the original work by Niggeman, Kl\"umper, and Zittartz\cite{NKZspin2} a
general two-site Hamiltonian, invariant under spin flips and $S_z$ rotations
(that is, the same $O(2)$ symmetry of the deformed-AKLT model) as well as
spatial symmetries, is presented in Eqs.~(3)-(5). The valence-bond construction
they proceed to define in Eq.~(8)-(10) is equivalent to our deformed-AKLT state.
There the parameters $a$ and $b$ correspond to $a_1$ and $a_2$, respectively,
in our work, and $\sigma$ specifies the bond state: We will set $\sigma=1$,
corresponding to the antiferromagnet singlet bond state $\ket{\psi^-}$.
The 22-parameter Hamiltonian of Eq.~(7) then, as in Eqs.~(12) and (13),
becomes the seven-parameter Hamiltonian
\begin{align}
\label{eq:NKZham}
h_{ij} =& \lambda_4\left(\ket{v_4}\bra{v_4}+\ket{v_{-4}}\bra{v_{-4}}\right)\notag\\
&+\lambda_3^+\left(\ket{v_3^+}\bra{v_3^+}+\ket{v_{-3}^+}\bra{v_{-3}^+}\right)\notag\\
&+\lambda_{22}^+\left(\ket{v_{22}^+}\bra{v_{22}^+}+\ket{v_{-22}^+}\bra{v_{-22}^+}\right)\notag\\
&+\lambda_{12}^+\left(\ket{v_{12}^+}\bra{v_{12}^+}+\ket{v_{-12}^+}\bra{v_{-12}^+}\right)\notag\\
&+\lambda_{03}^+\left(\ket{v_{03}^+}\bra{v_{03}^+}\right)\\
\ket{v_{\pm 4}}&=\ket{\!\pm\!2,\pm2}\\
\ket{v_{\pm 3}^+}&=\ket{\!\pm\!1,\pm1}+\ket{\!\pm\!2,\pm1}\\
\ket{v_{\pm 22}^+}&=\frac{a_2}{a_1^2}\ket{\!\pm\!1,\pm2}-(\ket{0,\pm2}+\ket{\!\pm\!2,0})\\
\ket{v_{\pm 12}^+}&=a_2(\ket{0,\pm1}+\ket{\!\pm\!1,0})-(\ket{\!\mp\!1,\pm2}+\ket{\!\pm\!2,\mp1})\\
\ket{v_{03}^+}&=\ket{0,0}+\frac{1}{a_1^2}(\ket{\!+\!1,-1}+\ket{\!-\!1,+1})\notag\\
&+\frac{1}{a_2^2}(\ket{\!+\!2,-2}+\ket{\!-\!2,+2})
\end{align}
Of these seven parameters, two are the deformation parameters $a_1$ and $a_2$.
Meanwhile a rescaling of the remaining five,
$\lambda_4,\lambda_3^+,\lambda_{22}^+,\lambda_{12}^+,\lambda_{03}^+$, is
simply an energy rescaling. Noting that energy shifts are already accounted
for, as the minimal eigenvalue of $h_{ij}$ is fixed at 0, we find that, in
this formulation, we have a four-parameter family of distinct parent
Hamiltonians for \textit{every} deformed-AKLT state $\ket{\Psi(a_1,a_2)}$
(with $0<a1,a2<\infty$).

Comparing this with the form of the Hamiltonian in \eqref{eqn:Ha}, we determine
that, for a given value of the deformation parameters $a_1$ and $a_2$, these
two Hamiltonians will be equal when we set
\begin{align}
\lambda_{03}^+ &= \frac{18}{35}\notag\\
\lambda_{12}^+ &= \frac{9}{14a_1^2a_2^2}\notag\\
\lambda_{22}^+ &= \frac{9}{7a_2^2}\notag\\
\lambda_3^+ &= \frac{9}{2a_1^2a_2^2}\notag\\
\lambda_4 &= \frac{36}{a_2^4}.
\end{align}

\subsection{Explanation of the XY phase}
\label{app:XY-theory}

In the continuum path integral approach Haldane adopts for
antiferromagnets\cite{Haldane2D}, we
note that we can emulate the deformation by inserting a zero-time
``projection'' factor into the semiclassical AKLT partition function:
\begin{align}
Z_{(a_1,a_2)} \simeq& \int \mathcal{D}[\vec{\Omega}]e^{i\Upsilon[\vec\Omega]}e^{-\mathcal{S}_\text{AKLT}[\vec{\Omega},\vec{L}]}P_{(a_1,a_2)}[\vec\Omega]\\
P_{(a_1,a_2)}[\vec\Omega] \equiv& \prod_{v\in V}p_{(a_1,a_2)}\!\left(\theta(t=0,\vec{r}=v)\right)
\end{align}
where $\vec\Omega$ is the N\'eel field, $\vec{L}$ is the net spin density,
$\Upsilon$ is the Berry phase arising from $\vec\Omega$,
$\mathcal{S}_\text{AKLT}$ is the AKLT action (as a continuum approximation),
$\theta$ is the zenith angle corresponding to the unit vector $\Omega$, and
$p_{\vec{a}}(\theta)$ is a function of $\theta$ (even under
$\theta\to \pi-\theta$) which represents the deformation $D(\vec{a})$ as
follows:
\begin{equation}
D(\vec{a})^2 = \int d\Omega\ p_{\vec{a}}(\theta)\ket{\Omega}\bra{\Omega}.
\end{equation}
This does not uniquely specify $p_{\vec{a}}$; in fact, any suitable
three-parameter ansatz should provide a solution within some region of phase
space. For example, if we anticipate a low-order polynomial in $\cos\theta$, we
find
\begin{align}
p_{(a_1,a_2)}(\theta) =&\frac{5}{4}(19-168\cos\theta+189\cos^2\theta)\\
&+(-15+175\cos\theta-210\cos^2\theta)a_1^2\notag\\
&+\frac{5}{8}(1-70\cos\theta+105\cos^2\theta)a_2^2\notag
\end{align}
Although this particular ansatz is not especially illustrative, we
hope that we can represent the deformation in some way such that, for
$a_1,a_2 \to 0$, $P$ restricts zero-time configurations towards the easy plane
$\theta=\frac{\pi}{2}$, so that valid configurations primarily remain in a band around the
equator. In this case the azimuthal angle $\phi$ at zero time should resemble
a classical XY rotor. Should that rotor belong to the low-temperature phase, 
described in the infrared by a compactified-free-boson CFT with some coupling
$g > 4$, then its most relevant scaling operator will be a vertex operator
$V_{\pm 1,0}(z)\propto\ :\mathrel{e^{\pm i\phi(z)}}:$. The
latter expression translates into the coherent-spin framework as
\begin{equation}
\int d\Omega\;e^{\pm i \phi}\ket{\Omega}\bra{\Omega} = \eta\frac{4\pi}{15}(S_x\pm iS_y).
\end{equation}
where $\eta = \pm 1$ is a sign that indicates alternating sublattices. 

\subsection{XY vortices and tunneling processes}
We refer to our hypothesis in Sec.~\ref{sec:XY-properties} that the XY phase
can be explained as coherent-spin rotors at an imaginary-time 0 surface at which
the deformation restricts spins away from the poles.
We then consider the Kosterlitz-Thouless transition,
effected by the condensation of vortices in these rotors.
Given such a $\tau=0$ vortex
configuration, we suggest that the dominant coherent-spin configurations
contributing to it will be tunneling processes, short-lived discontinuities in
the spin-wave configurations which are integrated over in the coherent-spin path
integral. We can view such configurations as vortices being created and then
dissipating; thus our claim is that that a vortex in the azimuthal angle
on the $\tau=0$ plane corresponds to a tunneling process occurring at
$\tau\sim 0$ with the configuration returning to a smoothly-varying spin wave
sufficiently far from $\tau\sim 0$.

Now we refer to Haldane's analysis\cite{Haldane2D} of topological effects in
$SU(2)$ antiferromagnets in 2D. His conclusion is that tunneling processes
provide the basis for the following distinction between even-integer and
odd-integer spin systems: In odd-integer spin systems, tunneling processes
change the sign of the Berry-phase factor when moved by one site; thus,
when one such process is sufficiently isolated from all others that this
change will not significantly affect the action, that configuration will cancel
with one in which the tunneling process in question has been moved by one site.
Thus, when we see the kind of XY-like behavior observed in the deformed-AKLT
system, if the spin is \textit{odd}-integer, then we expect the winding-number-1
vortices that become relevant at coupling $g=4$ to correspond to tunneling
processes isolated either in space or imaginary time and therefore prohibited
by these topological considerations. Instead, only even-winding-number vortices,
corresponding to bound pairs of tunneling processes, will contribute to the
path integral. As the first of these, $V_{0,\pm 2}$, becomes relevant at
$g = 1$, the XY-like phase will remain stable for all $g>1$ and the
Kosterlitz-Thouless transition will only occur at $g=1$. For
\textit{even}-integer systems, meanwhile, the Berry phase is trivial for
all mostly-smooth configurations, so no tunneling processes will be excluded.

\subsection{Explaining pseudo-quasi-long-range ordered behavior}
\label{app:spin1}
The square-lattice deformed-AKLT system studied herein is manifestly spin-2;
in particular, as described above, we can express it
in path-integral form as the spin-2, $SU(2)$-invariant AKLT path integral
with a planar deformation operator inserted at imaginary time $\tau=0$. Here
we suggest that, along the $a_2=0$ boundary of the
phase diagram, the system may be similarly approximated
as a deformation of a spin-1 antiferromagnet; in particular, that, when we
examine the $a_2\to 0$ limit of the Hamiltonian and extract the effective
Hamiltonian at leading order in perturbation theory, for some value of $a_1$
this will be a spin-1 $SU(2)$-invariant antiferromagnet.
Though it is not immediately clear how, or if, this may best be done rigorously,
we state it as an expectation based on intuitive observations of,
for example, the ``spin-1'' $O(2)$ transformation properties obeyed along
this line, and confirmed by the system's behavior.

If we can find such a relationship to a spin-1 system, Haldane's nonlinear
sigma model argument discussed above implies that this system will lack
isolated tunneling processes in its path integral, suppressing any isolated
vortices that would arise in the $\tau=0$ rotor configuration, as above.
In the deformed-AKLT model, therefore, we anticipate an \textit{approximate}
relationship between the $a_2=0$ limit and this system which demonstrates
vortex suppression, and, therefore, a likewise approximate suppression of
isolated vortices. Thus, while we find that XY vortices still become
\textit{relevant} when $g<4$, they are approximately suppressed and therefore
only become apparent at large length scales. In particular, we propose this
approximate spin-1 physics as an explanation for the
``pseudo-quasi-long-range order'' we observe near the $a_2=0$ axis, in that
very large but finite correlation lengths are caused by relevant
perturbations that remain very small far from the phase transition.

\section{Phase transitions from tracking critical data}
\label{app:KT}

\begin{figure}
\includegraphics[width=.5\textwidth]{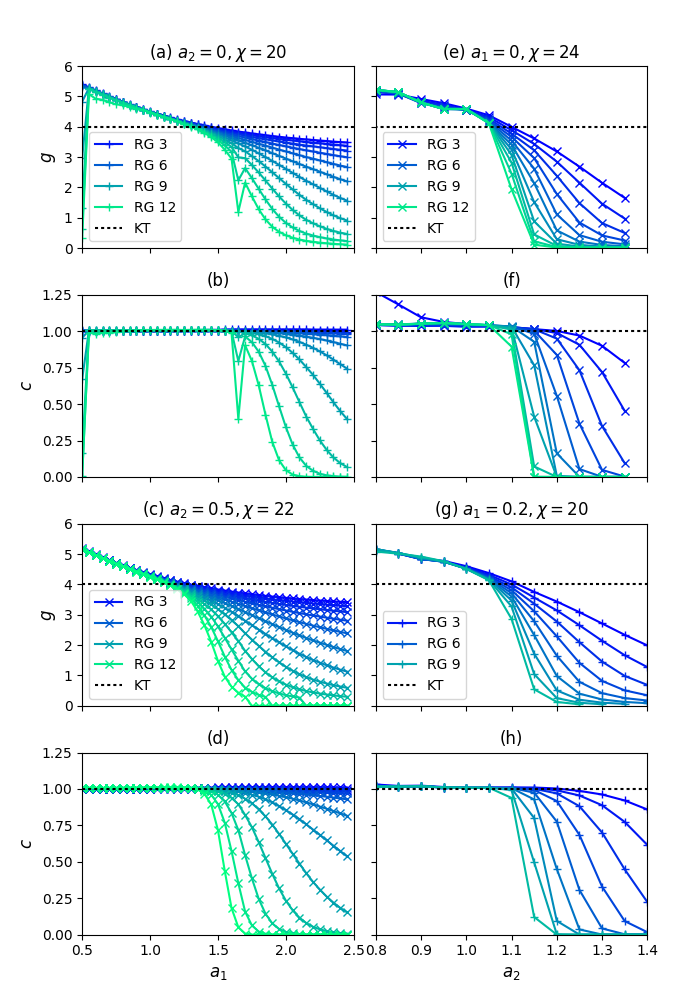}
\caption{We follow $c$ and $g$ along lines of the phase diagram which
pass through the Kosterlitz-Thouless-like transition from the XY phase to the
AKLT phase.
(a),(b) On the axis $a_2=0$, $g$ steadily decreases with increasing $a_1$ until 
$a_1 \simeq 1.25$, where a KT transition appears to occur. (c),(d) We observe
much the same behavior on the $a_2=0.5$ line, although here the transition
is closer to $a_1 = 1.15$. (Here we see some instability, characteristic of
the choice to preserve $\mathbb{Z}\times\mathbb{Z}$ rather than some larger
subgroup of $O(2)$). (e),(f) We
see a similar transition occur when we increase $a_2$ along the $a_1=0$
axis, with $g$ dropping below 4 by $a_2 = 1.1$. Here, however, the
transition is much sharper, indicating that the correlation length when
displaced by
$\Delta a_2 \simeq 0.05$ from the transition in (f) - that is, roughly
$\xi \sim 300$ at $\aparam{0}{1.25}$ - is as much or less than that at
$\Delta a_1 \simeq 1.0$ from the transition in (b) or (d), as for
example $\xi \sim 300$ at $\aparam{2.2}{0.5}$. This conclusion
is consistent with Figs.~\ref{fig:squarephase} and \ref{fig:VBScorr} in
its implications for a pseudo-quasi-long-range ordered region.
(g),(h) If we move slightly, to $a_1=0.2$, we see a much clearer
picture of a similarly sharp transition at
approximately the same value of $a_2$.}
\label{fig:KTlines}
\end{figure}

\begin{figure}
\includegraphics[width=.5\textwidth]{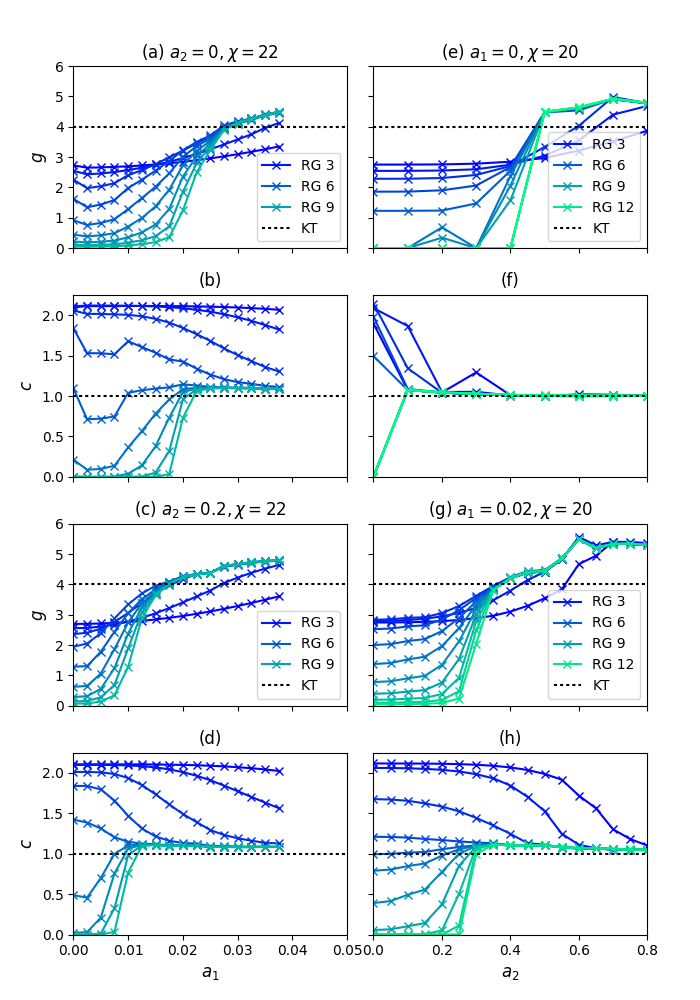}
\caption{As we approach the origin with fixed bond dimension, we find that
estimates of $c$ and $g$ behave much like they do in Fig.~\ref{fig:KTlines}
when they cross the KT
transition into the AKLT phase, although here the data are somewhat noisier and
harder to discern patterns from. (a),(b) On the $a_2=0$ axis, with $\chi=22$, we
observe a transition at approximately $a_1 = 0.028$. (c),(d) When we move to
$a_2=0.2$, this transition shifts to about $a_1 = 0.017$. (e),(f) Conversely,
on the $a_1=0$ axis with $\chi=20$, we see evidence of a transition around
$a_2 = 0.45$. (g),(h) But when we move to just $a_1 = 0.02$, it shifts closer
to $a_2 = 0.23$.}
\label{fig:prodKT}
\end{figure}

As the data summarized in Fig.~\ref{fig:XYmaps20} suggests, 
we find that the most distinctive way to identify and track the
Berezinskii-Kosterlitz-Thouless-like transition(s) observed at the boundary
or boundaries of the XY phase is to follow TNR estimates of $c$ and $g$ along
lines in the phase diagram which cross these boundaries. By plotting these
estimates at successive coarse-graining steps, we find that inside the XY
region, estimates of $c$ and $g$ both converge to nontrivial asymptotic
values: $c \simeq 1$, while $g \geq 4$ but varies continuously with the
deformation parameters.
Outside of that region, estimates for $g$ fall towards 0 and
estimates for $c$ eventually follow suit. By determining the parameters at
which $g$, at some length scale, drops below 4, we can find the boundaries of
the XY phase. In Fig.~\ref{fig:KTlines} we see that we can use this to define
the XY-AKLT transition as well as the pseudo-quasi-long-range ordered region.
(Some points of the transition explored in Fig.~\ref{fig:KTlines} have been
probed in more detail in Fig.~\ref{fig:KTcomp}.) In Fig.~\ref{fig:prodKT},
meanwhile, we see that a similar transition occurs as we approach the
origin from within the XY phase; it is almost as clear as the
transition into the AKLT phase, but it recedes to the origin as
bond dimension is increased.

\section{Using bond-dimension comparisons to analyze the product-state region}
\label{app:linecomp}

\begin{figure}
\includegraphics[width=.5\textwidth]{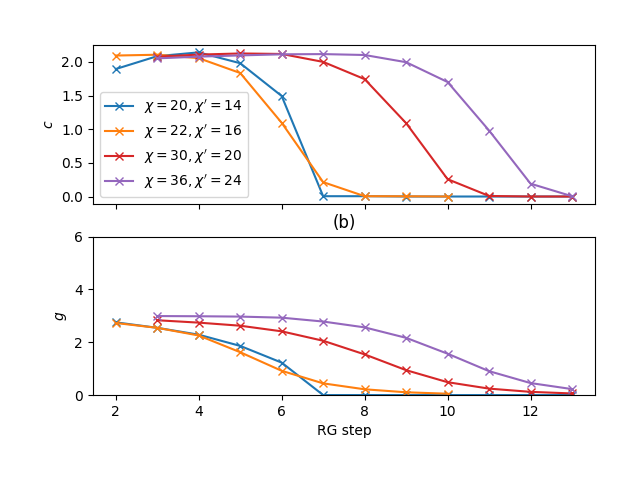}
\caption{$\aparam{0}{0}$: At the origin, we know that, theoretically,
$c=2$ and $g=3$
exactly. While increasing the bond dimension allows us to approximate these
numbers at larger length scales, even with $\chi=36$, TNR
estimates do not remain stable.}
\label{fig:comp0_0}
\end{figure}
\begin{figure}
\includegraphics[width=.5\textwidth]{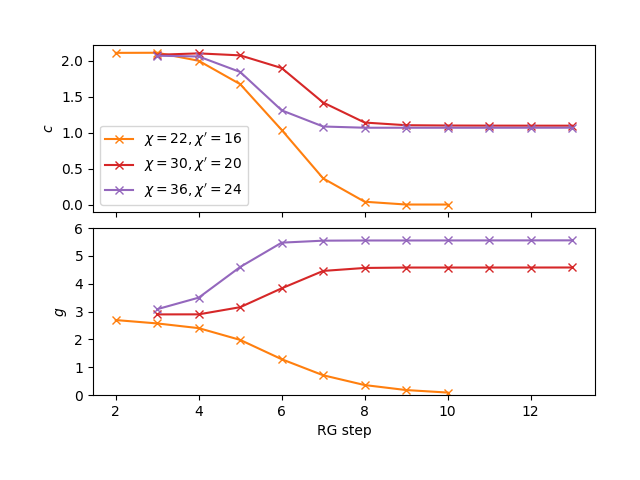}
\caption{$\aparam{0.01}{0}$: Here the system exhibits XY-like behavior at
$\cchip{30}{20}$; raising the bond dimension to $\cchip{36}{24}$ increases
the estimated value of $g$ substantially.}
\label{fig:comp01_0}
\end{figure}
\begin{figure}
\includegraphics[width=.5\textwidth]{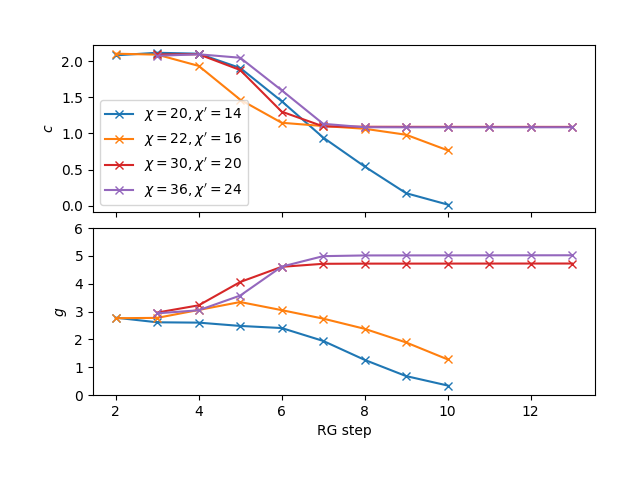}
\caption{$\aparam{0.01}{0.2}$: Here the system exhibits XY-like behavior at
$\cchip{30}{20}$; raising the bond dimension to $\cchip{36}{24}$ again
increases the estimated value of $g$.}
\label{fig:comp01_2}
\end{figure}
\begin{figure}
\includegraphics[width=.5\textwidth]{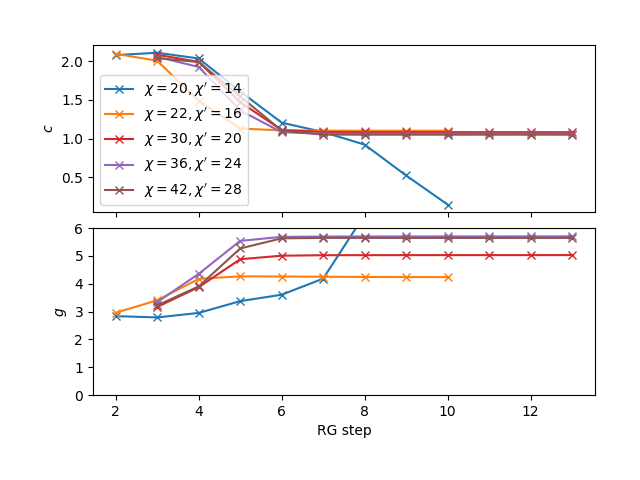}
\caption{$\aparam{0.02}{0.2}$: Here the system exhibits XY-like behavior with
bond dimension as low as $\cchip{22}{16}$; the estimated asymptotic value
of $g$ increases when raising the bond dimension to $\cchip{30}{20}$ and again
in raising it to $\cchip{36}{24}$; however, it appears to be stable when the
bond dimension increases to $\cchip{42}{28}$.}
\label{fig:comp02_2}
\end{figure}
\begin{figure}
\includegraphics[width=.5\textwidth]{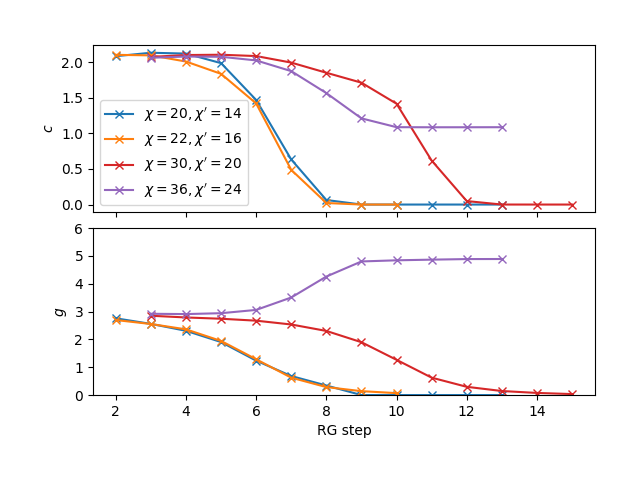}
\caption{$\aparam{0}{0.2}$: While the system does not exhibit XY-like behavior
for bond dimension $\cchip{30}{20}$, this changes when we increase the bond
dimension to $\cchip{36}{24}$. Tracking $g$ and $c$, we find that they
diverge at the crossover from $c=2$ behavior: with lower bond dimension, the
$c$ large, $g$ small behavior draws out until larger length scales before
proving unstable, while for higher bond dimension, $c$ more quickly falls to 1
and $g$ more quickly rises to about 5.}
\label{fig:comp0_2}
\end{figure}
\begin{figure}
\includegraphics[width=.5\textwidth]{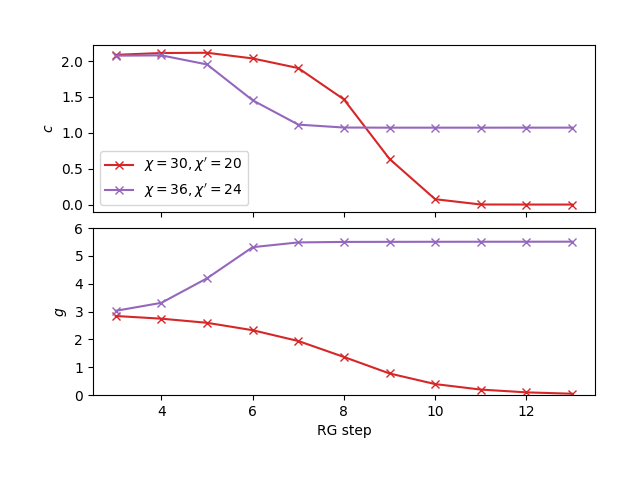}
\caption{$\aparam{0.001}{0.001}$: Here, very close to the origin, we see behavior
similar to that in Fig.~\ref{fig:comp0_0} for $\cchip{30}{20}$, but raising
the bond dimension to $\cchip{36}{24}$ we again see that a crossover into
XY-like behavior appears. As in Fig.~\ref{fig:comp0_2}, this suggests that the
effects that make the difference in these cases are short to medium-range, in
particular less than 100 sites.}
\label{fig:comp001_001}
\end{figure}
\begin{figure}
\includegraphics[width=.5\textwidth]{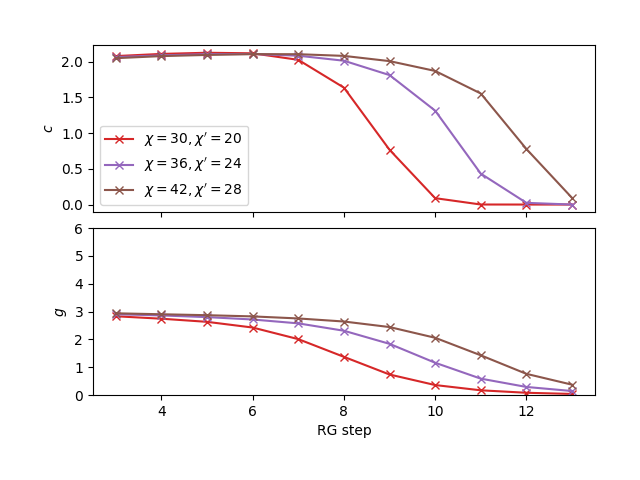}
\caption{$\aparam{0}{0.1}$: At this point relatively close to the origin along the
$a_1=0$ line, we observe behavior similar to that of Fig.~\ref{fig:comp0_0}
as we increase the bond dimension to $\cchip{42}{28}$. At no point do we see
behavior consistent with the XY phase.}
\label{fig:comp0_1}
\end{figure}
\begin{figure}
\includegraphics[width=.5\textwidth]{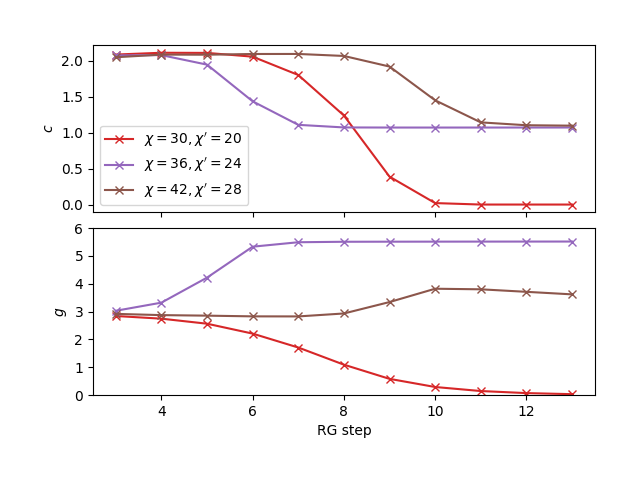}
\caption{$\aparam{0.001}{0.05}$: Here we have estimates from $\cchip{30}{20}$,
$\cchip{36}{24}$, and $\cchip{42}{28}$, none of which are mutually consistent.
$\cchip{30}{20}$ suggests product-state behavior; $\cchip{36}{24}$ suggests
XY-like behavior; and $\cchip{42}{28}$ demonstrates a crossover from $c=2$ to
$c=1$ behavior with an unstable value of $g$. We conclude that the bond
dimensions which we are capable of employing are inadequate at this point.}
\label{fig:comp001_05}
\end{figure}
By comparing estimates of $c$ and $g$ at each coarse-graining step,
we observe how increasing the bond dimension used during TNR affects our
estimates of the system's behavior near the origin.
In Fig.~\ref{fig:comp0_0} we see that even
the exactly-understood $c=2$ behavior at the origin is difficult to replicate
with TNR. Similar behavior also appears in Fig.~\ref{fig:comp0_1}, although
then it is at a point spaced reasonably far along the $a_1=0$ axis.
Figures~\ref{fig:comp01_0}, \ref{fig:comp01_2}, and \ref{fig:comp02_2}
demonstrate that, where we see $g$ take an asymptotic value in the
thermodynamic limit,
that limiting value will typically increase as we raise the bond dimension.
Figures~\ref{fig:comp0_2} and \ref{fig:comp001_001} suggest that numerical
inadequacies at relatively short length scales may sharply influence what
kind of infrared behavior we observe. Finally, through
Fig.~\ref{fig:comp001_05}, we conclude that the methods used in this work are
inconclusive as to the system's behavior in the innermost part of this
region, when $a_2 < 0.2$ and $a_1 \ll 0.01$.

\section{The critical exponent $\delta$}
\label{app:delta}

\begin{figure}
\includegraphics[width=.5\textwidth]{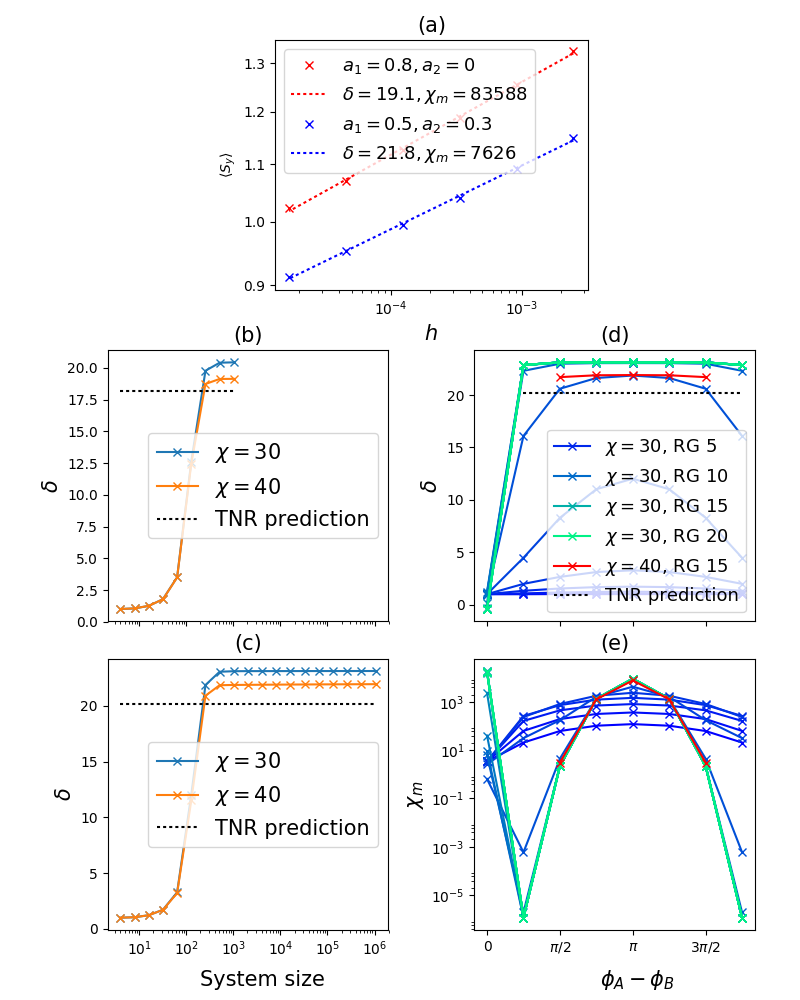}
\caption{Using the critical  exponent $\delta$, describing response to a
symmetry-breaking perturbation, to examine quasi-long-range order.
(a) Sample fitting of the ansatz $m^\delta = \chi_mh$, with $\chi=40$ and
linear system size $2^{10}$, perturbations on different sublattices being
separated by relative angle $\pi$. (b),(c) As we increase the system size, the
fitted value of $\delta$ increases to an asymptotic value. Then, as we increase
the bond dimension, this value approaches the $4g-1$ we expect from TNR.
(d) When we vary the angle between sublattice perturbations, we find that
$\delta$ approaches the same value - quickly enough that the data from most
coarse-graining steps are indistinguishable -
for any relative angle except 0, where
the ansatz does not fit as well.\footnote{Note in particular that the amplitude
of the response does not ``blow up'' at $\phi=0$ as it may appear from
(e); for $m = \chi^{1/\delta}h^{1/\delta}$ and an anomalous value $\delta\sim0$,
a very large $\chi$ corresponds to a more tempered response.}
(e) Additionally, at angles where the ansatz
does fit, the response coefficient that we call $\chi_m$ is greatest 
when $\phi_A~-~\phi_B~=~\pi$ and falls off dramatically approaching
$\phi_A~-~\phi_B~=~0$.}
\label{fig:deltareg}
\end{figure}

\begin{figure}
\includegraphics[width=.5\textwidth]{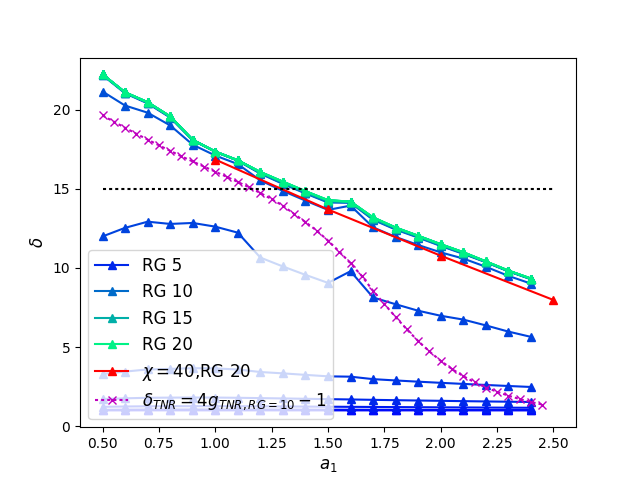}
\caption{We use HOTRG with $\chi=30$ to estimate the value of the critical
exponent $\delta$ on both sides of the KT transition, along the line
$a_2 = 0.5$. On the critical side,
its asymptotic value follows roughly what we would expect from the TNR data of
Fig.~\ref{fig:KTlines}, and approaches that value when we
increase $\chi$ to 40. On the AKLT side of the transition, however, we find
that the $\delta$ versus $a_1$ curve appears to remain approximately straight
rather than falling to a trivial value, and that corrections from increasing
the bond dimension only shift the line rather than changing its behavior.}
\label{fig:deltaline}
\end{figure}

At a point of the phase diagram we believe to be critical,
we may investigate another critical exponent:
Given a sufficiently small applied field $h$, we expect to find
magnetization $m$ such that \cite{Cardy}
\begin{align}
\label{eqn:delta}
h \propto m^\delta, && \delta = \frac{2-\Delta}{\Delta} = 4g-1.
\end{align}
We will implement the field $h$ by perturbing the wavefunction, simultaneously
applying $\exp(hS_{\phi_A})$
to every site of sublattice $A$ and $\exp(hS_{\phi_B})$
to every site of
sublattice $B$. Here $\phi_A$ and $\phi_B$ may in principle differ and the
difference $\phi_A - \phi_B$ may control the magnitude of the effect. Then
$m$ is the expectation value of $S_{\phi_A}$ at a site in sublattice $A$.
This changes the weight matrix of the doubled vertex model in a manner we
might expect from a magnetic field; as for the quantum model, though,

this reflects not one-site perturbations of the form $hS^{(i)}_\phi$ as we
might expect, but rather $h\{S^{(i)}_\phi,H_{i,j}\}$ obtained by deforming
the Hamiltonian as in \eqref{eqn:Ha}. It is, nonetheless, physical, and can
therefore be used to confirm quasi-long-range order of the quantum state.
When we investigate this quantity (using HOTRG),
we find in Fig.~\ref{fig:deltareg} that the
response of an order parameter to a ``magnetizing'' perturbation fits a
power-law curve as expected. From this curve we can obtain a value of $\delta$
through linear regression. Then, as we increase the system size, this value
rapidly converges.  When the perturbation
is at the same angle on both sublattices, the order parameter
responds less predictibly to perturbations. Otherwise it behaves in a
way indicative of
criticality, and as we increase bond dimension the limiting value of $\delta$
approaches the expected $4g-1$, with $g$ obtained through TNR.

We can also use $\delta$ to study the Kosterlitz-Thouless transition.
Following it as in Fig.~\ref{fig:deltaline},
we find that it appears to be smooth across the phase transition, approximating
the TNR value of $g$ on the critical side while taking a value of uncertain
significance on the AKLT side. We can use this to find the transition by looking
for $g_\delta=4$. Admittedly, this only indicates a transition if we accept
\textit{a priori} that the transition will occur at $g=4$.

\section{ The corner entropy}
\label{app:corner}

\begin{figure}
\includegraphics[width=0.4\textwidth]{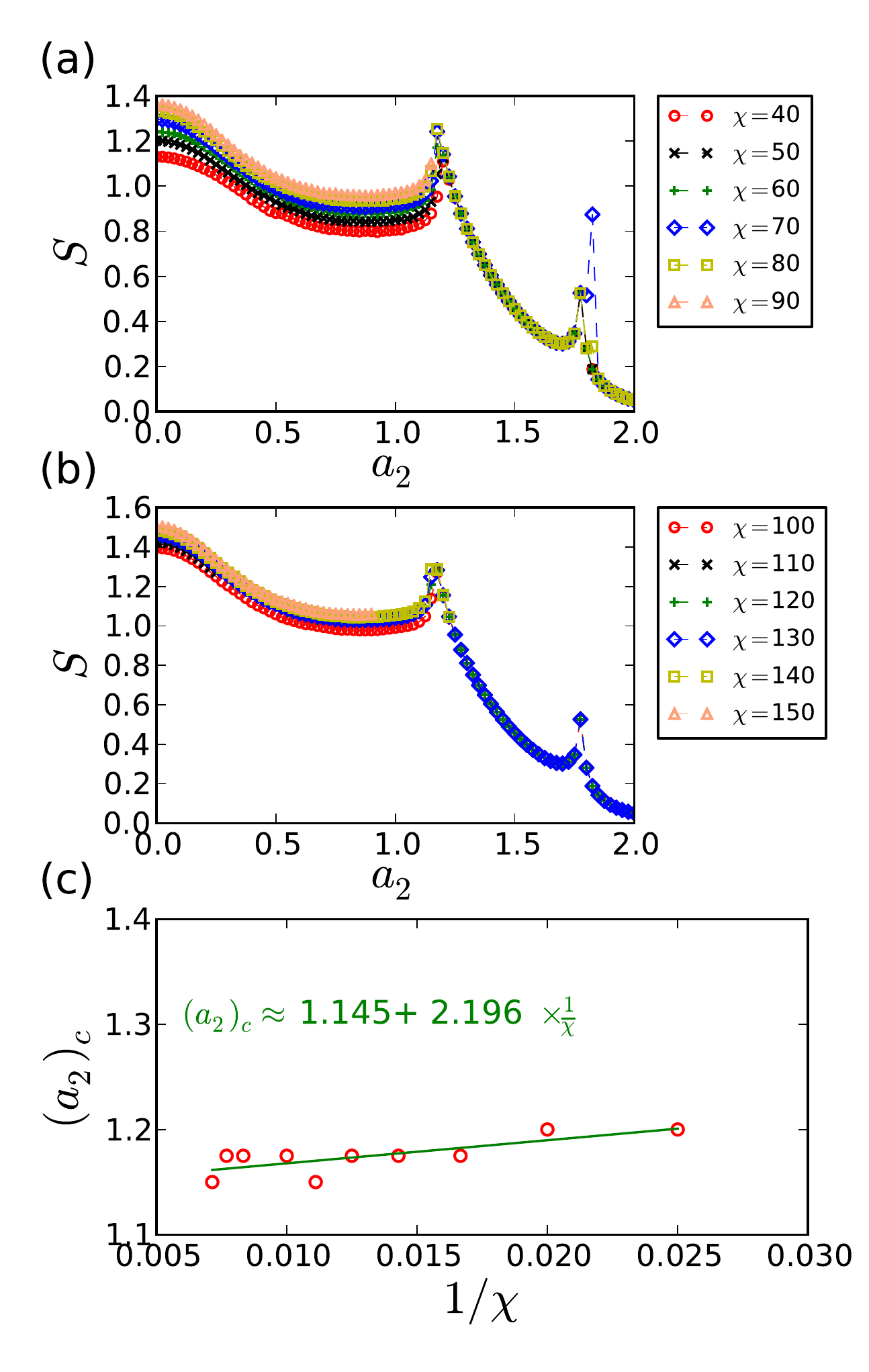}
\caption{
We extract the corner entropy for the norm of the deformed-AKLT state on the
square lattice as a function of the parameter $a_2$ on the $a_1=0.0$  axis,
varying the bond dimension (a) from 40 to 100  (b) from  110 to 150. 
(c)  We extrapolate the location $(a_2)_c$ of the KT transition on this line
in the limit of large bond dimension $\chi^{-1} \to 0$.
Extrapolation from a linear fit suggests that the critical point is at
$(a_2)_c \simeq 1.145$, a reasonable approximation of the TNR estimate
$(a_2)_c \simeq 1.10$. We note generally that both
critical points are fairly robust under increases in bond dimension, although
the corner entropy in the XY region generally keeps increasing
as the bond dimension increases, as expected since the true value
should not be finite.}
\label{fig:corner_entropy_a1_00}
\end{figure}

\begin{figure}
\includegraphics[width=0.4\textwidth]{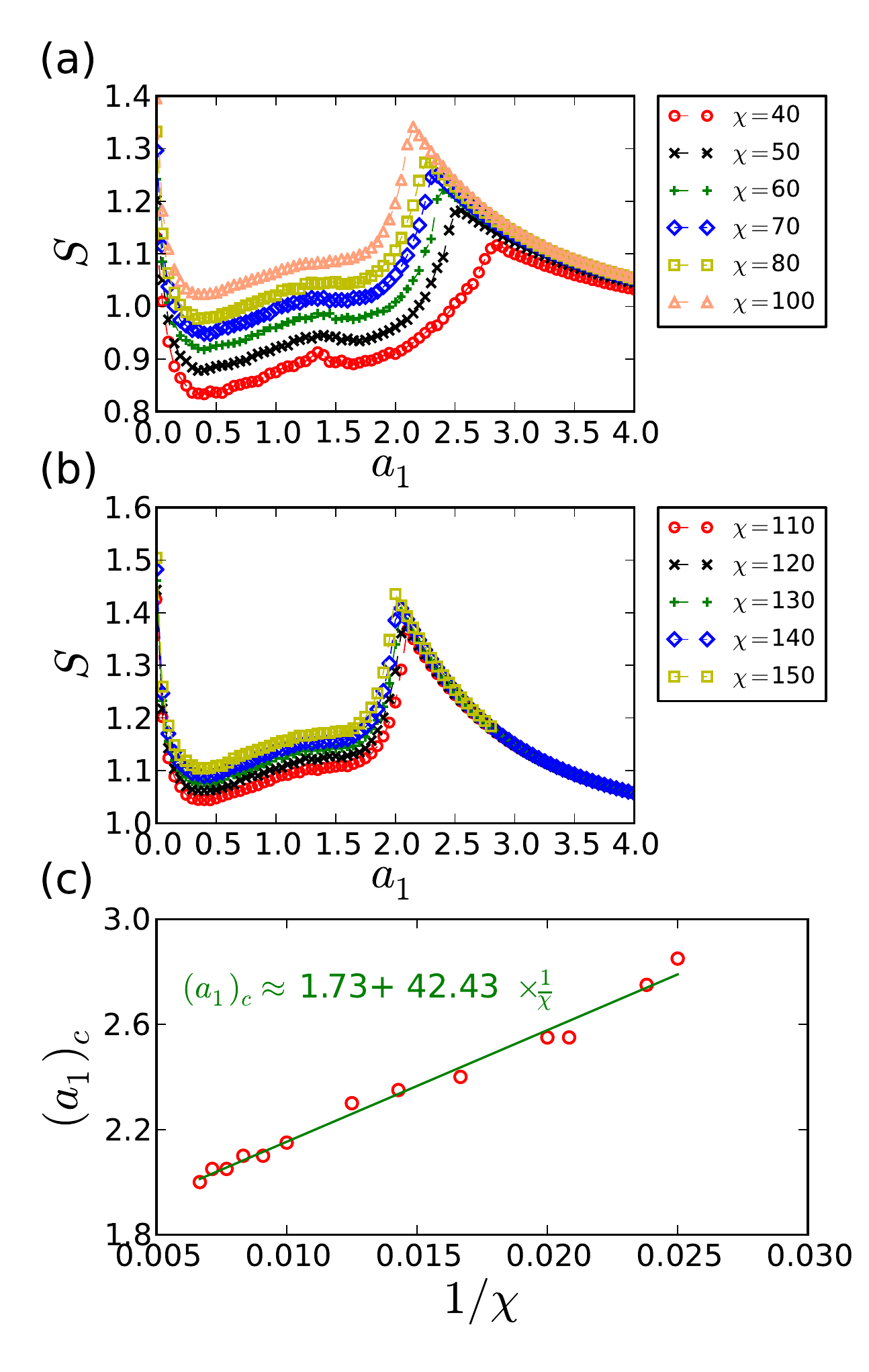}
\caption{
We extract the corner entropy as a function of the parameter
$a_1$ on the $a_2=0.0$  axis,
varying the bond dimension (a) from 40 to 100  (b) from  110 to 150. 
(c) We extrapolate the location $(a_1)_c$ of the KT transition along this line
in the limit of large bond dimension $\chi^{-1} \to 0$.
Extrapolating from a linear fit suggests that
the critical point is at $(a_1)_c \simeq 1.73$.
This is not a particularly reasonable approximation of the TNR estimate
$a_1 \simeq 1.30$;
in fact, we note that the estimates of $(a_1)_c$ using this method have
not begun to converge, even with high bond dimension $\chi=150$. As we do not
expect CTMRG to accurately approximate systems with the very large correlation
length observed in the pseudo-quasi-long-range ordered region, this is also not
surprising.}
\label{fig:corner_entropy_a2_00}
\end{figure}

We here use CTMRG to extract the corner entropy described in Appendix
\ref{app:CTMRG}.
We expect that this quantity will be singular at phase transitions and therefore
can be used to predict the boundary of phases. However, we note that, as
the doubled vertex model is critical throughout the XY region,
the corner entropy should diverge throughout this region. Since these estimates
are limited by bond dimension, they will not be precise at any point there.

\begin{figure*}
\includegraphics[width=0.9\textwidth]{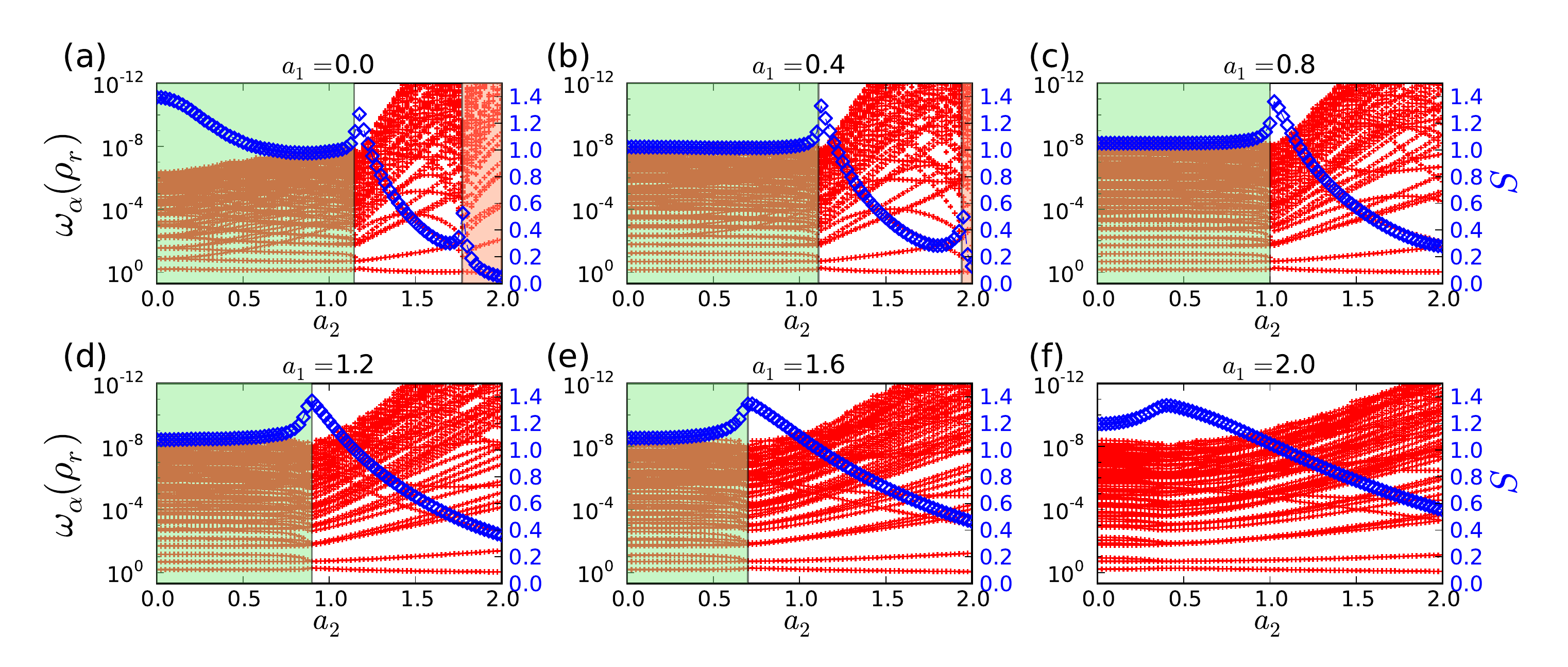}
\caption{
We plot the corner entanglement spectra $\omega_\alpha$ (red) and the
corner entropy (blue) for the norm of the square-lattice deformed-AKLT state as
a function of the parameter $a_2$  with 
(a) $a_1=0.0$,  (b) $a_1=0.4$, (c) $a_1=0.8$, (d) $a_1=1.2$, (e) $a_1=1.6$, and (f) $a_1=2.0$, obtained using CTMRG with bond dimension $\chi=100$. 
We find sharp phase transitions from the AKLT phase into the XY and
N\'eel-ordered phases. Close to the $a_2$ axis, in (a) and (b), we see
agreement with the phase boundaries obtained elsewhere. However, (d)-(f)
demonstrate that, in the pseudo-quasi-long-range ordered region, this method
does not accurately reflect the results obtained with TNR. This is somewhat
expected as the bond dimension of CTMRG effectively limits the correlation
length of systems it can effectively simulate.  }
\label{fig:corner_entropy}
\end{figure*}

In Fig.~\ref{fig:corner_entropy}, we see how we can extract the corner entropy
and use it to get
estimates for the transitions both into the XY and N\'eel-ordered phases.
Fig.~\ref{fig:corner_entropy_a1_00} demonstrates that these estimates are
fairly robust along the $a_2$ axis.
In Fig.~\ref{fig:corner_entropy_a2_00}, however, we see by increasing bond
dimension that the phase boundary obtained through this method is much less
robust in the pseudo-quasi-long-range ordered region.  In
particular we suggest that the boundary at some fixed bond dimension may
roughly demark the pseudo-quasi-long-range ordered region.

\FloatBarrier
\section{Operator content of critical points}
\label{app:tower}

\begin{figure*}
\includegraphics[width=.9\textwidth]{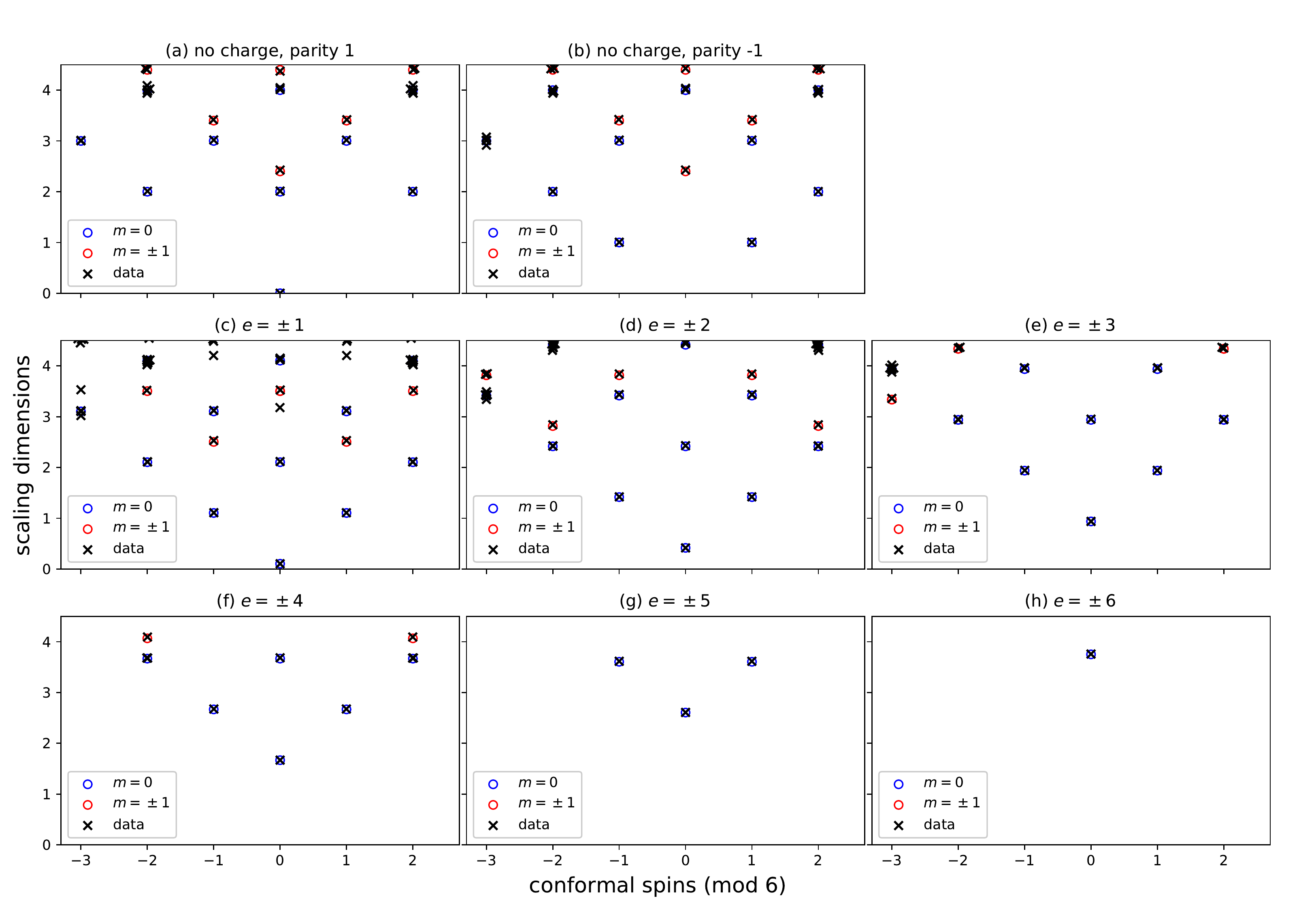}
\caption{
The conformal tower at $a_1=.8,a_2=0$, estimated from TNR at
$\cchip{26}{16}$, after the twelfth coarse-graining step. We estimate
$g = 4.80 \pm 0.01$.}
\label{fig:XYtower80}
\end{figure*}

\begin{figure*}
\includegraphics[width=.9\textwidth]{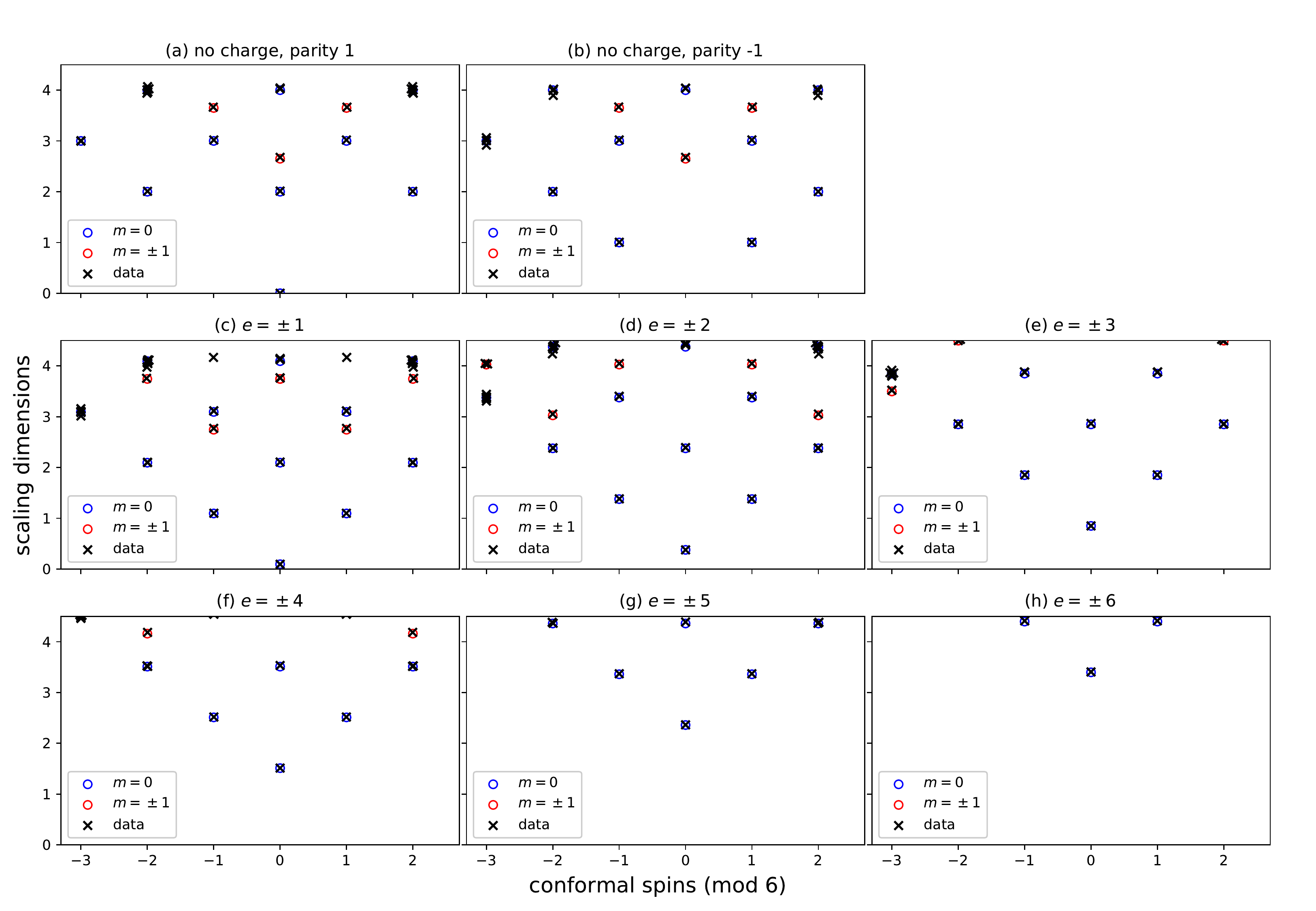}
\caption{The conformal tower at $a_1=.5,a_2=.3$, estimated from TNR at
$\cchip{26}{16}$, after the twelfth coarse-graining step.
We estimate $g = 5.30 \pm 0.01$.}
\label{fig:XYtower53}
\end{figure*}

When we find using TNR that a 2D classical theory reaches approximate scale
invariance under coarse-graining, we can estimate the operator content of the
CFT that may arise in the infrared limit.  In Figs.~\ref{fig:XYtower80} and
\ref{fig:XYtower53}, we use data from TNR with $\cchip{26}{16}$ after the
twelfth coarse-graining step. Data (black
$\times$s) is extracted by diagonalizing a transfer matrix of three coarse-grained
sites; the range of conformal spins is increased by approximating a 6-site
transfer matrix as in Appendix B of Hauru et al.\cite{TNRHauru}
Since the latter approximation
yields much less accuracy for scaling dimensions, we attempt to plot scaling
dimensions from 3-site data with conformal spins from the 6-site approximation,
but these datasets are occasionally mismatched.

Preserving $D_{2N}$ symmetry allows us to separately diagonalize blocks of
the transfer matrix which correspond to irreducible representations of
the on-site $O(2)$ symmetry.\footnote{The 2D irreps of
both $O(2)$ and $D_{2N}$ are indexed by $k$ (which corresponds to the absolute
value of $U(1)$ charge). An object that transforms under the representation $k$
of $O(2)$ for $k<N/2$ also transforms under the representation $k$ of its
$D_{2N}$ subgroup. We primarily use
$N=40$, where a general absence of $k>15$ irreps tells us that we should
expect no $k>15$ irreps of $O(2)$, and in particular that the $D_{2N}$ irreps
which we do observe correspond exactly to the $O(2)$ irreps with matching $k$.}
We then construct separate plots, for each irrep, of the scaling dimension and
conformal spin of scaling operators that transform under it (excluding
$k>6$ irreps, for which the data would not appear within the axes we have
chosen).

We also use the
smallest scaling dimension of charge $k=1$ to estimate $g$, as expressed in
Fig.~\ref{fig:XYcompare-interior}, and from there, we
plot our estimates of the conformal tower based on that value of $g$:
Blue circles correspond to scaling operators with no ``magnetic'' charge,
whereas red circles correspond to scaling operators with magnetic charge
$m = \pm 1$ (scaling operators with $|\!m\!|>1$ should have
$\Delta \geq 8$, and therefore should lie outside the range of these plots).
In doing so we find excellent agreement from the data:
In (a), we plot data from states that transform under the trivial
representation of $O(2)$ and find the descendants of the identity and of
the first ``vortex'' operator $V_{0,1} + V_{0,-1}$. In (b), we plot data from
states which have no $U(1)$ charge but have odd parity under $O(2)$ reflections,
and see the $\partial,\bar\partial$ operators and their descendants, as well as
the vortex operator $V_{0,1} - V_{0,-1}$. In (c)-(h), we see the
``electric'' operators $V_{\pm e,0}$ as well as the operators
$V_{\pm e,\pm 1}$ which have both ``electric'' and ``magnetic'' charge.

\makeatletter\onecolumngrid@push\makeatother

\makeatletter\onecolumngrid@pop\makeatother

\bibliography{bibs}	

\end{document}